\documentclass[aps,prb,twocolumn,groupedaddress]{revtex4-2}

\usepackage[T1]{fontenc}
\usepackage[utf8]{inputenc}
\usepackage{lmodern}
\usepackage{setspace}
\usepackage[english]{babel}

\makeatletter
\@ifundefined{l@en}{\expandafter\let\csname l@en\endcsname\l@english}{}
\@ifundefined{captionsen}{\expandafter\let\csname captionsen\endcsname\captionsenglish}{}
\@ifundefined{dateen}{\expandafter\let\csname dateen\endcsname\dateenglish}{}
\@ifundefined{extrasen}{\expandafter\let\csname extrasen\endcsname\extrasenglish}{}
\@ifundefined{noextrasen}{\expandafter\let\csname noextrasen\endcsname\noextrasenglish}{}
\makeatother

\usepackage{amsmath,amssymb,amsthm,mathtools}
\usepackage{amsfonts,amsthm,bm} 

\usepackage{graphicx}
\usepackage{xcolor}
\usepackage[normalem]{ulem}
\usepackage{tikz}
\usepackage{booktabs}
\usepackage{siunitx}
\usepackage{braket}
\usetikzlibrary{arrows.meta,calc,positioning}
\usepackage{macros}

\usepackage{listings}
\usepackage[hidelinks]{hyperref}
\usepackage[nameinlink,noabbrev]{cleveref}

\newcommand*\diff{\mathop{}\!\mathrm{d}}

\newcommand{\pSigma}{\hat \Sigma}
\newcommand{\pG}{\hat{\mathcal G}}

\begin{document}

\title{Accelerating a Strong-Coupling Non-Equilibrium Steady-State Impurity Solver using (Quantics) Tensor Trains}

\author{Bastian Schindler}
\email{bastian.schindler@uni-hamburg.de}
\affiliation{I. Institute for Theoretical Physics, University of Hamburg, Notkestraße 9-11, 22607 Hamburg, Germany}
\affiliation{The Hamburg Centre for Ultrafast Imaging, Hamburg, Germany}
\author{Martin Eckstein}
\affiliation{I. Institute for Theoretical Physics, University of Hamburg, Notkestraße 9-11, 22607 Hamburg, Germany}
\affiliation{The Hamburg Centre for Ultrafast Imaging, Hamburg, Germany}

\date{August 13, 2026}

\begin{abstract}
Including higher order diagrammatic corrections to the strong-coupling expansion is mainly limited by the evaluation of high-dimensional, time-ordered integrals. In this work we present and compare four different parametrizations of the integrands in order to obtain a low-rank (quantics) tensor-train representation using tensor cross interpolation. Particular emphasis is placed on a quantics time-difference formulation in which the required retarded convolutions are performed directly in quantics tensor-train form. Using controlled Gaussian benchmarks, we analyze the accuracy, bond dimensions, and computational scaling of the different approaches. We then validate the most promising formulations in self-consistent equilibrium and nonequilibrium DMFT calculations and demonstrate calculations up to the third order in the strong-coupling expansion. Finally, we extend the solver to impurity models with retarded density-density interactions and apply it within nonequilibrium extended DMFT. Our results show that tensor cross interpolation substantially reduces the cost of evaluating higher-order diagrams and provides a controlled, systematically improvable framework for nonequilibrium quantum impurity calculations.
\end{abstract}

\maketitle

\section{Introduction}

Quantum impurity models play a central role in the theory of strongly correlated electron systems. Originally introduced to describe magnetic impurities in metals \cite{Anderson-1961}, they became one of the key building blocks of modern many-body theory with the development of dynamical mean-field theory (DMFT), where an interacting lattice model is mapped onto a self-consistent quantum impurity problem \cite{Georges-etal-1996b}.  In equilibrium, impurity problems can be solved by a variety of complementary techniques. Hamiltonian-based approaches, such as exact diagonalization, 
numerical renormalization group (NRG), and density matrix renormalization group (DMRG)
\cite{Georges-etal-1996b,Lu-Haverkort-2017,Wilson-1975,Bulla-Costi-Pruschke-2008,Wolf2015,Bauernfeind2017}, 
provide direct access to real-frequency quantities but rely on a bath discretization. Diagrammatic methods, particularly continuous-time quantum Monte Carlo (QMC) 
\cite{Gull2011,Werner-etal-2006,Rubtsov2005}, 
instead treat continuum baths and provide numerically exact solutions up to statistical errors. Out of equilibrium, the situation is much more challenging. Real-time QMC suffers from the dynamical sign problem \cite{Muehlbacher2008,Werner2009}, while approaches based on a Hamiltonian  formulation  are more strongly limited by finite system size effects \cite{Gramsch-etal-2013,Wolf2014}. Although promising alternatives such as the auxiliary master equation approach \cite{Arrigoni-Knap-VonDerLinden-2013} and influence-functional methods \cite{Chen2024,Thoenniss2023,Nayak2025} have been developed, no general-purpose impurity solver comparable to equilibrium CT-QMC is available. Consequently, nonequilibrium DMFT simulations \cite{Aoki-etal-2014,Murakami2025} largely rely on perturbative impurity solvers.

A particularly versatile approach is the self-consistent strong-coupling expansion, which partially resums the hybridization expansion through a self-consistent Dyson equation \cite{Keiter-Kimball-1970,Coleman-1984,Bickers1987}. In equilibrium, its rapid convergence has enabled vertex-resummation techniques \cite{Haule2001,Kim2023,Kim2022}, a  diagrammatic Monte Carlo formulation \cite{Haule-2023}, and efficient intermediate-order implementations based on the discrete Lehmann representation \cite{Kaye-Strand-Wentzell-2024,Kaye-Chen-Parcollet-2022,Huang-etal-2025,Kaye-etal-2024}. In contrast, nonequilibrium applications remain largely restricted to the first order (noncrossing approximation, NCA) and second order (one-crossing approximation,  OCA) \cite{Eckstein-Werner-2010}, with few exceptions at short simulation times \cite{Eckstein2011}. The self-consistent diagrammatic Monte Carlo (``inchworm'') approach \cite{Cohen-etal-2015,Erpenbeck-Gull-Cohen-2023,Erpenbeck-etal-2024} alleviates this problem, but it has so far remained too computationally demanding for use as a general impurity solver in self-consistent real-time nonequilibrium DMFT.

The principal computational bottleneck of high-order strong-coupling methods is the evaluation of high-dimensional time-ordered integrals. In recent years, tensor-network techniques originating from numerical mathematics \cite{Savostyanov-2014, Oseledets-2011, Oseledets-Tyrtyshnikov-2010} have emerged as powerful tools for addressing this challenge. In particular, tensor cross interpolation (TCI) efficiently constructs low-rank approximations of the high-dimensional tensors arising in diagrammatic many-body calculations \cite{NúñezFernández-etal-2022,Dolgov-Savostyanov-2020}. Combined with the quantics tensor-train (QTT) representation \cite{Khoromskij-2011,Fernández-etal-2025,Shinaoka-etal-2023}, which factorizes tensor indices into their binary degrees of freedom, TCI has enabled highly compressed representations of vertex functions and nonequilibrium Green's functions on the Keldysh contour \cite{Frankenbach-etal-2025,Rohshap-etal-2025,Murray-Shinaoka-Werner-2024,Środa-etal-2025}.

For the solution of quantum impurity models, TCI-based techniques have been applied to imaginary-time-ordered integrations in the weak and strong coupling expansion \cite{Erpenbeck-etal-2023,Matsuura-etal-2025}, and for the bare weak-coupling expansion on the Keldysh contour \cite{Matsuura2026}.
For the nonequilibrium strong-coupling expansion, tensor-network approaches have so far focused on nonequilibrium steady states (NESS), where time-translational invariance reduces the complexity of the problem while still describing a wide range of applications, including quantum transport and quasi-steady photoexcited states \cite{Li-Eckstein-2021,Li2020,Ray2023}. A useful technical step  is the Keldysh parametrization, which expands the contour integrations as real-time integrations supplemented by a summation over Keldysh indices \cite{Eckstein-2024}. Although this introduces an additional summation over contour configurations, it typically leads to substantially lower tensor-train bond dimensions. 
The resulting convolution structure can be evaluated  using fast Fourier transforms (FFT) \cite{Eckstein-2024}, or QTT techniques can be used to construct the self-energies directly in the frequency domain \cite{Kim-Werner-2025,Geng-Kim-Werner-2025}.

The efficiency of tensor-network algorithms depends critically on the chosen parametrization, and both the FFT-based algorithm of Ref.~\onlinecite{Eckstein-2024} and the frequency-domain QTT approach of Refs.~\onlinecite{Kim-Werner-2025,Geng-Kim-Werner-2025} have distinct advantages. In this work, we introduce a real-time QTT formulation in which the convolution operations arising from the Keldysh parametrization are represented directly in the tensor-train language. This combines the favorable compression properties of QTT 
for dense grids and long times
with the 
real-time formulation of Ref.~\onlinecite{Eckstein-2024}, eliminating the need for 
Fourier
transforms. We systematically compare the efficiency of four quasi-linear algorithms for evaluating the time-ordered integrals of the nonequilibrium strong-coupling expansion, including the FFT-based approach of Ref.~\onlinecite{Eckstein-2024} and the direct convolutional QTT formulation introduced here. Building on the resulting formulation, we construct a third-order nonequilibrium impurity solver for retarded density-density 
interactions,
which constitutes
the central ingredient of nonequilibrium EDMFT \cite{Golez-Eckstein-Werner-2015} and GW+EDMFT \cite{Golez2017}.
Previous studies 
of impurity models with retarded interactions  %
based on different low-order approximations \cite{Werner-Eckstein-2013,Golez-Eckstein-Werner-2015} have reached qualitatively different conclusions \cite{Paprotzki-Eckstein-2025,Chen-etal-2016}, underscoring the need for systematically improvable impurity solvers.

The remainder of this paper is organized as follows. Section~\ref{ch:StrongCoupling} reviews the strong-coupling expansion and the steady-state Keldysh formalism. Section~\ref{ch:TensorTrains} introduces the tensor-train and quantics tensor-train representations together with tensor cross interpolation. Section~\ref{sec:decomps} develops four quasi-linear tensor representations for the strong-coupling diagrams, followed by benchmarks for model Gaussian integrals in Sec.~\ref{ch:Gaussian}. The most promising approaches are then validated within equilibrium DMFT (Sec.~\ref{ch:BetheEquil}) and subsequently applied to photodoped nonequilibrium steady states and to nonequilibrium EDMFT with retarded interactions (Secs.~\ref{ch:BetheTFLA} and \ref{ch:EDMFT}).

\section{Formalism}

This  section provides the technical background for the remainder of the paper: We recapitulate  the self-consistent strong-coupling expansion for  quantum impurity problems, and the use of 
tensor-cross interpolation for the solution of high-dimensional integrals. 

\subsection{Strong-Coupling Expansion}
\label{ch:StrongCoupling}

\subsubsection{Impurity action}

We consider a general impurity model, described by some action $S$.  The action will be generally formulated on the general L-shaped Keldysh contour $\mathcal C$, ranging from $t_{\rm min}^+$ on the forward real-time branch to some maximum time $t_{\rm max}^+$, back from $t_{\rm max}^-$ to $t_{\rm min}^-$ on the backward real time branch, and to $t_{\rm min}-i\beta$ in imaginary time. (See, e.g., Ref.~\cite{Aoki-etal-2014} for an introduction to the Keldysh formalism; $t^+$ and $t^-$ will denote times on the forward and backward real-time branch of $\mathcal C$, respectively.) The contour $\mathcal C$ will later be  reduced to the two branch contour to describe nonequilibrium steady states. Given $S$, expectation values and correlation functions are given in terms of a contour-ordered  trace
\begin{equation}
  \label{eq:time_ordered_trace}
 \langle \mathcal{T}_\mathcal{C} A(t) B(t') \rangle = \frac{1}{Z} \text{tr} ( \mathcal{T}_\mathcal{C}  e^{iS} A(t) B(t') ),
\end{equation}
with $Z = \text{tr} ( \mathcal{T}_\mathcal{C} e^{iS})$ and the contour time ordering  $\mathcal{T}_\mathcal{C} $. We decompose the action as $S=S_{\rm loc}+S_{\rm nloc}$, with
\begin{align}
  S_\text{loc} &= - \int_\mathcal{C} \diff t\, \hat H_\text{loc} 
    \label{Sloc}
\\
  S_\text{nloc} &= - \sum_{\alpha, \beta} \int_\mathcal{C} \diff t \diff t' \,\hat v_{\alpha} \Delta_{\alpha, \beta} (t,t') \hat w_\beta.
  \label{Snloc}
\end{align}
Here $H_\text{loc}$ is the Hamiltonian of the isolated impurity site, and $  S_\text{nloc}$ describes a coupling to the environment via pairs of operators $(\hat v_{\alpha},\hat w_{\beta})$ and a time-nonlocal function $ \Delta_{\alpha, \beta}(t,t')$. The latter has the symmetry of a bosonic (fermionic) correlation function if the operators $\hat v_\alpha$ and $\hat w_\beta$ contain an even (odd) number of Fermi operators.
 
For illustration we can consider the single-orbital Anderson model, where the local impurity Fock space is spanned by four states $\{\ket{0}, \ket{\uparrow}, \ket{\downarrow}, \ket{\uparrow \downarrow}\}$, and $\hat H_\text{loc} = U \hat n_{\uparrow}\hat n_{\downarrow} - \mu (\hat n_{\uparrow}+\hat n_{\downarrow} )$. The coupling to the environment is described by the pairs $(\hat v_{\alpha},\hat w_{\beta}) \equiv (c_\sigma^\dagger,c_\sigma)$ for $\sigma=\uparrow,\downarrow$, 
and
$ \Delta_{\alpha, \beta}\equiv \Delta_\sigma$ is the hybridization function which arises from integrating out a fermion reservoir. Within the EDMFT solution of the extended Hubbard model (Sec.~\ref{ch:EDMFT}), an additional nonlocal density-density interaction channel appears in  $S_{\rm nloc}$, where $\hat v =\hat w = n_\uparrow + n_\downarrow$, and $\Delta_{\rm nn}(t,t')$ corresponds to a retarded density-density interaction. 

\subsubsection{Pseudo-particle propagators and self-energies}

The strong-coupling expansion
 \cite{Keiter-Kimball-1970,Coleman-1984,Bickers1987,Eckstein-Werner-2010} 
 is a systematic expansion in the nonlocal part \eqref{Snloc} of the action. Below we summarize the equations needed for the subsequent discussion of the computational approach; 
a more detailed derivation is found in Refs.~\cite{Eckstein-Werner-2010} and \cite{Aoki-etal-2014}.

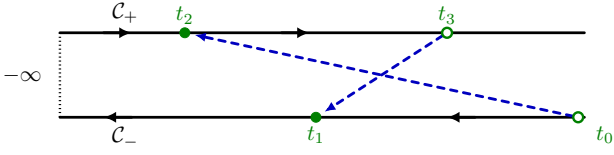
\begin{figure}[tbp]
  \centering
  \resizebox{\columnwidth}{!}{\begin{tikzpicture}[
  font=\normalsize,
  >={Latex[length=2mm,width=1.8mm]},
  contour/.style={
    line width=1.25pt,
    line cap=round,
    line join=round
  },
  arr/.style={
    contour,
    -{Latex[length=2mm,width=1.8mm]}
  },
  dot/.style={
    circle,
    draw=green!50!black,
    fill=green!50!black,
    inner sep=0pt,
    minimum size=4.5pt
  },
  opendot/.style={
    circle,
    draw=green!50!black,
    fill=white,
    line width=1.1pt,
    inner sep=0pt,
    minimum size=4.5pt
  },
  lab/.style={
    inner sep=1pt
  },
  zlabel/.style={
    text=green!50!black,
    inner sep=1pt
  },
  vdot/.style={
    line width=1.1pt,
    dash pattern=on 0.5pt off 1.3pt
  },
  connection/.style={
    blue!75!black,
    line width=1.3pt,
    dashed,
    line cap=round,
    -{Latex[length=1.8mm,width=1.5mm]},
    shorten <=2.5pt,
    shorten >=4.5pt
  }
]

\def\xL{0}
\def\xR{8.4}
\def\yU{1.35}
\def\yL{0}

\def\xthree{2.0}
\def\xtwo{4.1}
\def\xone{6.2}
\def\xzero{8.3}

\path[use as bounding box]
  (-1.20,-1.35) rectangle (9.15,2.35);

\draw[contour]
  (\xL,\yU) -- (\xR,\yU);

\draw[contour]
  (\xL,\yL) -- (\xR,\yL);

\draw[vdot]
  (\xL,\yU) -- (\xL,\yL);

\draw[arr]
  (0.70,\yU) -- (1.15,\yU);

\draw[arr]
  (3.55,\yU) -- (4.00,\yU);

\draw[arr]
  (1.15,\yL) -- (0.70,\yL);

\draw[arr]
  (6.65,\yL) -- (6.20,\yL);

\node[lab,anchor=east]
  at (\xL-0.25,{(\yU+\yL)/2})
  {$-\infty$};

\node[lab]
  at (1.05,\yU+0.32)
  {$\mathcal{C}_{+}$};

\node[lab]
  at (1.05,\yL-0.32)
  {$\mathcal{C}_{-}$};

\coordinate (t2c) at (\xthree,\yU);
\coordinate (t1c) at (\xtwo,\yL);
\coordinate (t3c) at (\xone,\yU);
\coordinate (t0c) at (\xzero,\yL);

\draw[connection]
  (t0c) -- (t2c);

\draw[connection]
  (t3c) -- (t1c);

\node[dot] (t2) at (t2c) {};
\node[dot] (t1) at (t1c) {};
\node[opendot] (t3) at (t3c) {};
\node[opendot] (t0) at (t0c) {};

\node[
  zlabel,
  anchor=south,
  yshift=3pt
] at (t2) {$t_2$};

\node[
  zlabel,
  anchor=north,
  yshift=-3pt
] at (t1) {$t_1$};

\node[
  zlabel,
  anchor=south,
  yshift=3pt
] at (t3) {$t_3$};

\node[
  zlabel,
  anchor=north west,
  xshift=7pt,
  yshift=-3pt
] at (t0) {$t_0$};

\end{tikzpicture}}
  \caption{Steady-state Keldysh contour and an example OCA self-energy diagram. The imaginary-time branch has been omitted and the initial time has been shifted to $t_{\rm min}\to-\infty$, where the forward and backward real-time branches $\mathcal C_+$ and $\mathcal C_-$ are joined as indicated by the dotted vertical line. The branch arrows show the contour orientation, and the marked arguments are in cyclic order $t_3\succ t_2\succ t_1\succ t_0$. The contour interval $[t_2\succ t_1]$ crosses the 
  initial point $t_{\rm min}$; hence $\pG(t_2,t_1)=\pG^<(t_2-t_1)$. The intervals $[t_3\succ t_2]$ and $[t_1\succ t_0]$ do not cross $t_{\rm min}$ and give $\pG^>(t_3-t_2)$ and $\pG^>(t_1-t_0)$, respectively; see Eq.~\eqref{pGlesgtr}. Directed hybridization events $\Delta_i$ are shown as dashed blue arrows, with open green starting vertices and filled green endpoints.}
  \label{fig:steady_state_diagram}
\end{figure}

To formulate the strong-coupling equations on the Keldysh contour, it is convenient to regard $\mathcal C$ as a closed cyclic contour by formally identifying its endpoint, $t_{\rm min}-i\beta$, with its starting point, $t_{\rm min}^+$. Correspondingly, we introduce a cyclic ordering of contour times: $t_3\succ t_2\succ t_1$ indicates that the sequence $(t_3,t_2,t_1)$ appears 
anti-clockwise from $t_3$ to $t_1$ along the oriented contour
(see Fig.~\ref{fig:steady_state_diagram}), $[t\succ t']$ denotes the contour interval containing all times $\bar t$ with $t\succ \bar t\succ t'$, and
\begin{align}
(A \circledast B)(t,t')=\int_{\bar t\in[t\succ t']} d\bar t\, A(t,\bar t)B(\bar t,t')
\end{align}
defines a cyclic convolution of two contour functions.

The central object of the strong-coupling expansion is the resolvent operator $\pG(t,t')$, a matrix in the impurity Hilbert space which represents a dressed evolution operator along the contour segment $[t\succ t']$. Following the terminology of the auxiliary-particle formulation of the strong-coupling expansion \cite{Coleman-1984}, we will also refer to $\pG$ as the pseudo-particle propagator. The pseudo-particle propagator is related to the reduced impurity density matrix through the relation
\begin{align}
\label{wnsskszmain}
\hat \rho(t)=i\hat \xi \pG(t^+,t^-),
\end{align}
with $\mathrm{Tr}[\hat\rho(t)]=1$, which provides access to all local expectation values. Here $\hat\xi$ is the fermion-parity operator, whose appearance is partly a matter of convention in the precise definition of $\pG$ \cite{Eckstein-Werner-2010,Aoki-etal-2014}. The bare propagator $\pG_0$ is proportional to the evolution operator of the isolated impurity, and corrections to $\pG_0$ are collected in the pseudo-particle self-energy $\pSigma(t,t')$, from which $\pG$ is recovered from the Dyson equation
\begin{equation}
\pG
=\pG_0+\pG_0\circledast\pSigma\circledast\pG
=\pG_0+\pG\circledast\pSigma\circledast\pG_0.
\label{Dyson}
\end{equation}
Conversely, $\pSigma$ is obtained from $\pG$ through the diagrammatic expressions $\pSigma[\pG]$  discussed below. Apart from the cyclic ordering of the convolution integrals, the  self-consistent solution of the integral equation \eqref{Dyson} together with the evaluation of the diagrammatic functional $\pSigma[\pG]$ is mathematically analogous to conventional self-consistent weak-coupling perturbation theory.

The main numerical effort lies in the evaluation of the self-energy $\pSigma$. The diagrammatic rules are as follows (an example is shown in 
Fig.~\ref{fig:steady_state_diagram}):
An $n$th-order diagram $\mathcal D$ for $\pSigma(t,t')$ consists of a sequence of $2n-1$ solid lines ($\pG$-lines), connecting the cyclically ordered contour times
\begin{align}
\label{wnsskorder}
t=t_{2n-1}\succ t_{2n-2}\succ\cdots\succ t_0=t'.
\end{align}
Pairs of contour times are connected by directed hybridization lines $\Delta$ (dashed lines),  such that the diagram is $\pG$-irreducible and skeleton-like, i.e., no $\pG$-line contains a self-energy insertion. Each  hybridization line carries an interaction flavor $(\alpha,\beta)$, and we associate the operators $\hat v_\alpha$ and $\hat w_\beta$ with the endpoint and starting point of a hybridization line $\Delta_{\alpha,\beta}$, respectively. The corresponding analytical expression is
\begin{equation}
\label{eq:selfenergy_diagram}
\pSigma^{(\mathcal D)}(t,t')
=
C^{(\mathcal D)}
\int_{\mathcal C^\succ}
\diff\bm t\,
F^{(\mathcal D)}(\bm t)
W^{(\mathcal D)}(\bm t),
\end{equation}
where $\bm t=(t_0,\ldots,t_{2n-1})$, and $\int_{\mathcal C^\succ} \diff\bm t\,$ denotes integration over the internal times $t_1,\ldots,t_{2n-2}$ with the ordering \eqref{wnsskorder}. The function
\begin{equation}
F^{(\mathcal D)}(\bm t)
=
\prod_{j=1}^{n}
\Delta_{\alpha_j,\beta_j}(t_{v_j},t_{w_j})
\label{mssllssss}
\end{equation}
collects the hybridization lines, where we denote by $w_j,v_j\in\{0,\ldots,2n-1\}$ the starting- and end time-indices of the $j$th hybridization line $\Delta_{\alpha_j,\beta_j}$. The factor
\begin{align}
W^{(\mathcal D)}(\bm t)
=
\hat\psi_{2n-1}^{(\mathcal D)}
\pG(t_{2n-1},&t_{2n-2})
\hat\psi_{2n-2}^{(\mathcal D)}
\,\,\,\cdots
\nonumber\\
&\cdots\,\,\,
\hat\psi_1^{(\mathcal D)}
\pG(t_1,t_0)
\hat\psi_0^{(\mathcal D)}
\end{align}
is a time-ordered matrix product of the propagators $\pG$ and the vertex  operators connected to the hybridization lines ($\hat\psi^{(\mathcal D)}_{v_j}=\hat v_{\alpha_j}$ and $\hat\psi^{(\mathcal D)}_{w_j}=\hat w_{\beta_j}$). The overall prefactor is $C^{(\mathcal D)}=(-1)^{f_\mathcal D}i^n$, where $f_\mathcal D$ is the number of crossings between fermionic hybridization lines plus the number of fermionic hybridization lines whose direction is opposite to that of the $\pG$ propagator. The $n$th-order contribution to the self-energy is obtained by summing over all corresponding diagram topologies, and each assignment of flavors $(\alpha,\beta)$ to the hybridization lines. 

In Fig.~\ref{fig:steady_state_diagram} 
we exemplarily show an OCA diagram
containing two directed hybridization lines. Their starting and endpoints determine both the associated vertex operators and the greater or lesser component of each hybridization function.
An open green circle marks the starting $w$-vertex and a filled green circle the endpoint $v$-vertex.
 The resulting contribution to the self-energy is
\begin{align}
  \label{eq:oca_selfenergy_diagram}
  &\pSigma^{(\mathcal{D})}(t_3, t_0)
  =C^{(\mathcal{D})}
  \int_{\bm{t}\succ}\mathrm{d}\bm{t}\;
  \Delta^{<}_{\alpha_1,\beta_1}(t_2, t_0)
  \Delta^{>}_{\alpha_2,\beta_2}(t_1, t_3)
 \nonumber\\
  &{}\times
  \hat w_{\beta_2}\pG^>(t_3, t_2)
  \hat v_{\alpha_1}\pG^<(t_2, t_1)
  \hat v_{\alpha_2}\pG^>(t_1, t_0)
  \hat w_{\beta_1}.
\end{align}
  The line $t_0\to t_2$ therefore represents
  $\Delta^{<}_{\alpha_1,\beta_1}(t_2,t_0)$, whereas the oppositely directed   line $t_3\to t_1$ represents   $\Delta^{>}_{\alpha_2,\beta_2}(t_1,t_3)$.

\subsubsection{Steady-state formulation}

In this work, we focus on nonequilibrium steady states (NESS). Such states are obtained by taking the initial time to the remote past, $t_{\rm min}\to-\infty$. Assuming that the transient dynamics on the impurity is damped by the coupling to the environment, all correlation functions become time-translationally invariant. The imaginary-time branch of the Keldysh contour can then be eliminated, leaving only the forward and backward real-time branches \cite{Eckstein-2024, Li-Eckstein-2021, Kamenev-2023}. The pseudo-particle propagator, and analogously $\pSigma$, can then be reconstructed from their greater and lesser components, where the lesser component corresponds to the case in which the contour interval $[t\succ t']$ wraps around the initial time $t_{\rm min}$ (Fig.~\ref{fig:steady_state_diagram}),
\begin{equation}
\label{pGlesgtr}
  \pG(t,t')=
  \begin{cases}
    \pG^>(t-t'), & \text{if } t_{\rm min}^+\notin [t\succ t'],\\
    \pG^<(t-t'), & \text{if } t_{\rm min}^+\in [t\succ t'].
  \end{cases}
\end{equation}
In the numerical implementation, we fix one time argument to $0^-$ and evaluate $\pSigma(t,0^-)$ for $t\in[0^+\succ0^-]$ along the closed contour, i.e., for physical times $t\le0$, so that
\begin{align}
\pSigma^>(t)=\pSigma(t^-,0^-),\qquad
\pSigma^<(t)=\pSigma(t^+,0^-).
\end{align}
The corresponding functions for $t>0$ are obtained from the Hermitian symmetry $\pSigma^{>,<}(t)= -\pSigma^{>,<}(-t)^\dagger$.

The Dyson equation \eqref{Dyson} can likewise be reformulated as a closed set of equations for $\pG^{>}$ and $\pG^{<}$ in terms of $\pSigma^{>}$ and $\pG_0$. 
 Since the cyclic convolution integrals for the lesser components extend to $-\infty$, it can be helpful to stabilize the numerical solution of the 
 steady-state %
 Dyson equation by introducing a small exponential damping factor $e^{-\eta|t-t'|}$ into the convolution integrals, in particular if the propagators $\pG$ decay slowly in time. Convergence is then verified by extrapolating to the limit $\eta\to0$ (see App.~\ref{app:dyson} for details). 
 The resulting self-consistent equations $\pSigma$ and $\pG$  are solved by fixed-point iteration: Starting from an initial guess $\pG^{(0)}=\pG_0$, we iteratively evaluate the self-energy $\pSigma^{(n+1)}=\pSigma[\pG^{(n)}]$, and solve the Dyson equation to obtain the updated propagator $\pG^{(n+1)}$, repeating the procedure until convergence. In some cases, linear mixing between successive iterations is employed to stabilize convergence.

\subsubsection{Physical correlation functions}

From the self-consistent pseudo-particle propagators, physical two-point correlation functions 
\begin{align}
C_{AB}(t, t') = - i \langle T_\mathcal C A(t) B(t') \rangle 
\end{align}
must be obtained from a separate diagrammatic expansion. In a steady state, $C_{AB}(t, t')$ depends only on time difference. With the conventional definition of lesser and greater contour-ordered correlation functions, we again set the second argument  $t'=0^-$, and evaluate
\begin{align}
C_{AB}^>(t)=C_{AB}(t^-,0^-),\,\,
C_{AB}^<(t)=C_{AB}(t^+,0^-)
\end{align}
 for $t<0$; the case $t>0$ is recovered from the Hermitian symmetry $C_{AB}^{<,>}(-t)=-C_{B^\dagger A^\dagger}^{<,>}(t)^*$. Diagrams for $C$ and $\pSigma$ are closely related: To get all diagrams for $C_{AB}(t_e,0^-)$, take all diagrams for $\pSigma(t_{2n-1},t_0=0^-)$, and attach an external line $\pG(0^+,t_{2n-1})$ to the left. The hybridization line connected to $t_0$ is  a {\em source line} taking the value unity, directed from $t_0$ to the external time argument  $t_{e}$ with operators $\hat B$ and $\hat A$ at its beginning and end. All other hybridization lines can take any flavors $(\alpha,\beta)$ as in the $\pSigma$ diagrams. The analytical expression is 
\begin{equation}
\label{eq:C_diagram}
C_{AB}^{(\mathcal D)}(t_e,0^-)
=
C^{(\mathcal D)}
\int_{\mathcal C^\succ}
\diff\bm t\,
F^{(\mathcal D)}(\bm t)
W^{(\mathcal D)}(\bm t),
\end{equation}
where now $\int_{\mathcal C^\succ} \diff\bm t$ denotes integration over all  internal times $\{t_1,\ldots,t_{2n-1}\}/\{t_e\}$, keeping the ordering $0^+\succ t_{2n-1}\succ t_{2n-2}\succ\cdots\succ t_0=0^-$. The function $F^{(\mathcal D)}(\bm t)$ and the sign $C^{(\mathcal D)}$ are determined as for $\pSigma$, while the factor $W^{(\mathcal D)}(\bm t)$ becomes a trace
\begin{align}
W^{(\mathcal D)}(\bm t)
=
\text{Tr}\big(
\hat \xi
\pG(0^+, &t_{2n-1})
\hat\psi_{2n-1}^{(\mathcal D)}
\pG(t_{2n-1},t_{2n-2})
\,\,\,\cdots
\nonumber\\
&
\cdots\,\,\,
\hat\psi_1^{(\mathcal D)}
\pG(t_1,t_0)
\hat\psi_0^{(\mathcal D)}\big).
\end{align}

\begin{figure}[tbp]
  \centering
  \resizebox{\columnwidth}{!}{\begin{tikzpicture}[
  font=\normalsize,
  >={Latex[length=2mm,width=1.8mm]},
  contour/.style={
    line width=1.25pt,
    line cap=round,
    line join=round
  },
  arr/.style={
    contour,
    -{Latex[length=2mm,width=1.8mm]}
  },
  dot/.style={
    circle,
    draw=green!50!black,
    fill=green!50!black,
    inner sep=0pt,
    minimum size=4.5pt
  },
  opendot/.style={
    circle,
    draw=green!50!black,
    fill=white,
    line width=1.1pt,
    inner sep=0pt,
    minimum size=4.5pt
  },
  externaldot/.style={
    circle,
    draw=red!75!black,
    fill=red!75!black,
    inner sep=0pt,
    minimum size=4.5pt
  },
  externalopendot/.style={
    circle,
    draw=red!75!black,
    fill=white,
    line width=1.1pt,
    inner sep=0pt,
    minimum size=4.5pt
  },
  lab/.style={
    inner sep=1pt
  },
  tlabel/.style={
    text=green!50!black,
    inner sep=1pt
  },
  vdot/.style={
    line width=1.1pt,
    dash pattern=on 0.5pt off 1.3pt
  },
  connection/.style={
    blue!75!black,
    line width=1.3pt,
    dashed,
    line cap=round,
    -{Latex[length=1.8mm,width=1.5mm]},
    shorten <=2.5pt,
    shorten >=4.5pt
  },
  externalconnection/.style={
    red!75!black,
    line width=1.3pt,
    dashed,
    line cap=round,
    -{Latex[length=1.8mm,width=1.5mm]},
    shorten <=2.5pt,
    shorten >=4.5pt
  }
]

\def\xL{0}
\def\xR{8.4}
\def\yU{1.35}
\def\yL{0}

\def\xthree{2.0}
\def\xtwo{4.1}
\def\xone{6.2}
\def\xzero{8.3}

\path[use as bounding box]
  (-1.20,-1.35) rectangle (9.15,2.35);

\draw[contour]
  (\xL,\yU) -- (\xR,\yU);

\draw[contour]
  (\xL,\yL) -- (\xR,\yL);

\draw[vdot]
  (\xL,\yU) -- (\xL,\yL);

\draw[vdot]
  (\xR,\yU) -- (\xR,\yL);

\draw[arr]
  (0.70,\yU) -- (1.15,\yU);

\draw[arr]
  (3.55,\yU) -- (4.00,\yU);

\draw[arr]
  (1.15,\yL) -- (0.70,\yL);

\draw[arr]
  (6.65,\yL) -- (6.20,\yL);

\node[lab,anchor=east]
  at (\xL-0.25,{(\yU+\yL)/2})
  {$-\infty$};

\node[lab]
  at (1.05,\yU+0.32)
  {$\mathcal{C}_{+}$};

\node[lab]
  at (1.05,\yL-0.32)
  {$\mathcal{C}_{-}$};

\coordinate (t2c) at (\xthree,\yU);
\coordinate (t1c) at (\xtwo,\yL);
\coordinate (t3c) at (\xone,\yU);
\coordinate (t0c) at (\xzero,\yL);

\draw[connection]
  (t3c) -- (t1c);

\draw[externalconnection]
  (t0c) -- (t2c);

\node[externaldot] (t2) at (t2c) {};
\node[dot] (t1) at (t1c) {};
\node[opendot] (t3) at (t3c) {};
\node[externalopendot] (t0) at (t0c) {};

\node[
  tlabel,
  anchor=south,
  yshift=3pt
] at (t2) {$t_2$};

\node[
  tlabel,
  anchor=north,
  yshift=-3pt
] at (t1) {$t_1$};

\node[
  tlabel,
  anchor=south,
  yshift=3pt
] at (t3) {$t_3$};

\node[
  tlabel,
  anchor=north west,
  xshift=7pt,
  yshift=-3pt
] at (t0) {$t_0$};

\end{tikzpicture}}
  \caption{Illustration of an OCA diagram contribution to the physical correlator $C_{AB}^{(\mathcal{D})}$. The conventions and contour geometry are the same as in Fig.~\ref{fig:steady_state_diagram}. The additional dotted connection at the right edge closes the two contour branches and indicates the trace with the $\xi$ factor. The dashed blue arrow denotes the remaining hybridization line, while the dashed red arrow from $t_0$ to $t_2$ connects the red external vertices and represents the unit source.}
  \label{fig:steady_state_correlator_diagram}
\end{figure}
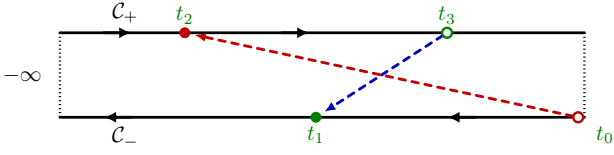

  An exemplary OCA contribution to the correlation function 
  is illustrated in
  Fig.~\ref{fig:steady_state_correlator_diagram}. 
  Here the directed line
  $t_0\to t_2$ is the unit source line, with $\hat B$ at its starting point
  and $\hat A$ at its endpoint, while $t_3\to t_1$ is the remaining
  hybridization line. The resulting expression is
\begin{align}
  \label{eq:oca_correlator_diagram}
  &C_{AB}^{(\mathcal D)}(t_2,t_0)
  = C^{(\mathcal D)}
  \int_{\bm t\succ}\mathrm{d}\bm t\;
  \Delta^{>}_{\alpha_2,\beta_2}(t_1,t_3)
 \nonumber\\
  &{}\times\mathrm{Tr}\!\left(
  \hat\xi
  \hat w_{\beta_2}\pG^>(t_3,t_2)
  \hat A\pG^<(t_2,t_1)
  \hat v_{\alpha_2}\pG^>(t_1,t_0)
  \hat B
  \right).
\end{align}

\subsection{Tensor Cross Interpolation and Tensor Trains}
\label{ch:TensorTrains}
\begin{figure}[tbp]
  \centering
  \begin{tikzpicture}[
  Fstyle/.style={ppscWideTensor, minimum width=31mm},
  Mstyle/.style={ppscCore},
  bond/.style={ppscBond},
  phys/.style={ppscBond},
  leglab/.style={ppscLegLabel}
  ]
  \def\xF{0}
  \def\xap{3.45}
  \def\xMone{3.85}
  \def\xMtwo{5.25}
  \def\xMD{7.25}
  \def\dotStub{1.8mm}

  \node[Fstyle, anchor=west] (F) at (\xF,0) {$F$};
  \node[ppscSym] at (\xap,0) {$\approx$};

  \node[Mstyle, anchor=west] (M1) at (\xMone,0) {$M^{(1)}$};
  \node[Mstyle, anchor=west] (M2) at (\xMtwo,0) {$M^{(2)}$};
  \node[Mstyle, anchor=west] (MD) at (\xMD,0) {$M^{(D)}$};
  \coordinate (Md) at ($($(M2.east)+(\dotStub,0)$)!0.5!($(MD.west)+(-\dotStub,0)$)$);

  \draw[bond] (M1.east) -- (M2.west);
  \draw[bond] (M2.east) -- ++(\dotStub,0);
  \draw[bond] (MD.west) -- ++(-\dotStub,0);
  \node[ppscDots] at (Md) {$\cdots$};

  \draw[phys] (M1.north) -- ++(0,4.5mm) node[leglab, above] {$s_1$};
  \draw[phys] (M2.north) -- ++(0,4.5mm) node[leglab, above] {$s_2$};
  \draw[phys] (MD.north) -- ++(0,4.5mm) node[leglab, above] {$s_D$};

  \coordinate (F1) at ($(F.north west)!0.13!(F.north east)$);
  \coordinate (F2) at ($(F.north west)!0.42!(F.north east)$);
  \coordinate (FD) at ($(F.north west)!0.84!(F.north east)$);
  \draw[phys] (F1) -- ++(0,4.5mm) node[leglab, above] {$s_1$};
  \draw[phys] (F2) -- ++(0,4.5mm) node[leglab, above] {$s_2$};
  \draw[phys] (FD) -- ++(0,4.5mm) node[leglab, above] {$s_D$};

  \node[leglab] at
    ($(F.north west)!0.63!(F.north east)+(0,6.5mm)$)
    {$\cdots$};

  \pgfresetboundingbox
  \path[use as bounding box] (-0.05,-0.62) rectangle (8.65,1.45);

\end{tikzpicture}
  \caption{Illustration of a general decomposition of a $D$-dimensional tensor $F$ (brown) as a TT with three leg cores $M$ (blue), for example achievable via TCI.
  }
  \label{fig:tci_factorization}
\end{figure}
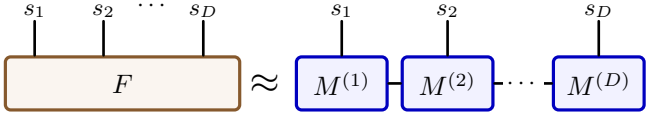

\begin{figure}[tbp]
  \centering
  \begin{tikzpicture}[
  Fstyle/.style={ppscWideTensor, minimum width=31mm},
  Mstyle/.style={ppscCore},
  Wstyle/.style={ppscWeight},
  bond/.style={ppscBond}
  ]
  \def\wHeight{12.5mm}

  \node[Fstyle, anchor=west] (F) at (0,0) {$F$};

  \coordinate (F1) at ($(F.north west)!0.13!(F.north east)$);
  \coordinate (F2) at ($(F.north west)!0.42!(F.north east)$);
  \coordinate (FD) at ($(F.north west)!0.84!(F.north east)$);

  \node[Wstyle] (W1) at ($(F1 |- 0,\wHeight)$) {$w$};
  \node[Wstyle] (W2) at ($(F2 |- 0,\wHeight)$) {$w$};
  \node[Wstyle] (WD) at ($(FD |- 0,\wHeight)$) {$w$};

  \draw[bond] (W1.south) -- (F1);
  \draw[bond] (W2.south) -- (F2);
  \draw[bond] (WD.south) -- (FD);

  \coordinate (Fdots) at
    ($(F.north west)!0.63!(F.north east)$);
  \node[ppscDots]
    at (Fdots |- W1.center)
    {$\cdots$};

  \node[ppscSym] at (3.45,0) {$\approx$};

  \def\xMone{3.85}
  \def\xMtwo{5.25}
  \def\xMD{7.25}
  \def\dotStub{1.8mm}
  \node[Mstyle, anchor=west] (M1) at (\xMone,0) {$M^{(1)}$};
  \node[Mstyle, anchor=west] (M2) at (\xMtwo,0) {$M^{(2)}$};
  \node[Mstyle, anchor=west] (MD) at (\xMD,0) {$M^{(D)}$};
  \coordinate (Md) at ($($(M2.east)+(\dotStub,0)$)!0.5!($(MD.west)+(-\dotStub,0)$)$);

  \draw[bond] (M1.east) -- (M2.west);
  \draw[bond] (M2.east) -- ++(\dotStub,0);
  \draw[bond] (MD.west) -- ++(-\dotStub,0);
  \node[ppscDots] at (Md) {$\cdots$};

  \node[Wstyle] (w1) at (M1.north |- 0,\wHeight) {$w$};
  \node[Wstyle] (w2) at (M2.north |- 0,\wHeight) {$w$};
  \node[Wstyle] (wD) at (MD.north |- 0,\wHeight) {$w$};

  \draw[bond] (w1.south) -- (M1.north);
  \draw[bond] (w2.south) -- (M2.north);
  \draw[bond] (wD.south) -- (MD.north);

  \pgfresetboundingbox
  \path[use as bounding box] (-0.05,-0.62) rectangle (8.65,1.75);

\end{tikzpicture}
  \caption{Hypercubic integration of TT via contraction with independent quadrature weight tensors $w$ (green).}
  \label{fig:hypercube_contraction}
\end{figure}
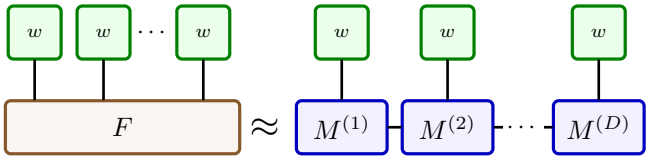

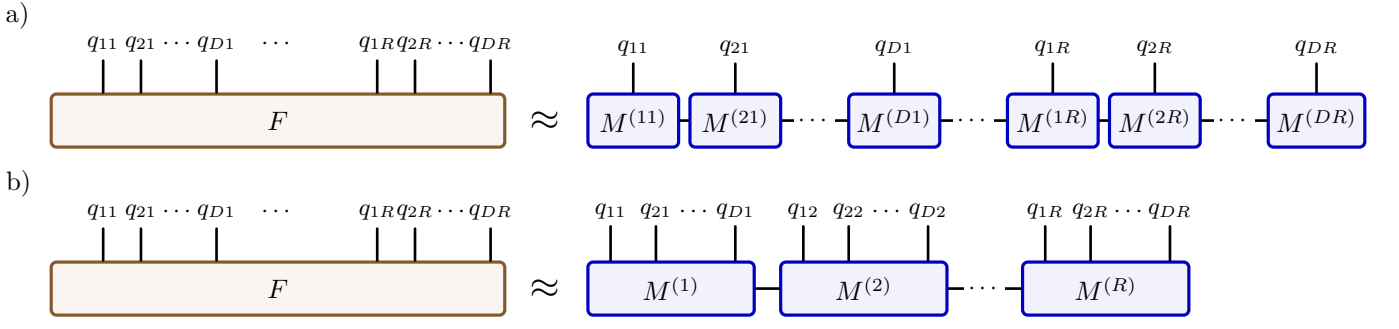
\begin{figure*}[tbp]
  \centering

  \begin{minipage}{\textwidth}%
    \raggedright%
    \makebox[\ppscPanelGutter][l]{\ppscPanel{a}}%
    \raisebox{\ppscPanelPlotRaise}{%
      \begin{minipage}[t]{\dimexpr\textwidth-\ppscPanelGutter\relax}%
        \vspace{0pt}\raggedright\begin{tikzpicture}[
	Fbig/.style={
		ppscWideTensor,
		minimum width=60mm
	},
	Mcore/.style={
		ppscCore,
		minimum width=12mm
	},
	bond/.style={ppscBond},
	leglab/.style={ppscLegLabel}
]

\def\legH{3.8mm}
\def\labH{5.9mm}
\def\qFleg{4.2mm}
\def\qFlab{6.4mm}
\def\qFdot{6.4mm}
\def\qFxA{-5mm}
\def\qFxB{ 0mm}
\def\qFxC{10mm}
\def\dotStub{2.5mm}

\node[Fbig, anchor=west] (F) at (0,0) {$F$};

\coordinate (g1) at ($(F.north west)!0.20!(F.north east)$);
\coordinate (gR) at ($(F.north west)!0.80!(F.north east)$);

\newcommand{\ShowInterleavedGroup}[4]{%
	\draw[bond] ($(#1)+(\qFxA,0)$) -- ++(0,\qFleg);
	\draw[bond] ($(#1)+(\qFxB,0)$) -- ++(0,\qFleg);
	\draw[bond] ($(#1)+(\qFxC,0)$) -- ++(0,\qFleg);
	\node[leglab] at ($(#1)+(\qFxA,\qFlab)$) {$#2$};
	\node[leglab] at ($(#1)+(\qFxB,\qFlab)$) {$#3$};
	\node[leglab] at ($(#1)+(5mm,\qFdot)$) {$\cdots$};
	\node[leglab] at ($(#1)+(\qFxC,\qFlab)$) {$#4$};
}

\ShowInterleavedGroup{g1}{q_{11}}{q_{21}}{q_{D1}}

\node[leglab] at ($(g1)!0.5!(gR)+(0mm,\qFlab)$) {$\cdots$};

\ShowInterleavedGroup{gR}{q_{1R}}{q_{2R}}{q_{DR}}

\node[ppscSym] at (6.55,0) {$\approx$};

\node[Mcore, anchor=west] (M11) at (7.10,0) {$M^{(11)}$};
\node[Mcore, anchor=west] (M21) at (8.45,0) {$M^{(21)}$};
\node[Mcore, anchor=west] (MD1) at (10.55,0) {$M^{(D1)}$};
\node[Mcore, anchor=west] (M1R) at (12.65,0) {$M^{(1R)}$};
\node[Mcore, anchor=west] (M2R) at (14.00,0) {$M^{(2R)}$};
\node[Mcore, anchor=west] (MDR) at (16.10,0) {$M^{(DR)}$};

\coordinate (D1dots) at ($($(M21.east)+(\dotStub,0)$)!0.5!($(MD1.west)+(-\dotStub,0)$)$);
\coordinate (Rdots) at ($($(MD1.east)+(\dotStub,0)$)!0.5!($(M1R.west)+(-\dotStub,0)$)$);
\coordinate (DRdots) at ($($(M2R.east)+(\dotStub,0)$)!0.5!($(MDR.west)+(-\dotStub,0)$)$);

\draw[bond] (M11.east) -- (M21.west);
\draw[bond] (M21.east) -- ++(\dotStub,0);
\draw[bond] (MD1.west) -- ++(-\dotStub,0);
\node[ppscDots] at (D1dots) {$\cdots$};
\draw[bond] (MD1.east) -- ++(\dotStub,0);
\draw[bond] (M1R.west) -- ++(-\dotStub,0);
\node[ppscDots] at (Rdots) {$\cdots$};
\draw[bond] (M1R.east) -- (M2R.west);
\draw[bond] (M2R.east) -- ++(\dotStub,0);
\draw[bond] (MDR.west) -- ++(-\dotStub,0);
\node[ppscDots] at (DRdots) {$\cdots$};

\draw[bond] (M11.north) -- ++(0,\legH);
\node[leglab] at ($(M11.north)+(0,\labH)$) {$q_{11}$};

\draw[bond] (M21.north) -- ++(0,\legH);
\node[leglab] at ($(M21.north)+(0,\labH)$) {$q_{21}$};

\draw[bond] (MD1.north) -- ++(0,\legH);
\node[leglab] at ($(MD1.north)+(0,\labH)$) {$q_{D1}$};

\draw[bond] (M1R.north) -- ++(0,\legH);
\node[leglab] at ($(M1R.north)+(0,\labH)$) {$q_{1R}$};

\draw[bond] (M2R.north) -- ++(0,\legH);
\node[leglab] at ($(M2R.north)+(0,\labH)$) {$q_{2R}$};

\draw[bond] (MDR.north) -- ++(0,\legH);
\node[leglab] at ($(MDR.north)+(0,\labH)$) {$q_{DR}$};

\pgfresetboundingbox
\path[use as bounding box] (-0.05,-0.62) rectangle (17.25,1.45);

\end{tikzpicture}%
      \end{minipage}%
    }%
  \end{minipage}%

  \begin{minipage}{\textwidth}%
    \raggedright%
    \makebox[\ppscPanelGutter][l]{\ppscPanel{b}}%
    \raisebox{\ppscPanelPlotRaise}{%
      \begin{minipage}[t]{\dimexpr\textwidth-\ppscPanelGutter\relax}%
        \vspace{0pt}\raggedright\begin{tikzpicture}[
	Fbig/.style={ppscWideTensor, minimum width=60mm},
	Mcore/.style={ppscWideCore, minimum width=22mm},
	bond/.style={ppscBond},
	sym/.style={ppscSym},
	leglab/.style={ppscLegLabel}
]

\def\yLeg{4.5mm}
\def\yLab{6.5mm}
\def\qFleg{4.2mm}
\def\qFlab{6.4mm}
\def\qFdot{6.4mm}

\def\qFxA{-5mm}
\def\qFxB{ 0mm}
\def\qFxC{10mm}
\def\xA{-8mm}
\def\xB{-2mm}
\def\xDot{3.25mm}
\def\xC{8.5mm}
\def\dotStub{2.5mm}

\node[Fbig, anchor=west] (F) at (0,0) {$F$};

\coordinate (g1) at ($(F.north west)!0.20!(F.north east)$);
\coordinate (gR) at ($(F.north west)!0.80!(F.north east)$);

\newcommand{\ShowQFGroup}[4]{%
	\draw[bond] ($(#1)+(\qFxA,0)$) -- ++(0,\qFleg);
	\draw[bond] ($(#1)+(\qFxB,0)$) -- ++(0,\qFleg);
	\draw[bond] ($(#1)+(\qFxC,0)$) -- ++(0,\qFleg);
	\node[leglab] at ($(#1)+(\qFxA,\qFlab)$) {$#2$};
	\node[leglab] at ($(#1)+(\qFxB,\qFlab)$) {$#3$};
	\node[leglab] at ($(#1)+(5mm,\qFdot)$) {$\cdots$};
	\node[leglab] at ($(#1)+(\qFxC,\qFlab)$) {$#4$};
}

\ShowQFGroup{g1}{q_{11}}{q_{21}}{q_{D1}}
\ShowQFGroup{gR}{q_{1R}}{q_{2R}}{q_{DR}}
\node[leglab] at ($(g1)!0.5!(gR)+(0,\qFlab)$) {$\cdots$};

\node[sym] (approx) at (6.55,0) {$\approx$};

\node[Mcore, anchor=west] (M1) at (7.10,0) {$M^{(1)}$};
\node[Mcore, anchor=west] (M2) at (9.65,0) {$M^{(2)}$};
\node[Mcore, anchor=west] (MR) at (12.85,0) {$M^{(R)}$};
\coordinate (Mdots) at ($($(M2.east)+(\dotStub,0)$)!0.5!($(MR.west)+(-\dotStub,0)$)$);

\draw[bond] (M1.east) -- (M2.west);
\draw[bond] (M2.east) -- ++(\dotStub,0);
\draw[bond] (MR.west) -- ++(-\dotStub,0);
\node[ppscDots] at (Mdots) {$\cdots$};

\newcommand{\ShowGroupTop}[4]{%
	\draw[bond] ($(#1.north)+(\xA,0)$) -- ++(0,\yLeg);
	\draw[bond] ($(#1.north)+(\xB,0)$) -- ++(0,\yLeg);
	\draw[bond] ($(#1.north)+(\xC,0)$) -- ++(0,\yLeg);
	\node[leglab] at ($(#1.north)+(\xA,\yLab)$) {$#2$};
	\node[leglab] at ($(#1.north)+(\xB,\yLab)$) {$#3$};
	\node[leglab] at ($(#1.north)+(\xDot,\yLab)$) {$\cdots$};
	\node[leglab] at ($(#1.north)+(\xC,\yLab)$) {$#4$};
}

\ShowGroupTop{M1}{q_{11}}{q_{21}}{q_{D1}}
\ShowGroupTop{M2}{q_{12}}{q_{22}}{q_{D2}}
\ShowGroupTop{MR}{q_{1R}}{q_{2R}}{q_{DR}}

\pgfresetboundingbox
\path[use as bounding box] (-0.05,-0.62) rectangle (15.15,1.45);

\end{tikzpicture}%
      \end{minipage}%
    }%
  \end{minipage}%

  \vspace{-1ex}
  \caption{Quantics tensor-train illustrations, where a) depicts the interleaved and b) the fused quantics decomposition of a quantized tensor.}
  \label{fig:tt_notation}
\end{figure*}

TCI is a protocol to decompose a general $D$-dimensional tensor into a TT based on successive matrix CI \cite{Oseledets-Tyrtyshnikov-2010, Oseledets-2011}
\begin{equation}
  T_{s_1, s_2, \dots, s_{D}} = M^{(1)}_{s_1} \cdot M^{(2)}_{s_2} \cdots M^{(D)}_{s_{D}},
\end{equation}
where $[M^{(d)}_{s_d}]_{\alpha_{d}, \alpha_{d+1}}$ is a $\chi_{d} \times \chi_{d+1}$ matrix. The matrix dimensions are called bond dimensions; if not stated otherwise, we will refer to $\chi = \max_{d} (\chi_d)$ as the  bond dimension of the whole TT.
Such a decomposition can be represented visually using standard tensor network notation as seen in Fig. \ref{fig:tci_factorization}.
While structurally very similar to the MPS construction using a SVD, the CI algorithm does not need access to all entries of the initial tensor.
This is achieved by an iterative pivot search algorithm and successive approximation of the intermediate matrices via a CI decomposition \cite{NúñezFernández-etal-2022}.
A newer version of the algorithm using a partially rank-revealing LU-decomposition is presented in \cite{Fernández-etal-2025}, which can be shown to be equivalent to the matrix CI approach via the Schur complement.
Both algorithms are implemented in the \texttt{C++} library \texttt{xfac} \cite{Fernández-etal-2025}. The approximation accuracy is mainly controlled by the maximally allowed bond dimension and the local pivot error \cite{Fernández-etal-2025}. If $N$ is the dimension of the physical indices $s_d$, the computational effort scales with $\mathcal{O}(\chi^3 N^2 D)$ when using the \texttt{full-pivot} mode, which searches for a suitable pivot in the full intermediate matrix. This is improved by \texttt{rook-pivoting} to linear behavior $\mathcal{O}(\chi^3 N n_{\text{rook}} D)$, albeit special care about the ergodicity may be needed in some cases \cite{Fernández-etal-2025}.

This procedure can readily be generalized to the decomposition of a continuous multivariate function $f(\bm x)$ on a region $\bm x\in V\subset \mathbb R^D$, if we discretize the domain $V$  by  a predefined grid $\{ \bm  x_{\bm  s} \}$ with $N_d$ points  $s_d \in {1, \dots, N_d}$ along axis $d$. The discretized function $f(\bm s)$ can then be interpreted as a $D$-dimensional tensor, and the  TCI algorithm can provide an effective factorization of the function, which allows efficient further manipulation.  For example, the integration of $f(\bm x)$ on a hypercubic domain $\mathcal{V}$ is given by \cite{Fernández-etal-2025}
\begin{equation}
    I_\mathcal{V} = \int_\mathcal{V} \mathrm{d} \bm x \; f(\bm x) \approx \sum_{\bm s} \Big (\prod_{d=1}^{D} w^{(d)}_{s_d} \Big) \, f(\bm s),
\end{equation}
where we have the choice to either apply the quadrature weights $w^{(d)}$ after the sampling of the discretized function $f(\bm s) \approx M^{(1)}_{s_1} \cdots M^{(D)}_{s_{D}}$ or to sample the weighted integrand $\Big (\prod_{d=1}^{D} w^{(d)}_{s_d} \Big) \, f(\bm s) \approx \tilde{M}^{(1)}_{s_1} \cdots \tilde{M}^{(D)}_{s_{D}}$ instead. Both approaches yield an effective factorization of the integral
\begin{equation}
  \label{eq:hypercube_tt}
  I_\mathcal{V} \approx \prod_{d=1}^{D} \Big ( \sum_{s_d = 1}^{N_d} w^{(d)}_{s_d} M^{(d)}_{s_d} \Big ) \approx \prod_{d=1}^{D} \Big ( \sum_{s_d = 1}^{N_d} \tilde{M}^{(d)}_{s_d} \Big ),
\end{equation}
which reduces the numerical effort from $\mathcal{O} (N^D)$ to $\mathcal{O} (\chi^3 N D)$. The tensor contraction for the unweighted sampling is illustrated in Fig. \ref{fig:hypercube_contraction}. This approach is efficient if the integrand can be represented with a low-rank $\chi \ll N$, which in general depends on the parametrization of the discretized function.

One more recent addition to the TT toolbox is given by the quantics representation, which decomposes a uniform discrete index as an array of bits $q_{d,r} \in \{0, 1\}$
\begin{equation}
  s_d = (q_{d, 1}, \dots, q_{d, R})_2 = \sum_{r = 1}^{R} 2^{R-r} q_{d,r},
  \label{eq:quantics_representation}
\end{equation}
which yields a reparametrization of the discrete function $f(\underline{\bm q})$ in terms of the individual bits $\underline{\bm{q}} = \{ q_{d, r} | d = \{1, D\}, r = \{1, R\} \}$.
A TCI factorization in terms of the bits  now amounts to a factorization in terms of the bits and thus in terms of the different length scales.
This has been shown in \cite{Khoromskij-2011} to be an effective decomposition for many common functions, leading to a compression of order $\mathcal{O}(2 \log_2 (N))$, and was explored in Ref.~\cite{Shinaoka-etal-2023} in the context of diagrammatic many body perturbation theory. 

Within the quantics representation, there is a certain degree of freedom to rearrange the bits in order to obtain a lower bond dimension: In particular, the interleaved representation groups the bits corresponding to a single length scale across different dimension together
\begin{equation}
  f(\underline{\bm q}) = M_{q_{1,1}} \cdots M_{q_{D,1}} \cdots M_{q_{1,R}} \cdots M_{q_{D,R}}
\end{equation}
which can be obtained at a $\mathcal{O}(\chi^3 R D)$ effort \cite{Fernández-etal-2025} and is shown schematically in Fig.~\ref{fig:tt_notation}(a). Alternatively, one can also merge the $D$ adjacent cores resulting in the fused representation with $R$ cores
\begin{equation}
  \label{eq:fused_quantics_representation}
  f(\underline{\bm q}) = \tilde{M}_{\bm q_{1}} \cdots \tilde{M}_{\bm q_{R}},
\end{equation}
where $\bm q_{r}=(q_{1,r},...,q_{D,r})$ (see Fig.~\ref{fig:tt_notation}(b)).  This approach can yield a slightly more accurate description at a lower bond dimension for multivariate functions with correlated dimensions than the interleaved ordering, albeit at a slightly higher computational effort with the number of dimensions $\mathcal{O} (\chi^3 R 2^D)$ during sampling. In practice, the fused ordering is also often more stable, as the strong factorization of the interleaved ordering may enhance ergodicity problems within the sampling routine.

\section{TCI-accelerated evaluation of strong-coupling diagrams}
\label{sec:decomps}

Since high-order integrals are the main bottleneck in the evaluation of $\Sigma$ and $C$ diagrams, it is natural to accelerate their evaluation using TCI by decomposing the integrand into a TT representation. However, the rank of this decomposition depends crucially on the chosen parametrization. In addition, for integration over a time-ordered domain, the summation boundaries and the quadrature weights along the different axes are not independent, and the simple evaluation via Eq.~\eqref{eq:hypercube_tt} is no longer possible. Possible strategies to overcome this problem include mapping the time-ordered domain onto a hypercube (which may require interpolation) \cite{Erpenbeck-etal-2023, Matsuura-etal-2025}, or incorporating the triangular time-ordering mask into the sampling. In the following section, we explain the parametrization and decomposition schemes used to evaluate the diagrammatic expressions for $\pSigma$ and $C$.

\subsection{Keldysh parametrization}
\label{ch:keldysh}

The diagrammatic expressions for $\pSigma$ [Eq.~\eqref{eq:selfenergy_diagram}] and $C$ [Eq.~\eqref{eq:C_diagram}] can be written in the unified form ($X=\pSigma$ or $X=C$)
\begin{equation}
\label{eq:X_diagram}
X^{(\mathcal D)}(t,0^-)
=
\int_{\mathcal C^\succ}
\diff\bm t\,
\delta(t,t_{e(\mathcal D)})
I^{(\mathcal D)}(\bm t),
\end{equation}
where $I^{(\mathcal D)}(\bm t)\equiv C^{(\mathcal D)}F^{(\mathcal D)}W^{(\mathcal D)}$ is the respective integrand, and $\bm t=(t_0,\ldots,t_{D})$ are the contour times, where here and in the following we denote $D=2n-1$ for an $n$th-order diagram. The symbol $\int_{\mathcal C^\succ}$ denotes integration over the times $t_1,\ldots,t_{D}$ with the contour ordering $0^+\succ t_{D}\succ t_{D-1}\succ\cdots\succ t_1\succ t_0=0^-$. The factor $\delta(t,t_{e(\mathcal D)})$ constrains the external time argument to the external vertex of the diagram, whose index $e(\mathcal D)$ is uniquely determined by the topology: $e=D$ for $\pSigma$ diagrams, while for $C$ diagrams, $e$ is the endpoint of the source line connected to $t_0=0^-$.

For the numerical integration, one could  parametrize $\bm t$ in terms of successive time differences along the cyclic contour. In this parametrization, the factor $W^{(\mathcal D)}(\bm t)$ is already in a factorized TT form, and only $F^{(\mathcal D)}(\bm t)$ needs to be decomposed. However, as demonstrated in Ref.~\cite{Eckstein-2024}, this cyclic parametrization can lead to rather large TT bond dimensions 
when vertices lie on the both upper and lower branch. 
Throughout this manuscript, we therefore instead use the \emph{Keldysh parametrization} 
\cite{Eckstein-2024}, where each contour time $t_j$ is represented by the pair $(\tilde t_j^{\,\sigma_j})$ consisting of a real physical time $\tilde t_j$ and a Keldysh index $\sigma_j\in\{\pm1\}$. For ordered physical times $(\tilde t_{D}<\cdots <\tilde t_{1}< 0)$, the configuration $\bm\sigma=(\sigma_{D},...,\sigma_1)$ uniquely determines the ordering of the contour-time arguments. For example, physical times $(\tilde t_5<\ldots<\tilde t_1)$ with Keldysh indices $(\sigma_{5},\ldots,\sigma_1)=(+,+,-,+,-)$ correspond to the contour ordering $0^+\succ \tilde t_2^+\succ \tilde t_4^+ \succ \tilde t_5^+ \succ t_3^-\succ t_1^-\succ 0^-$. Equation~\eqref{eq:X_diagram} therefore becomes
\begin{align}
&X^{(\mathcal D)}(t,0^-)
=
\sum_{\bm\sigma}
X^{(\mathcal D,\bm \sigma)}(t,0^-)
\\
\label{eq:X_diagram_K}
&X^{(\mathcal D,\bm \sigma)}(\tilde t^\sigma,0^-)
=\int_{>}
\diff \tilde{\bm t}\,
\delta_{\sigma,\sigma_e}\delta(\tilde t- \tilde t_e)
I^{(\mathcal D,\bm \sigma)}(\bm t),
\end{align}
with  a standard time-ordered integral,
\begin{equation}
\int_{>}
\diff \tilde{\bm t}\,
=
\int_{-\infty}^{0} \diff\tilde t_{D}
\int_{\tilde t_{D}}^{0} \diff\tilde t_{D-1}
\cdots
\int_{\tilde t_2}^0 \diff\tilde t_1,
\end{equation}
and a sum
$\sum_{\bm\sigma}$
over all Keldysh configurations $\bm\sigma=(\sigma_{D},\ldots,\sigma_1)$, 
with
$\tilde t_0^{\,\sigma_0}=0^-$. The index $e=e(\mathcal D,\bm \sigma)$ of the external time argument is uniquely determined by the diagram topology $\mathcal D$ and the Keldysh configuration $\bm\sigma$: for $C$ diagrams, $e$ is the endpoint of the source line connected to $t_0=0^-$, while for $\pSigma$ diagrams, $e$ corresponds to the last contour-time argument in the cyclic ordering ($e=D$ if all 
vertices are on $\mathcal C_-$;
 if at least one argument is on $\mathcal C_+$, $e$ is given by the smallest $d$ with $\sigma_d=+$).

In the Keldysh parametrization, the numerical evaluation of $\pSigma$ and $C_{AB}$ involves a sum over all diagrams $(\mathcal D,\bm\sigma)$, defined by the diagram topology, the assignment of hybridization flavors $(\alpha,\beta)$, and the Keldysh configuration $\bm\sigma$. Although this introduces an additional summation over $2^{D}$ configurations $\bm\sigma$ compared to the cyclic parametrization in Eq.~\eqref{eq:X_diagram}, previous studies \cite{Eckstein-2024, Kim-Werner-2025, Geng-Kim-Werner-2025}, as well as our own tests indicate that the improved efficiency of the TT representation of the integrand more than compensates for this additional computational cost. Therefore, in the remainder of this paper we restrict our discussion of evaluation strategies to the Keldysh parametrization. Within the Keldysh parametrization, we explore four different decompositions of the integral $\int_{>} \diff \tilde{\bm t}\, \delta_{\sigma,\sigma_e}\delta(t-\tilde t_e)\cdots$, in which, for simplicity, we drop the tilde over the physical time arguments 
from now on.%

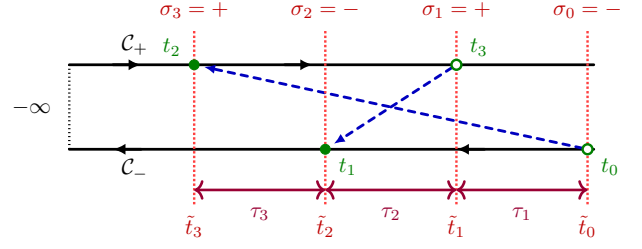
\begin{figure}[tbp]
  \centering
  \resizebox{\columnwidth}{!}{%
    \begin{tikzpicture}[
  font=\normalsize,
  contour/.style={
    line width=1.25pt,
    line cap=round,
    line join=round
  },
  arr/.style={
    contour,
    -{Latex[length=2mm,width=1.8mm]}
  },
  dot/.style={
    circle,
    draw=green!50!black,
    fill=green!50!black,
    inner sep=0pt,
    minimum size=4.5pt
  },
  opendot/.style={
    circle,
    draw=green!50!black,
    fill=white,
    line width=1.1pt,
    inner sep=0pt,
    minimum size=4.5pt
  },
  cycliclabel/.style={
    text=green!50!black,
    inner sep=1pt
  },
  lab/.style={
    inner sep=1pt
  },
  vdot/.style={
    line width=1.1pt,
    dash pattern=on 0.5pt off 1.3pt
  },
  connection/.style={
    blue!75!black,
    line width=1.3pt,
    dashed,
    line cap=round,
    -{Latex[length=1.8mm,width=1.5mm]},
    shorten <=2.5pt,
    shorten >=4.5pt
  },
  guide/.style={
    red!75,
    line width=1.1pt,
    densely dotted
  },
  redlabel/.style={
    text=red!75!black,
    inner sep=1pt
  },
  interval/.style={
    purple!80!black,
    line width=1.2pt,
    <->
  }
]

\def\xL{0}
\def\xR{8.4}
\def\yU{1.35}
\def\yL{0}

\def\xthree{2.0}
\def\xtwo{4.1}
\def\xone{6.2}
\def\xzero{8.3}

\path[use as bounding box]
  (-1.20,-1.35) rectangle (9.15,2.35);

\draw[contour] (\xL,\yU) -- (\xR,\yU);
\draw[contour] (\xL,\yL) -- (\xR,\yL);
\draw[vdot] (\xL,\yU) -- (\xL,\yL);

\draw[arr] (0.70,\yU) -- (1.15,\yU);
\draw[arr] (3.45,\yU) -- (3.90,\yU);
\draw[arr] (1.15,\yL) -- (0.70,\yL);
\draw[arr] (6.65,\yL) -- (6.20,\yL);

\node[lab,anchor=east]
  at (\xL-0.25,{(\yU+\yL)/2}) {$-\infty$};
\node[lab] at (1.05,\yU+0.32) {$\mathcal C_+$};
\node[lab] at (1.05,\yL-0.32) {$\mathcal C_-$};

\coordinate (t2c) at (\xthree,\yU);
\coordinate (t1c) at (\xtwo,\yL);
\coordinate (t3c) at (\xone,\yU);
\coordinate (t0c) at (\xzero,\yL);

\draw[connection] (t0c) -- (t2c);
\draw[connection] (t3c) -- (t1c);

\foreach \x in {\xthree,\xtwo,\xone,\xzero}{
  \draw[guide] (\x,\yL-0.88) -- (\x,\yU+0.55);
}

\node[dot] (t2) at (t2c) {};
\node[opendot] (t3) at (t3c) {};
\node[dot] (t1) at (t1c) {};
\node[opendot] (t0) at (t0c) {};

\node[cycliclabel,anchor=south east,xshift=-5pt,yshift=3pt]
  at (t2) {$t_2$};
\node[cycliclabel,anchor=south west,xshift=5pt,yshift=3pt]
  at (t3) {$t_3$};
\node[cycliclabel,anchor=north west,xshift=5pt,yshift=-3pt]
  at (t1) {$t_1$};
\node[cycliclabel,anchor=north west,xshift=5pt,yshift=-3pt]
  at (t0) {$t_0$};

\node[redlabel,anchor=south]
  at (\xthree,\yU+0.70) {$\sigma_3=+$};
\node[redlabel,anchor=south]
  at (\xtwo,\yU+0.70) {$\sigma_2=-$};
\node[redlabel,anchor=south]
  at (\xone,\yU+0.70) {$\sigma_1=+$};
\node[redlabel,anchor=south]
  at (\xzero,\yU+0.70) {$\sigma_0=-$};

\draw[interval]
  (\xthree,\yL-0.65) -- (\xtwo,\yL-0.65)
  node[midway,below=4pt] {$\tau_3$};
\draw[interval]
  (\xtwo,\yL-0.65) -- (\xone,\yL-0.65)
  node[midway,below=4pt] {$\tau_2$};
\draw[interval]
  (\xone,\yL-0.65) -- (\xzero,\yL-0.65)
  node[midway,below=4pt] {$\tau_1$};

\node[redlabel,anchor=north]
  at (\xthree,\yL-1.02) {$\tilde t_3$};
\node[redlabel,anchor=north]
  at (\xtwo,\yL-1.02) {$\tilde t_2$};
\node[redlabel,anchor=north]
  at (\xone,\yL-1.02) {$\tilde t_1$};
\node[redlabel,anchor=north]
  at (\xzero,\yL-1.02) {$\tilde t_0$};

\end{tikzpicture}%
  }
  \caption{Keldysh reparametrization of an OCA self-energy diagram. Green labels denote the cyclically ordered contour arguments $t_j$. Open and filled green circles mark the starting and endpoint vertices of the directed dashed hybridization lines, respectively. Red guides associate the vertices with the ordered physical times $\tilde t_j$ and the Keldysh indices $\sigma_j$, while the purple arrows indicate successive physical-time intervals.}
  \label{fig:keldysh_reparametrization_oca}
\end{figure}

In Fig.~\ref{fig:keldysh_reparametrization_oca} we show the Keldysh reparametrization of the OCA example in Fig.~\ref{fig:steady_state_diagram}.
For the configuration shown, the external contour argument $t_3$ corresponds to the physical time $\tilde t_1$. Introducing the positive successive time differences $\tau_j=\tilde t_{j-1}-\tilde t_j>0$, we have $\tilde t_1=-\tau_1$ because $\tilde t_0=0$. Hence $\tau_1$ is fixed by the external time, while $\tau_2$ and $\tau_3$ are integrated:
\begin{align}
  &\pSigma^{(\mathcal{D},\bm{\sigma})}(-\tau_1)
  = \mathcal{C}^{(\mathcal{D},\bm{\sigma})}
  \int_{0}^{\infty}\mathrm{d}\tau_2
  \int_{0}^{\infty}\mathrm{d}\tau_3\;
  \nonumber\\
  &{}\times
  \hat w_{\beta_2}\pG^>(\tau_2+\tau_3)
  \hat v_{\alpha_1}\pG^<(-\tau_3)
  \hat v_{\alpha_2}\pG^>(-\tau_1-\tau_2)
  \hat w_{\beta_1}
  \nonumber\\ 
  &{}\times\Delta^{<}_{\alpha_1,\beta_1}
    (-\tau_1-\tau_2-\tau_3)
  \Delta^{>}_{\alpha_2,\beta_2}(-\tau_2).
\end{align}

\subsection{Time Decomposition}
\label{ch:TimeDecomposition}

\begin{figure}[tbp]
  \centering
  \resizebox{\columnwidth}{!}{\begin{tikzpicture}[
  S/.style={ppscWideTensor, minimum width=31mm},
  M/.style={ppscCore},
  W/.style={ppscConstraint},
  bond/.style={ppscBond},
  sym/.style={ppscSym}
  ]

  \def\dotStub{2.5mm}

  \node[S, anchor=west] (Sig) at (0,0) {$\Sigma^{(D)}_{s}$};

  \node[sym] (eqtop) at (3.45,0) {$\approx$};
  \node[M, anchor=west] (M1) at (3.85,0) {$M^{(1)}$};
  \node[M, anchor=west] (M2) at (5.25,0) {$M^{(2)}$};
  \node[M, anchor=west] (Md) at (7.25,0) {$M^{(D)}$};
  \coordinate (dots) at ($($(M2.east)+(\dotStub,0)$)!0.5!($(Md.west)+(-\dotStub,0)$)$);

  \draw[bond] (M1.east) -- (M2.west);
  \draw[bond] (M2.east) -- ++(\dotStub,0);
  \draw[bond] (Md.west) -- ++(-\dotStub,0);
  \node[ppscDots] at (dots) {$\cdots$};

  \def\wY{0.8}%
  \def\wMargin{1mm}%

  \path let
  \p1 = ($(Md.north east)+( \wMargin,0)$),
  \p2 = ($(M1.north west)+(-\wMargin,0)$),
  \n1 = {\x1-\x2}
  in
  node[W, minimum width=\n1] (Wbar)
  at ($(M1.north west)!0.5!(Md.north east)+(0,\wY)$)
  {$s_1 \le s_2 \le \cdots \le s_D = s$};

  \draw[bond] (Wbar.south -| M1.north) -- (M1.north);
  \draw[bond] (Wbar.south -| M2.north) -- (M2.north);
  \draw[bond] (Wbar.south -| Md.north) -- (Md.north);

  \pgfresetboundingbox
  \path[use as bounding box] (-0.05,-0.62) rectangle (8.65,1.45);

\end{tikzpicture}}
  \caption{TT-factorized representation of the integrand with its time-ordering constraint.}
  \label{fig:realtime_recursion}
\end{figure}
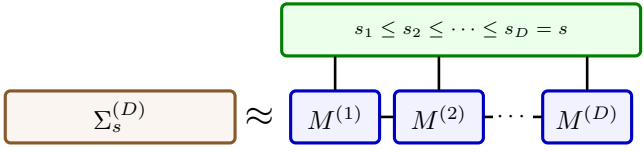

In the time decomposition, we attempt a decomposition of the discretized integrand in Eq.~\eqref{eq:X_diagram_K} in terms of the discretized physical time indices $t_d \equiv -s_d \Delta t$, with $s_d\in(0,1,\ldots,s_{\rm max})$, where $-t_c=- s_{\rm max}\Delta t$ is a lower time cutoff for the infinite time interval:
\begin{equation}
\label{decomp-time}
  I^{(\mathcal{D},\bm \sigma)}_{s_1, \dots, s_{D}} \approx M^{(1)}_{s_1} M^{(2)}_{s_2} \cdots  M^{(D)}_{s_{D}}
\end{equation}
Notably, the integrand can be {\em continuously} extended to the rectangular domain $\{t_d\in [-t_c,0]  \text{~for all~} d=1,\ldots,D\}$ by fixing the contour components (lesser or greater) of all factors $\pG$ and $\Delta$ to the configuration they take within the time-ordered domain.
In Eq.~\eqref{decomp-time}, we therefore sample the integrand on this extended rectangular domain $s_d\in(0,\ldots,s_{\rm max})$. The discretized integral \eqref{eq:X_diagram_K} must ultimately be summed only over the time-ordered domain given by $0\le s_1\le\cdots \le  s_e\le\cdots \le s_D \le s_{\rm max}$, where $e(\mathcal D,\bm \sigma)$ corresponds to the external argument $t_e=-s_e\Delta t$. For example, if the external argument is $e=D$, as for the greater component of the self-energy, the discretized integral becomes 
\begin{equation}
  \label{eq:time_sum}
  \begin{aligned}
    X_{s_e} &= 
    \Big[ \sum_{s_{D - 1} = 0}^{s_e} \dots \Big[ \sum_{s_2 = 0}^{s_3} \Big[ \sum_{s_1 = 0}^{s_2} M^{(1)}_{s_1} \Big] M^{(2)}_{s_2} \Big] \\ \cdots &M^{(D - 1)}_{s_{D-1}}\Big] M^{(D)}_{s_{D} = s_e},
  \end{aligned}
\end{equation}
taking a simple Riemann integration with unity quadrature weights, as illustrated in Fig.~\ref{fig:realtime_recursion}. As the integrand is factorized, the sum can be efficiently evaluated using the recursive relation
\begin{equation}
  \label{eq:time_recursion_forwards}
  X^{(d)}_s = X^{(d)}_{s - 1} + X^{(d - 1)}_s \cdot M^{(d)}_s 
\end{equation}
for $s=0,1,\ldots$, with the starting conditions  $X^{(0)}_s = 1$ and $X^{(d)}_{-1}=0$; the final result is $X_{s_e}=    (\Delta t)^{D-1}X^{(D)}_{s_e}$. Since, for each new time slice, we perform only $D$ additions and multiplications, the formal runtime complexity is expected to be linear in the number of time steps, $\mathcal{O}(D N_t \chi^2)$. For the lesser component $\pSigma^<$ or for a typical $C$ diagram, the external time index does not correspond to the last time variable ($e\neq D$). In this case we split the ordered sum into contributions from below the external argument
\begin{equation}
X_s^{(L)} \!=\!\sum_{s_{e-1}=0}^{s} 
\cdots\Big[ \sum_{s_2=0}^{s_3}  \Big[\sum_{s_1=0}^{s_2} M^{(1)}_{s_1} \Big] M^{(2)}_{s_2} \Big]\cdots M^{(e-1)}_{s_{e-1}},
\end{equation}
and from above the external argument
\begin{equation}
 X_s^{(U)} \!=\!\sum_{s_{e+1}=s_{e}}^{s_{\rm max}} \!\!\!\! M^{(e+1)}_{s_{e+1}}
\cdots \Big[ \!\!\!\!\!\!\!\sum_{s_{D-1}=s_{D-2}}^{s_{\rm max}} \!\!\!\!\!\!\!\! M^{(D-1)}_{s_{D-1}}  
\Big[\!\!\!\! \sum_{s_D=s_{D-1}}^{s_{\rm max}} \!\!\!\! M^{(D)}_{s_D} \Big]  \Big],
\end{equation}
and combine
\begin{equation}
X_{s_e} = (\Delta t)^{D-1}
X_{s_e}^{(L)} M^{(e)}_{s_{e}} X_{s_{e}}^{(U)}.
\end{equation}
Here $X^{(L)}$ is given by the forward recursion Eq.~\eqref{eq:time_recursion_forwards}, while $X^{(U)}$ can be computed efficiently using the backward recursion
\begin{equation}
  \label{eq:time_recursion_backwards}
X^{(d)}_s = X^{(d)}_{s + 1} + M^{(d)}_s \cdot X^{(d + 1)}_s.
\end{equation}
for $s=s_{\rm max},s_{\rm max}-1,\ldots,0$, with the starting conditions $X^{(D+1)}_s = 1$ and $X^{(d)}_{s_{\rm max}+1}=0$.

The inclusion of higher-order integration weights in Eq.~\eqref{eq:time_sum} and the sums for $X^{(U)}$ and $X^{(L)}$ would be possible with a modified recursion or an explicit summation of boundary corrections.
However, since the required bond dimensions of the time-decomposition scheme 
turns out to be
rather high in our benchmarks below, we do not further investigate an efficient scheme for incorporating such integration weights.

\subsection{Quantics Time Decomposition}
\label{ch:QuanticsTime}

In the time decomposition \eqref{decomp-time}, one may ask whether one could instead of sampling the integrand on the extended rectangular domain directly decompose the product $ I^{(\mathcal{D},\bm \sigma)}_{s_1, \dots, s_{D}}  \Theta_{s_1, \dots, s_{D}}$, including the  time-ordering constraint 
\begin{align}
 \Theta_{s_1, \dots, s_{D}} = \begin{cases}
 1 & 0\le s_1\le\cdots \le s_D \le s_{\rm max}
 \\
 0 & \text{~else}
\end{cases}.
\label{Theta}
\end{align}
However, one can easily see that the matrix $ \Theta_{s_1,s_2}$ has full rank, 
and we therefore neither expect the full tensor $ \Theta_{s_1, \dots, s_{D}}$ to be directly decomposable.

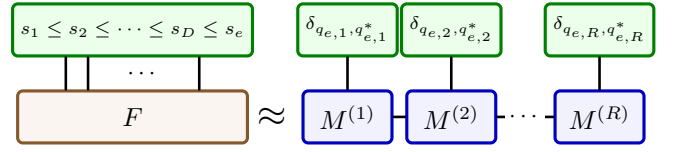
\begin{figure}[tbp]
  \centering
  \resizebox{\columnwidth}{!}{\begin{tikzpicture}[
  Fbig/.style={ppscWideTensor, minimum width=31mm},
  Wlhs/.style={ppscConstraint, minimum width=31mm},
  Mcore/.style={ppscCore},
  Wlocal/.style={ppscWeight},
  bond/.style={ppscBond},
  sym/.style={ppscSym},
  arr/.style={ppscArrow}
  ]

  \def\maskY{8mm}
  \def\dotStub{1.8mm}

  \def\tLone{-9mm}
  \def\tLtwo{-6mm}
  \def\tR{ 9mm}

  \node[Fbig, anchor=west] (F) at (0,0) {$F$};

  \node[Wlhs] (Mask)
  at ($(F.north)+(0,\maskY)$)
  {$s_1 \le s_2 \le \cdots \le s_D \le s_e$};

  \coordinate (mL1) at ($(Mask.south)+(\tLone,0)$);
  \coordinate (mL2) at ($(Mask.south)+(\tLtwo,0)$);
  \coordinate (mR)  at ($(Mask.south)+(\tR,0)$);

  \coordinate (fL1) at ($(F.north)+(\tLone,0)$);
  \coordinate (fL2) at ($(F.north)+(\tLtwo,0)$);
  \coordinate (fR)  at ($(F.north)+(\tR,0)$);

  \draw[bond] (mL1) -- (fL1);
  \draw[bond] (mL2) -- (fL2);
  \draw[bond] (mR)  -- (fR);

  \coordinate (timeX) at ($(fL2)!0.5!(fR)$);
  \coordinate (timeY) at ($(F.north)!0.5!(Mask.south)$);
  \node[ppscDots] at (timeX |- timeY) {$\cdots$};

  \node[sym] at (3.45,0) {$\approx$};
  \node[Mcore, anchor=west] (M1) at (3.85,0) {$M^{(1)}$};
  \node[Mcore, anchor=west] (M2) at (5.25,0) {$M^{(2)}$};
  \node[Mcore, anchor=west] (MR) at (7.25,0) {$M^{(R)}$};
  \coordinate (Mdots) at ($($(M2.east)+(\dotStub,0)$)!0.5!($(MR.west)+(-\dotStub,0)$)$);

  \draw[bond] (M1.east) -- (M2.west);
  \draw[bond] (M2.east) -- ++(\dotStub,0);
  \draw[bond] (MR.west) -- ++(-\dotStub,0);
  \node[ppscDots] at (Mdots) {$\cdots$};

  \def\sumY{8mm}
  \node[Wlocal] (W1) at ($(M1.north)+(0,\sumY)$)
    {$\delta_{q_{e,1},q^*_{e,1}}$};
  \node[Wlocal] (W2) at ($(M2.north)+(0,\sumY)$)
    {$\delta_{q_{e,2},q^*_{e,2}}$};
  \node[Wlocal] (WR) at ($(MR.north)+(0,\sumY)$)
    {$\delta_{q_{e,R},q^*_{e,R}}$};

  \draw[bond] (W1.south) -- (M1.north);
  \draw[bond] (W2.south) -- (M2.north);
  \draw[bond] (WR.south) -- (MR.north);

  \pgfresetboundingbox
  \path[use as bounding box] (-0.05,-0.62) rectangle (8.65,1.45);

\end{tikzpicture}}
  \caption{Illustration of the masked sampling employed for the quantics time decomposition, which results in factorized integration boundaries. The local Kronecker factors fix the external-time bits $q_{e,r}$ to their target values $q^*_{e,r}$.}
  \label{fig:quantics_time_contraction}
\end{figure}

This changes when the arguments $s_d$ are further split using the quantics representation \eqref{eq:quantics_representation}. It can be shown analytically (see App.~\ref{app:quantics}) that the Heaviside function $\Theta(t - t')$ is represented with a low bond dimension $\chi = 2$ in a fused quantics parametrization. 
 This
 motivates a decomposition of the masked integrand
\begin{equation}
 I^{(\mathcal{D},\bm \sigma)}_{s_1, \dots, s_{D}} \Theta_{s_1, \dots, s_{D}} 
 \approx
 M^{(1)}_{\bm q_1} M^{(2)}_{\bm q_2} \cdots M^{(R)}_{\bm q_{R}}
\end{equation}
using the fused quantics parametrization \eqref{eq:fused_quantics_representation}.
Fixing the target bits $q^*_{e,r}$ through
$s_e = (q^*_{e,1}, \cdots, q^*_{e,R})_2$, the time-ordered integration
in \eqref{eq:X_diagram_K} is given by the unrestricted, factorized sum
illustrated in Fig.~\ref{fig:quantics_time_contraction},
\begin{equation}
X_{s_e}^{(\mathcal{D},\bm \sigma)}
= ( \Delta t)^{D-1}
\prod_{r=1}^{R} \sum_{\bm{q}_r} M^{(r)}_{\bm{q}_r}
   \delta_{q_{e,r}, q^*_{e,r}},
\end{equation}
where the Kronecker factor $\delta_{q_{e,r},q^*_{e,r}}$ fixes the bit
associated with the external dimension at scale $r$, ensuring that the
correct external time $s_e$ is recovered. Reduction of the internal sums
for each $r$ corresponds to a total runtime complexity
$\mathcal{O}(2^{D}R\chi^2)$. The resulting TT with external arguments
$q^*_{e,r}$ can be evaluated into a dense vector
$X_{s_e}^{(\mathcal{D},\bm \sigma)}$ with a runtime complexity
$\mathcal{O}(2^R R \chi^3)
=\mathcal{O}(N_t \log_2(N_t)\chi^3)$.
If the quantity $X$ ($\pSigma$ or $C$) is not needed in the dense representation (i.e.,~in case one can work directly with the QTT representation), one could avoid the last contribution. The exponential contribution $\sim2^D$ is due to the fused index, which could be avoided by an interleaved quantics representation. We will, however, omit a detailed discussion of the interleaved representation, as preliminary tests in practice showed significantly worse decomposition performance already on the Heaviside mask, since analytically only the fused decomposition guarantees a low-rank (again see App.~\ref{app:quantics}). In the quantics time decomposition  scheme one can also include higher-order integration weights during the sampling process by replacing the uniform $\Theta_{s_1, \dots, s_{D}}$ by a weight tensor containing the boundary correction weights from the Gregory quadrature.

\subsection{Time Difference Decomposition}
\label{ch:TimeDifference}

\begin{figure}[tbp]
  \centering
  \begin{tikzpicture}[
  F/.style={ppscCore},
  S/.style={ppscTensor, minimum width=14mm},
  Op/.style={ppscOperator},
  Wbar/.style={ppscConstraint},
  arr/.style={ppscArrow}
  ]

  \def\hgap{1.8mm}
  \def\biggap{1.6mm}
  \def\rowgap{12mm}

  \node[F] (f1) {$M^{(1)}$};
  \node[Op, right=\hgap of f1] (c2) {$*$};
  \node[F, right=\hgap of c2] (f2) {$M^{(2)}$};
  \node[Op, right=\hgap of f2] (c3) {$*$};
  \node[right=\hgap of c3] (dotsT) {$\cdots$};
  \node[F, right=\hgap of dotsT] (fn) {$M^{(D - 1)}$};
  \node[S, right=\hgap of fn] (SigT) {$\Sigma^{(D)}$};

  \draw[arr] (f1) -- (c2);
  \draw[arr] (c2) -- (f2);
  \draw[arr] (f2) -- (c3);
  \draw[arr] (c3) -- (dotsT);
  \draw[arr] (dotsT) -- (fn);
  \draw[arr] (fn) -- (SigT);

  \def\wY{7.2mm}
  \def\wMargin{0.8mm}

  \path let
  \p1 = ($(fn.north east)+(\wMargin,0)$),
  \p2 = ($(f1.north west)+(-\wMargin,0)$),
  \n1 = {\x1-\x2}
  in
  node[Wbar, minimum width=\n1] (TauBar)
  at ($(f1.north west)!0.5!(fn.north east)+(0,\wY)$)
  {$\tau_j < -t - \sum_{k=0}^{j-1}\tau_k$};

  \draw[arr] (TauBar.south -| f1.north) -- (f1.north);
  \draw[arr] (TauBar.south -| f2.north) -- (f2.north);
  \draw[arr] (TauBar.south -| fn.north) -- (fn.north);

  \begin{scope}[yshift=-\rowgap]

    \node[F]  (Ff1) at (f1 |- 0,0) {$\hat M^{(1)}$};
    \node[Op] (m2)  at (c2 |- 0,0) {$\cdot$};
    \node[F]  (Ff2) at (f2 |- 0,0) {$\hat M^{(2)}$};
    \node[Op] (m3)  at (c3 |- 0,0) {$\cdot$};
    \node      (dotsW) at (dotsT |- 0,0) {$\cdots$};
    \node[F]  (Ffn) at (fn |- 0,0) {$\hat M^{(D-1)}$};
    \node[S]  (SigW) at (SigT |- 0,0) {$\hat\Sigma^{(D)}$};

    \draw[arr] (Ff1) -- (m2);
    \draw[arr] (m2) -- (Ff2);
    \draw[arr] (Ff2) -- (m3);
    \draw[arr] (m3) -- (dotsW);
    \draw[arr] (dotsW) -- (Ffn);
    \draw[arr] (Ffn) -- (SigW);

  \end{scope}

  \draw[arr]
  (f1.south) --
  node[ppscLegLabel,right,fill=white] {FFT}
  (Ff1.north);

  \draw[arr]
  (SigW.north) --
  node[ppscLegLabel,right,fill=white] {iFFT}
  (SigT.south);

\end{tikzpicture}
  \caption{Solution of the time-ordered integration in the time difference parametrization via successive convolution, which can be efficiently implemented in the frequency domain via a Fourier transform.}
  \label{fig:timefreq_fft_flow}
\end{figure}
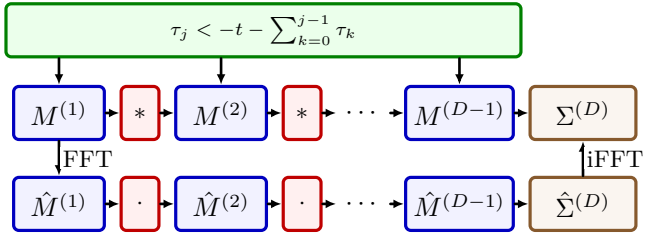

In this section we summarize the time difference decomposition used in Ref.~\cite{Eckstein-2024}.
The time difference decomposition is inspired by fact that in the steady-state we have time translational invariance and all functions only depend on the time difference $t - t'$; especially $\pG(t, t') = \pG(t - t')$ and $\Delta(t, t') = \Delta(t - t')$.
Hence, it is not far-fetched to expect that also products of such functions will have a low-rank representation in this parametrization. 
This approach is further motivated by the 
real-time generalization  \cite{Paprotzki-Eckstein-2025} of the sum of exponentials decomposition \cite{Kaye-etal-2024}.
Following the derivation in \cite{Eckstein-2024} we represent the discretized integrand in the physical time differences $t[s_d] - t[s_{d+1}] = (s_d - s_{d+1}) \Delta t \equiv \tau_{d} \Delta t \geq 0$
\begin{equation}
  I^{(\mathcal{D})}_{\tau_1, \dots, \tau_{D-1}} \approx M^{(1)}_{\tau_{D-1}} M^{(2)}_{\tau_{D-2}} \cdots  M^{(D-1)}_{\tau_1},
\end{equation}
with an inverted order of the time difference indices.
The integral can then be solved recursively by
\begin{equation}
  \label{eq:diff_recursion}
  \begin{aligned}
  L^{(1)} &= M^{(1)}, \quad L^{(d)} = L^{(d-1)}* M^{(d)}  \\
    R^{(D)} &= 1, \quad R^{(d)} = R^{(d+1)} * M^{(d)} \\
    \Sigma_{\tau_e}^{(\mathcal{D})} &\approx C^{(\mathcal{D})} R^{(e)}_{n_t - 1 - \tau_e} L^{(e - 1)}_{\tau_e},
  \end{aligned}
\end{equation}
where $*$ denotes the retarded convolution integral.
Structurally this is very similar to the forward and backward recursion obtained in the time decomposition (see Eqs.~\eqref{eq:time_recursion_forwards} and \eqref{eq:time_recursion_backwards}).
We are able to speed up the $\mathcal{O}(N_t^2)$ runtime complexity to $\mathcal{O}(N_t \log (N_t))$ using a fast Fourier transform (FFT) to evaluate the convolutions
(see Ref.~\cite{Eckstein-2024} for details, and Fig.~\ref{fig:timefreq_fft_flow} for illustration).

\subsection{Quantics Time Difference Decomposition}
\label{ch:QuanticsDifference}

In order to leverage the favorable sampling complexity of quantics tensor cross interpolation while retaining a low-rank approximation, we also explore a decomposition of the integrand as a function of the quantics bits $q_{d,r}$ corresponding to the time-difference indices $\tau_d$.
In general, this decomposition can be formulated using either a natural, interleaved, or fused ordering of the bits \cite{Fernández-etal-2025}.
An important observation is that the fused ordering enables an efficient evaluation of the retarded convolutions.
After obtaining the factorized integrand in terms of the fused indices $\bm q_r$,
\begin{equation}
  I^{(\mathcal{D})}_{\underline{\bm q}}
  \approx
  M^{(1)}_{\bm q_1} M^{(2)}_{\bm q_2} \cdots M^{(R)}_{\bm q_R},
\end{equation}
we proceed analogously to the time-difference decomposition and apply the recursion of
Eq.~\eqref{eq:diff_recursion}.
In principle, the convolutions could be reduced  to TT multiplications using the quantics Fourier transform, at the cost of a moderate increase in the bond dimension \cite{Fernández-etal-2025}. 
 Instead, we implement the convolutions directly in terms of the fused bits $\bm q_r$ using an auxiliary carry bit. As derived in detail in App.~\ref{app:quantics}, this operation can be formulated as a low-rank MPO. 
In this way, also
 trapezoidal endpoint corrections can be incorporated directly into the QTT convolution with 
 moderate overhead. Overall, the direct QTT convolution increases the bond dimension by only a factor of two for the Riemann sum and by at most a factor of four (with a block-diagonal structure) when trapezoidal weights are included. This compares favorably with the approximately $11$-fold increase incurred by the quantics Fourier transform \cite{Fernández-etal-2025}.

The direct QTT convolution can, in principle, be formulated for both the fused and the interleaved ordering (App.~\ref{app:quantics}). The primary difference between the two lies in the TCI sampling of the tensor train. In particular, for higher diagram orders, sampling in the interleaved ordering is expected to reduce the sampling cost from $\mathcal{O}(2^D R \chi^3)$ to $\mathcal{O}(D R \chi^3)$. However, special care is required, since the interleaved ordering may identify suboptimal pivots, terminate prematurely, and generally lead to larger bond dimensions $\chi$, as correlations between different scales of the same dimension may have to propagate through many tensor cores. Rather than reformulating the direct QTT convolution for the interleaved ordering, one can instead perform the sampling in the interleaved ordering and subsequently fuse the cores on each level $r$,
\begin{equation}
  M^{(r)}(\bm q_r)
  =
  \prod_{d=1}^{D}
  \tilde M^{(d,r)}(q_{d,r}),
\end{equation}
thereby reusing the convolution algorithm developed for the fused QTT representation.

\section{Benchmark and Comparison}

To assess the efficiency of the proposed algorithms, we 
first study synthetic diagrammatic kernels, allowing us to isolate the performance of the tensor decompositions independently of the impurity solver. We then investigate how the resulting tensor and quadrature errors propagate through equilibrium DMFT calculations before comparing the different algorithms within self-consistent nonequilibrium steady-state DMFT using the nonthermal distribution of Ref.~\onlinecite{Künzel-etal-2024}.

\subsection{Gaussian Scalar Test}
\label{ch:Gaussian}
In this synthetic benchmark, inspired by \cite{Eckstein-2024}, the pseudo-particle propagators and hybridization functions are given by
\begin{equation}
  \begin{aligned}
    \mathcal{G}^> &= -i e^{-0.01 t^2} (e^{i t} + e^{0.33 i t}) \\ 
    \mathcal{G}^< &= -i e^{-0.01 t^2} (e^{-0.14 i t} + e^{0.2 i t}) \\
    \Delta^> &= -i e^{-t^2} (e^{0.28 i t} + e^{0.23 i t}) \\ 
    \Delta^< &= -i e^{-t^2} (e^{0.2 i t} + e^{-0.44 i t}),
  \end{aligned}
\end{equation}
and the vertex operators are set to unity such that we work in a scalar local Hilbert space. This test evaluates the OCA self-energy topology of Eq.~\eqref{eq:oca_selfenergy_diagram} on the fixed interval $t\in[-8,0]$, discretized by a uniform grid of $N_t$ points, such that $\Delta t = t_c / (N_t - 1)$ with $t_c = 8$.
For each grid, the reference result is obtained by evaluating the discretized integral with the same quadrature rule as in the corresponding TCI calculation, which allows for an isolated analysis of the TCI error through the root mean square error (RMSE) between the two results. The convergence of the TCI approximation is then examined by varying the maximally allowed bond dimension $\chi$, while keeping the time grid and the remaining sampling parameters fixed.

\begin{figure}[t]
  \centering
  \includegraphics{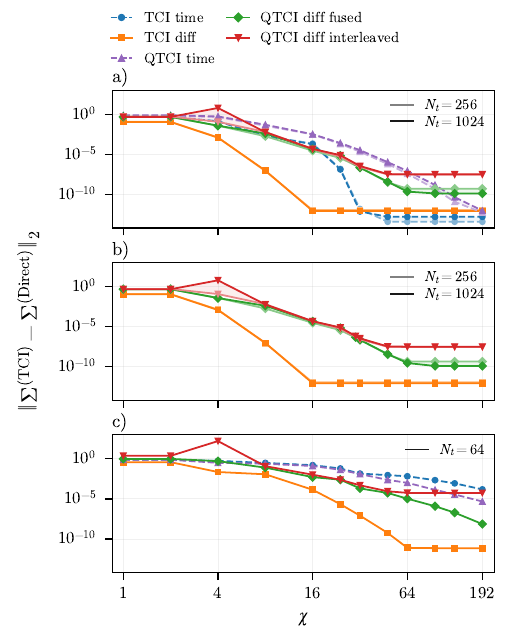}
  \caption{
  Gaussian benchmark efficiency for a) OCA, b) OCA with trapezoidal endpoint corrections and c) TOA self-energies as a function of the maximum allowed bond dimension. The error is calculated with respect to the respective direct quadrature.}
  \label{fig:gaussian_accuracy_vs_chi}
\end{figure}

In Fig.~\ref{fig:gaussian_accuracy_vs_chi}(a), we compare the error of the diagrammatic integral (more precisely, its discrete Riemann sum) obtained with the various TCI decomposition strategies against the direct evaluation. Among the tested parametrizations, the time difference representation requires the smallest bond dimension ($\chi=16$) to reach machine precision. The time decomposition ($\chi=48$) and quantics time decomposition ($\chi=192$) require substantially larger bond dimensions, while both quantics difference variants saturate at higher errors ($\sim10^{-9}$ for the fused and $\sim10^{-7}$ for the interleaved representation) at approximately $\chi=64$. The saturation behavior may be related to the difficulty of identifying suitable pivot points in sparse tensors \cite{Fernández-etal-2025,Geng-Kim-Werner-2025}. Nevertheless, the resulting TCI error of order $10^{-8}$ is already comparable to or smaller than the quadrature error encountered in typical calculations. Importantly, the required bond dimension is independent of the number $N_t$  of grid points in the integration interval, indicating that all four parametrizations capture a stable low-rank representation of the integrand.

The results shown in Fig.~\ref{fig:gaussian_accuracy_vs_chi}(a) deliberately exclude trapezoidal or higher-order quadrature weights in order to isolate the error introduced by the tensor decomposition itself. Including the integration weights modifies the approximation error only slightly, as shown in Fig.~\ref{fig:gaussian_accuracy_vs_chi}(b) for the time difference parametrization. 
We have implemented and tested higher-order quadrature weights only for the time difference representations  (Secs.~\ref{ch:TimeDifference} and \ref{ch:QuanticsDifference}), as the time parametrizations (Secs.~\ref{ch:TimeDecomposition} and \ref{ch:QuanticsTime}) ultimately prove less competitive. 
Finally, Fig.~\ref{fig:gaussian_accuracy_vs_chi}(c) shows the convergence for the next order in the expansion. Here, all methods require approximately a fourfold increase in bond dimension to achieve a comparable accuracy.

\begin{figure}[t]
  \centering
  \includegraphics{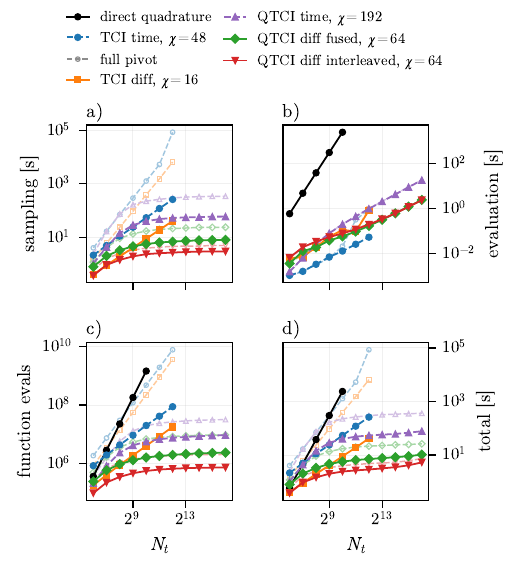}
  \caption{Runtime performance of the Gaussian benchmark for the OCA self-energies without integration weights. The timing components are a) sampling, b) evaluation, and d) total runtime, all reported in seconds; c) shows the number of function evaluations. Rook pivoting (bold lines) and full pivoting (shallow lines) are compared. The bond dimension per method is chosen such that the respective TCI error (see Fig. \ref{fig:gaussian_accuracy_vs_chi}) is at the minimal error plateau to ensure a fair comparison.}
  \label{fig:gaussian_oca_runtime}
\end{figure}

Figure~\ref{fig:gaussian_oca_runtime} shows an analysis of the actual runtime for the unweighted OCA integration on an Apple M3 laptop CPU. The runtime is analyzed as a function of the time grid size  $N_t$
with a bond dimension $\chi$ chosen sufficiently large for the  integration  error  to reach the respective error plateau at the minimal integration error (see legend).  For the TCI-based integration, we split the total time into the sampling time needed to construct the TT [Fig.~\ref{fig:gaussian_oca_runtime}(a)] and the evaluation of the integral from the TT 
representation of the integrand
[Fig.~\ref{fig:gaussian_oca_runtime}(b)]. The total time is shown in Fig.~\ref{fig:gaussian_oca_runtime}(d). 
The pivot search in the TCI sampling can be performed using either \texttt{rook-pivoting} or  \texttt{full-pivoting}, see full lines and shaded lines in  Fig.~\ref{fig:gaussian_oca_runtime}(a) and (c).  The comparison in Tab.~\ref{tab:gaussian_rook_pivot_comparison} shows that the largely improved performance of the \texttt{rook-pivoting} approach comes at no significant loss in accuracy, and we discuss only the \texttt{rook-pivoting} results in the following.

For the direct evaluation, there is no sampling, and only the evaluation time and total time are shown, which 
scale like $\mathcal{O}(N_t^3)$ with $N_t$.
For the TCI-based integration, the evaluation scales roughly linearly for all methods [see Fig.~\ref{fig:gaussian_oca_runtime}(b)], with a small prefactor. In comparison to Fig.~\ref{fig:gaussian_oca_runtime}(a), however, the runtime of all methods is largely dominated by the sampling time, which is proportional to the number of integrand evaluations [Fig.~\ref{fig:gaussian_oca_runtime}(c)].  Figures~\ref{fig:gaussian_oca_runtime}(a) and \ref{fig:gaussian_oca_runtime}(c) show that the plain TCI methods reduce the polynomial integration cost to roughly linear scaling during sampling (for \texttt{rook-pivoting}), while the QTCI methods exhibit logarithmic runtime behavior. Already at relatively small $N_t \approx 2^{10}$, this results in better overall runtime performance of the QTCI difference parametrizations, despite a significantly higher bond dimension. 

When comparing the TCI time and TCI difference methods, the lower bond dimension for the latter leads to a constant reduction in runtime. This overall analysis motivates restricting the further convergence analysis to the difference parametrization, with and without quantics.  

\begin{figure}[t]
  \centering
  \includegraphics{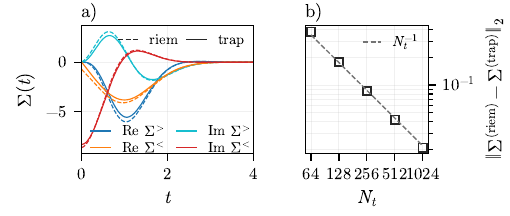}
  \caption{Direct Gaussian OCA quadrature comparison between Riemann and trapezoidal integration for a representative single run with $N_t = 256$ in a) showcasing the violation of the hermitian property, and the expected $1 / N_t$ scaling of the difference between the Riemann and trapezoidal rule in b).}
  \label{fig:gaussian_direct_trapz_single_run}
\end{figure}

At last, we use this Gaussian benchmark to emphasize the importance of 
trapezoidal endpoint corrections, which 
are
significant because the integrand has a high weight at the lower integration boundary. Through  the nested integration, errors propagate and  small inaccuracies at the boundary values can lead to significant deviations in the short time behavior. In  particular, the error can imply  a violation of the hermitian property of the lesser self-energy, as illustrated in Fig.~\ref{fig:gaussian_direct_trapz_single_run}(a) for a rather sparse grid. The slow $1/N_t$ decay of the integration error in the Riemann integration (Fig.~\ref{fig:gaussian_direct_trapz_single_run}(b)) would eventually require very dense grids, which become unfavorable even though the sampling  effort within the QTCI approach increases only logarithmically with $N_t$ (in particular for test cases where the self-energies decay more slowly than in the Gaussian test case).

\begin{table}[t]
  \centering
  \scriptsize
  \setlength{\tabcolsep}{3pt}
  \begin{tabular}{lcccc}
\hline
method & $\chi$ & $N_t$ & rel. RMSE & speedup \\
\hline
TCI time & 48 & 4096 & $3.1\times 10^{-13}$ & 326.9 \\
TCI diff & 16 & 4096 & $2.7\times 10^{-13}$ & 156.1 \\
QTCI time & 192 & 65536 & $4.4\times 10^{-12}$ & 4.7 \\
QTCI diff fused & 64 & 65536 & $2.5\times 10^{-10}$ & 2.5 \\
QTCI diff interleaved & 64 & 65536 & $6.6\times 10^{-10}$ & 1.4 \\
\hline
\end{tabular}

  \caption{Comparison of full-pivot and rook-pivot TCI runs for the representative Gaussian runtime curves at the largest common $N_t$ available for each method. The relative RMSE compares the resulting self-energies, and the speedup is the total runtime ratio.}
  \label{tab:gaussian_rook_pivot_comparison}
\end{table}

\subsection{Equilibrium DMFT and Convergence}
\label{ch:BetheEquil}

In this section we apply the 
TCI-based
impurity solver to the equilibrium DMFT solution of the Hubbard model
\begin{equation}
  \label{eq:hubbard_model}
  H_\text{hub} = t_h \sum_{\sigma, \langle i, j \rangle} c^\dagger_{i, \sigma} c_{j, \sigma} + U \sum_i n_{i, \uparrow} n_{i, \downarrow} + \mu \sum_i n_i
\end{equation}
with nearest-neighbour hopping amplitude $t_h$ and onsite interaction $U$ at half filling ($\mu = - U / 2$).
Here $c_{i, \sigma}$ and $c^\dagger_{i, \sigma}$ are annihilation and creation operators of an electron with spin $\sigma$ on site $i$ and $n_{i, \sigma}$ is the corresponding density while the total density per site is given by $n_i = n_{i, \uparrow} + n_{i, \downarrow}$.
The model is solved on the Bethe lattice, which  implies a simple close form self-consistency relation; in the steady state,
\begin{align}
\Delta^{R,<}(\omega) = t_{\rm h}^2 G^{R,<}(\omega).
\label{deltedmaft}
\end{align}

\begin{figure}[t]
  \centering
  \includegraphics{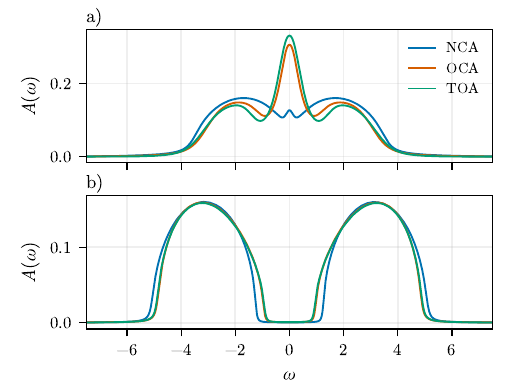}
  \caption{Spectral function of the half-filled Hubbard model in equilibrium on the Bethe lattice, comparing the NCA, OCA, and TOA strong-coupling expansion orders for a) a metallic solution at $U=3$ and b) a Mott-insulating solution at $U=6$. Both spectra are calculated at $\beta=7$ and $t_h=1$.}
  \label{fig:dmft_bethe_equil_order_spectrum}
\end{figure}

Figure~\ref{fig:dmft_bethe_equil_order_spectrum} illustrates the characteristic equilibrium DMFT spectra of the half-filled Hubbard model showcasing the dependence on the truncation order within a strong-coupling expansion around the atomic limit. The local interaction redistributes spectral weight from the noninteracting band into lower and upper Hubbard bands, which are associated with the removal or addition of an electron to a singly occupied site. In the metallic regime at $U=3$, these incoherent features coexist with a narrow quasiparticle resonance around the Fermi level $\omega=0$. By contrast, for the larger interaction $U=6$, the low-energy spectral weight is suppressed and the separation of the Hubbard bands produces the Mott gap of the insulating solution.
The comparison between NCA, OCA, and TOA also highlights the different convergence properties of the strong-coupling expansion in these two regimes. The diagrammatic series rapidly converges for 
large $U$
in the Mott insulating regime,
while
in the correlated metal higher-order diagrammatic contributions are necessary to accurately describe the coherent low-energy resonance.
Benchmarks below are for the metallic regime ($U = 3$, $t_{\rm h} = 1$, $\beta = 7$), where high-order corrections are significant. The large time cutoff $t_{\rm c}=327.67$ for the self-energy evaluation is fixed for all simulations, such that the pseudo-particle propagators are sufficiently decayed at $t_{\rm c}$ already for the NCA solution.

\begin{figure}[t]
  \centering
  \includegraphics{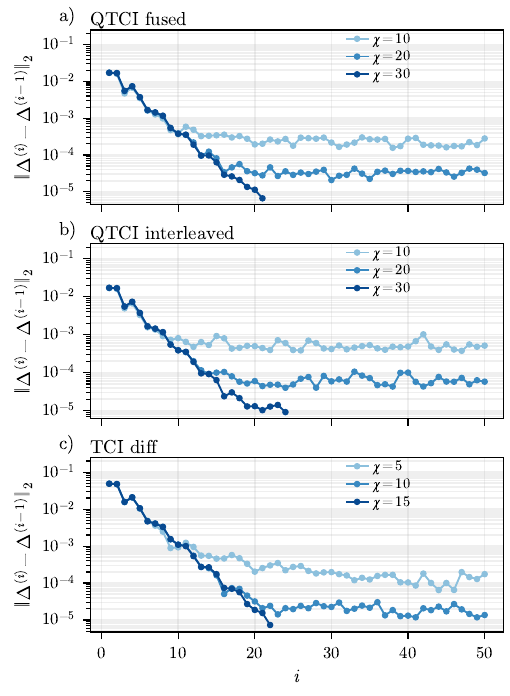}
  \caption{
  Convergence analysis plot of the self-consistent DMFT cycle showcasing the RMSE of two successive hybridization functions $\Delta$ as a function of the DMFT iteration $i$ for different values of the maximally allowed bond dimension $\chi$ for a) the fused QTCI difference parametrization, b) the interleaved QTCI difference parametrization, and c) the non-quantics TCI difference parametrization. The time grids are fixed at $\log_2(N_t)=15$, $15$, and $12$, respectively, and trapezoidal weights are used for the QTCI calculations while fifth-order Gregory weights are used for TCI.}
  \label{fig:dmft_bethe_equil_chi}
\end{figure}

\begin{figure}[t]
  \centering
  \includegraphics{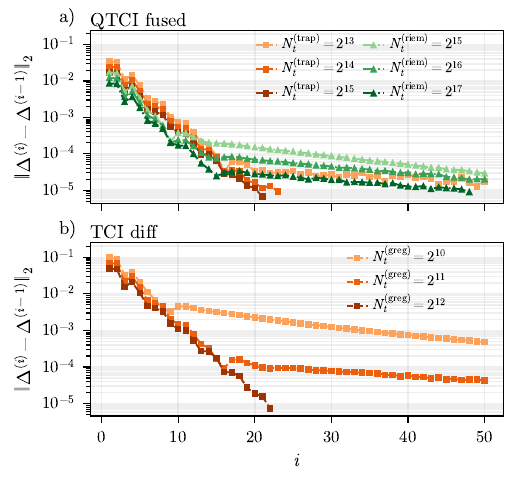}
  \caption{Convergence analysis plot as in Fig.~\ref{fig:dmft_bethe_equil_chi} for increasing number of time steps $N_t$ on the fixed cutoff time $t_c = 327.67$ (resulting in decreasing time steps $\Delta t$), for a) fused QTCI at fixed $\chi=30$ and b) non-quantics TCI difference parametrization at fixed $\chi=15$. For fused QTCI, trapezoidal (trap) and Riemann (riem) quadrature are compared, whereas the TCI calculations use fifth-order Gregory (greg) quadrature, which illustrates the importance of including at least trapezoidal quadrature weights.}
  \label{fig:dmft_bethe_equil_nt}
\end{figure}

The goal is to analyze the influence of both integration quadrature inaccuracies and the TCI decomposition error on the fully  self-consistent DMFT solution.  
Using the TCI based solver, we find that the root mean square error (RMSE) of the hybridization function $\Delta$ between successive DMFT iterations first decreases exponentially with the DMFT iteration, but then saturates at an error plateau which is controlled by the accuracy of the solution (Figs.~\ref{fig:dmft_bethe_equil_chi} and \ref{fig:dmft_bethe_equil_nt}). This behavior is similar to what is found for stochastic impurity solvers. We can therefore take this saturated DMFT error as a diagnostic for the accuracy of the DMFT solution.

\begin{figure}[t]
  \centering
  \includegraphics{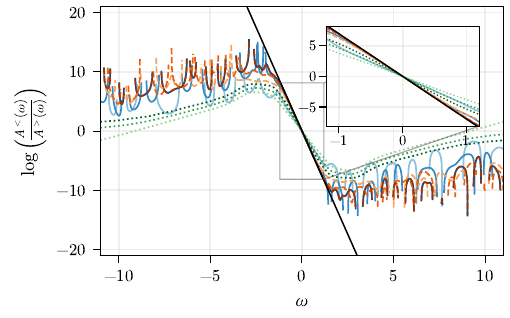}
  \caption{Equilibrium distribution diagnostic for the fused QTCI parametrization. The logarithm of the ratio of the lesser and greater spectral components is expected to follow $-\beta\omega$ (black line). The curves show the same bond-dimension and time-grid scans as in Figs.~\ref{fig:dmft_bethe_equil_chi}(a) and \ref{fig:dmft_bethe_equil_nt}(a), including the comparison between trapezoidal and Riemann quadrature. The inset resolves the low-frequency region around the Fermi level. (Only results for the fused QTCI time difference parametrization are shown;  results for the other methods look similar.)}
  \label{fig:dmft_bethe_equil_distribution}
\end{figure}

Figures~\ref{fig:dmft_bethe_equil_chi} and \ref{fig:dmft_bethe_equil_nt} separate the two numerical parameters that control the self-consistent solution, i.e., the maximum bond dimension $\chi$, and the grid size $N_t$, which controls the  time-step $\Delta t$ on the fixed time interval  $[0,t_{\rm c}]$ and thus  the quadrature error. We restrict the analysis to OCA, since the qualitative convergence behavior is expected to translate rather similarly to the third order. Figure~\ref{fig:dmft_bethe_equil_chi} shows the DMFT iteration error  in dependence  of $\chi$, which controls the TCI decomposition error. For all three decompositions, the final error plateau is systematically reduced with increasing bond dimension. 
As expected from the Gaussian benchmark, the non-quantics TCI difference parametrization reaches reliable convergence at the lowest maximum bond dimension, demonstrating the explicit efficiency of the low-rank factorization. Among the QTCI schemes, the interleaved representation exhibits a slightly higher error plateau at the smaller bond dimensions. 

The dependence on the grid size $N_t$ is analyzed in Fig.~\ref{fig:dmft_bethe_equil_nt}.
For fused QTCI, the comparison of trapezoidal and Riemann summation makes the importance of endpoint corrections particularly clear: even the Riemann calculation with $N_t=2^{17}$, corresponding to $\Delta t\simeq0.0025$, converges only after a pronounced slowdown, whereas the trapezoidal calculations reach the target error at substantially smaller grids. This behavior is consistent with Fig.~\ref{fig:gaussian_direct_trapz_single_run}, where the Riemann rule produces a leading short-time integration error. For the non-quantics TCI parametrization, the 
possibility to include higher-order Gregory quadrature weights allows for a much smaller $N_t$ at the same accuracy, and a more rapid convergence of the error with increasing $N_t$ (compare the separation between the error plateaus of successively doubled $N_t$ values).
Similar to the Gaussian benchmark, this improved scaling with $N_t$ does not necessarily imply a shorter total runtime, because the logarithmic sampling cost of the QTCI variants can compensate for their larger bond dimensions and the finer time grids.
For example on the M3 laptop chip a serial DMFT Bethe iteration for the QTCI difference method with $N_t = 2^{15}$ and $\chi = 30$ takes between a few seconds up to a minute while using non-quantics TCI difference for $N_t = 2^{12}$ and $\chi = 15$ takes several minutes.
The optimal method (time difference or quantics time difference) will depend on the physical regime and the maximal time $t_c$ needed for converged results, which in turn sets the necessary amount of points $N_t$ at a given required accuracy $\Delta t$. In cases where the time cutoff $t_{\rm c}$ can be chosen sufficiently low, or when the long-time tail of $\pSigma(t)$ can be reliably extrapolated,
the small discretization error and the low bond dimensions can however make the non-quantics approach favorable.

Figure~\ref{fig:dmft_bethe_equil_distribution} provides a complementary frequency-domain diagnostic for the fused QTCI method. We show the distribution function defined through the lesser Green's function $G^<(\omega)=2\pi i A(\omega) F(\omega)$ and the local spectral function $A(\omega)=-\frac{1}{\pi}\text{Im} G^R(\omega+i0)$. In the exact equilibrium solution, $F(\omega)$ should reduce to the Fermi function $f_\beta(\omega)$. While a known distribution function $F(\omega)$ could also simply be fixed in the ratio $\text{Im} \Delta^R(\omega)/\Delta^<(\omega)$, in the present case the fluctuation dissipation relation  is implicitly enforced through  a bath in the pseudo-particle Dyson equation (for further details, see App.~\ref{app:dyson} and Refs.~\cite{Li-Eckstein-2021, Eckstein-2024}), while both lesser and retarded components of the hybridization function are determined self-consistently from Eq.~\eqref{deltedmaft}.  As one generally aims to reduce the  bath coupling $\eta$, this scheme is very sensitive to numerical errors and the convergence of $F(\omega)$ to $f_\beta(\omega)$ provides a nontrivial benchmark for the self-consistent evaluation of  $G^<(\omega)$ within the different (Q)TCI schemes ($\eta = 0.01$ below).  In equilibrium, the fluctuation-dissipation relation requires the logarithmic ratio of lesser and greater spectral components to approach $-\beta\omega$. Figure~\ref{fig:dmft_bethe_equil_distribution} confirms that this is satisfied by the converged results. The deviations at high frequencies are strongest for small $\chi$ and coarse time grids, where the spectral weight entering the ratio is also small. The Riemann data exhibit more systematic frequency-dependent deviations, while the trapezoidal results approach the expected linear behavior as $\chi$ and $N_t$ are increased. The inset shows that the converged curves reproduce the equilibrium slope particularly well in the low-frequency region relevant to the metallic quasiparticle peak.

In summary, all three parametrizations recover the converged OCA solution within fewer than approximately twenty DMFT iterations once their respective resolution parameters are sufficiently large. The separate $\chi$ and $N_t$ scans demonstrate that the decomposition and quadrature errors can be controlled independently, while the distribution diagnostic confirms that self-consistency convergence is accompanied by the recovery of the equilibrium fluctuation-dissipation relation.

\begin{figure}[t]
  \centering
  \includegraphics{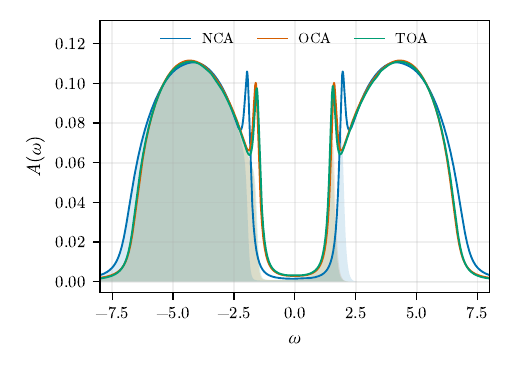}
  \caption{Spectral function of the photodoped Hubbard model on a Bethe lattice at $U = 8$, $t_{\rm h} = \sqrt{2}$, $\beta = 12.5$, and fixed $n_\mathrm{ex}=0.03$, comparing NCA, OCA, and TOA.}
  \label{fig:dmft_tfla_broadening}
\end{figure}

\subsection{Nonequilibrium DMFT}
\label{ch:BetheTFLA}

In this section we test the impurity solver at different levels of approximation (NCA, OCA, TOA) for nonequilibrium steady states within the Hubbard model.
To get a nonequilibrium steady-state representative of a photodoped state in the Mott-insulating regime, we 
follow Ref.~\cite{Künzel-etal-2024} and 
prescribe a 
fixed  distribution function
\begin{equation}
  \label{eq:tfla_distribution}
  F_{\beta, \gamma} (\omega) = \Theta_\alpha (\omega) f_\beta (\omega + \gamma) + (1 - \Theta_\alpha (\omega))  f_\beta (\omega - \gamma),
\end{equation}
where $f_\beta$ is the fermi function at inverse temperature $\beta$ and $\gamma$ is the chemical potential associated with the photoexcited states.
The interpolation function is set to be a smooth Heaviside $\Theta_\alpha (\omega) = 0.5 (1 - \tanh(\omega \alpha / 2))$, with $\alpha = \beta$ for convenience, as the result is expected to be roughly independent of the interpolation due to the missing spectral weight in the Mott gap \cite{Künzel-etal-2024}.
Results at different orders are compared at fixed  excitation density
\begin{equation}
  \label{eq:photodoped_density}
  n_\text{ex} (\beta, \gamma) = -\frac{1}{2} \frac{\int_0^\infty \diff \omega \, \mathrm{Im} (G^< (\omega))}{\int_{-\infty}^\infty \diff \omega \, \mathrm{Im} (G^\text{ret} (\omega))},
\end{equation}
and $\gamma$ is adapted separately for each approximation (NCA,OCA,TOA) to fix $ n_\text{ex}$.
Because we perform this analysis on the Bethe lattice, the solution at  fixed $F$ can be implemented by enforcing the DMFT self-consistency on the retarded (spectral) component of the Green's function, $\Delta^{\rm ret}(\omega) = t_h^2 G^{\rm ret} (\omega)$, while choosing the lesser component as 
\begin{equation}
  \Delta^< (\omega) = -2 i F_{\beta, \gamma} (\omega) \text{Im} \Delta^\text{ret} (\omega),
\end{equation}
which reproduces the setup given in \cite{Künzel-etal-2024}.

Figure~\ref{fig:dmft_tfla_broadening} shows the convergence of the photodoped spectral function with increasing diagrammatic order for NCA, OCA, and TOA at a fixed photodoping density of $n_{\mathrm{ex}} = 0.03$.
A self-convergence study with the maximally allowed bond dimension $\chi$ for the OCA and TOA self-energy diagrams is presented in Appendix~\ref{app:tfla_convergence}.
Overall we see the same qualitative behavior as in \cite{Künzel-etal-2024}, where the Mott gap remains stable against such a photodoping protocol, albeit the peak positions change with the renormalization of the Mott gap due to the higher-order corrections.
In the undoped parts of the 
Hubbard bands,
we see 
good
agreement between OCA and TOA, while for the holon and doublon quasiparticle peaks we see that TOA still provides a meaningful improvement over OCA.
In agreement with previous QTCI studies \cite{Kim-Werner-2025, Geng-Kim-Werner-2025} we find evidence for an overall convergence of the diagrammatic series in this case. 

Some remaining discrepancies to the QMC results from \cite{Künzel-etal-2024}, as already mentioned in \cite{Kim-Werner-2025}, may be due to statistical noise as well as the effective cutoff time $t_c$ imposed on the restricted propagator necessary to converge the Monte Carlo algorithm \cite{Erpenbeck-Gull-Cohen-2023}.

\section{Square Lattice EDMFT}
\label{ch:EDMFT}

To illustrate the versatility of the presented impurity solver, we apply it to the extended Hubbard model
\begin{equation}
  \label{eq:extended_hubbard_model}
  H_\text{ehub}
  =
  H_\text{hub}
  +
  V \sum_{\langle ij\rangle} \bar n_i \bar n_j,
\end{equation}
at half filling with $\bar{n}_i = n_i - 1$ on a square lattice, where the nearest-neighbour interaction gives rise to an additional retarded hybridization term in the impurity action, while the local part remains unchanged.
The EDMFT formulation thus requires solving an effective impurity model with a retarded bosonic interaction, whose accurate treatment remains challenging, particularly out of equilibrium.
This problem can be addressed either by an expansion in the retarded interaction \cite{Golez-Eckstein-Werner-2015} or by employing the polaron representation \cite{Werner-Eckstein-2013}. However, previous studies have demonstrated that low-order approximations are insufficient in physically relevant parameter regimes \cite{Chen-etal-2016, Paprotzki-Eckstein-2025}. Here we show that the TCI-based strong-coupling solver substantially improves the convergence of the hybridization expansion, yielding results that are already nearly converged at third order.

\begin{figure*}[t]
  \centering
  \includegraphics{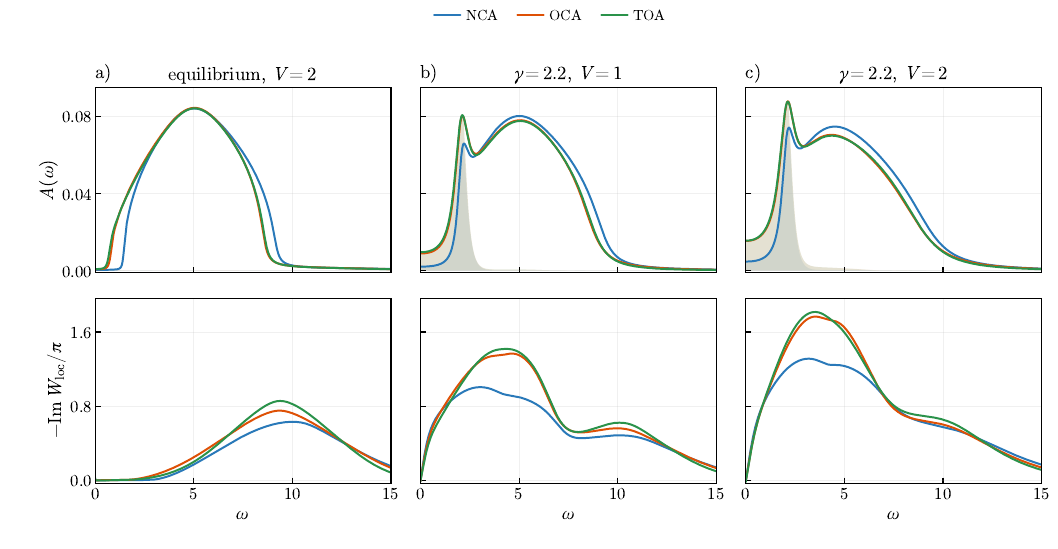}
  \caption{Positive-frequency square-lattice EDMFT spectra for $U=10$, $\beta=5$, and $t_h=1$ at different expansion orders. (a) Equilibrium result for $V=2$. Order convergence in the photodoped state at $\gamma=2.2$ for (b) $V=1$ and (c) $V=2$. The upper row depicts the fermionic spectral function and the lower row the local screened interaction.
  The corresponding effective photodoping densities, obtained from the occupied positive-frequency spectral weight, are given by $n_{\rm ex} = 0.030$ (NCA), $0.054$ (OCA), $0.056 $ (TOA)  for $V=1$ and $n_{\rm ex} =0.041$ (NCA), $0.067$ (OCA), $0.069$ (TOA) for $V=2$.
   }
  \label{fig:edmft_square_summary}
\end{figure*}

Following Ref.~\cite{Golez-Eckstein-Werner-2015} the action 
\begin{equation}
  \begin{aligned}
    \mathcal{S} = &-\mathrm{i} \int_\mathcal{C} \diff t \sum_{ij\sigma} c^*_{i\sigma} (t) [-\mathrm{i} \partial_t \delta_{ij} + t_h \delta_{\langle ij \rangle}] c_{j\sigma}(t) \\
                & +\frac{1}{2} \sum_{ij} \bar{n}_i(t) [U \delta_{ij} + V \delta_{\langle ij \rangle}] \bar{n}_j(t)
  \end{aligned}
\end{equation}
can be decoupled into local and hybridization terms using a Hubbard-Stratonovich transformation to obtain an effective impurity description after integrating out the bosonic fields
\begin{equation}
  \begin{aligned}
    \mathcal{S}_\text{imp} = &-\mathrm{i} \int_\mathcal{C} \diff t \diff t' \sum_{\sigma} c^*_{\sigma} (t) \mathcal{G}_{0 \sigma}^{-1} (t, t') c_{\sigma}(t') \\
    &+\frac{1}{2} \bar{n} (t) \, \mathcal{U} (t, t') \, \bar{n} (t') 
  \end{aligned}
\label{Simp-U}
\end{equation}
featuring fermionic $\mathcal{G}_{0\sigma}^{-1} (t, t') = \mathrm{i} \partial_t \delta_\mathcal{C} (t, t') - \Delta_\sigma (t, t')$ as well as bosonic $\mathcal{U}(t, t') = U \delta_\mathcal{C} (t, t') + \mathcal{D} (t, t')$ Weiss fields with fermionic $\Delta (t, t')$ as well as bosonic $\mathcal{D} (t, t')$ hybridization functions, which are found self-consistently.

The implementation of the EDMFT self-consistency, following Refs.~\cite{Golez-Eckstein-Werner-2015,Golež-Eckstein-Werner-2019}, is described in App.~\ref{app:edmft_lattice}. The fermionic self-consistency, which relates the hybridization function $\Delta(t)$ to the impurity Green's function $G_\text{imp}(t) = -i \langle \mathcal{T}_\mathcal{C} c(t) c^\dagger(0) \rangle$, is identical to that of the conventional Hubbard model. For the bosonic self-consistency, we first extract the impurity charge susceptibility, $\chi_\text{imp}(t)=-i\langle \mathcal{T}_\mathcal{C} \bar n(t)\bar n(0) \rangle$, which is evaluated analogously to the impurity Green's function by using $\bar n$ as the external vertex. From this quantity, we construct the impurity bosonic propagator
\begin{equation}
  W_\text{imp}
  =
  \mathcal{U}
  +
  \mathcal{U}
  *
  \chi_\text{imp}
  *
  \mathcal{U},
\end{equation}
which coincides with the local lattice bosonic Green's function upon convergence.

To solve the impurity model with action \eqref{Simp-U}, the strong-coupling expansion in the pseudo-particle space is augmented by an additional weak-coupling expansion in powers of the retarded density-density interaction.
Formally, for a consistent truncation of the double expansion a Lutting-Ward functional is constructed, whose functional derivative then yields the corresponding diagrammatic expressions \cite{Golez-Eckstein-Werner-2015}.
In practice this just amounts to adding additional bosonic hybridization lines connected via $\mathcal{D} (t, t')$ and $\bar{n}(t)$ as vertex operators in the pseudo-particle impurity model, which will significantly increase the number of diagrams at each order.
In Tab.~\ref{tab:diagram-counts} we summarize the exemplary diagram counts, which showcase the rapid increase of diagrammatic terms with the order in the Keldysh reparametrization.
This is particularly pronounced in the double expansion for EDMFT, which warrants a symmetry reduction by exploiting both the trivial endpoint symmetry of the bosonic lines as well as the spin symmetry between up and down spin components.

\begin{table}[t]
\centering
\scriptsize
\setlength{\tabcolsep}{3pt}
\begin{tabular}{ll|rrr|rrr}
\toprule
& & \multicolumn{3}{c|}{No symmetry reduction}
    & \multicolumn{3}{c}{Symmetry reduced} \\
\cmidrule(lr){3-5}\cmidrule(lr){6-8}
Model & Order
    & $U$ & $D$ & $K$
    & $U$ & $D$ & $K$ \\
\midrule
DMFT
    & NCA & 1 &   4 &     8 & 1 &   2 &    4 \\
    & OCA & 1 &  16 &   128 & 1 &   4 &   32 \\
    & TOA & 4 & 256 & 8\,192 & 4 &  32 & 1\,024 \\
\midrule
EDMFT
    & NCA & 1 &   6 &    12 & 1 &   3 &    6 \\
    & OCA & 1 &  36 &   288 & 1 &   9 &   72 \\
    & TOA & 4 & 864 & 27\,648 & 4 & 108 & 3\,456 \\
\bottomrule
\end{tabular}

\caption{Number of self-energy diagrams for the one-orbital Hubbard and
extended Hubbard models. Here, $U$, $D$, and $K$ denote undirected
topologies, directed topologies, and Keldysh-enumerated diagrams,
respectively.}
\label{tab:diagram-counts}
\end{table}

In Fig.~\ref{fig:edmft_square_summary} we show the convergence with diagrammatic order for selected values of $V$ and $\gamma$.
In equilibrium [Fig.~\ref{fig:edmft_square_summary}(a)], for the moderately strong nearest-neighbour interaction $V=2$, the fermionic spectral function converges rapidly, whereas the bosonic spectrum exhibits more pronounced changes upon including higher-order diagrams.
The photodoped spectra [Fig.~\ref{fig:edmft_square_summary}(b,c)] show a substantial increase in the photodoped carrier density at the fixed photodoping chemical potential $\gamma=2.2$ when going from NCA to OCA, while the TOA yields only a minor correction, consistent with the equilibrium results. 
The resulting effective photodoped densities [cf. Eq.~\eqref{eq:photodoped_density}], as stated in the figure caption, further reinforce the convergence with diagrammatic order. 
Increasing $V$ leads to a progressive filling of the Mott gap, signaling enhanced dynamical screening and a  destabilization of the Mott insulating state. This trend is more pronounced in the (almost identical) OCA and TOA results, whereas the NCA overestimates the stability of the Mott phase.
The bosonic spectra in the photodoped case develop pronounced low-frequency spectral weight, reflecting the emergence of low-energy charge excitations induced by photodoping. This low-energy response is noticeably more sensitive to higher-order corrections than the fermionic spectra, but also in this case OCA and TOA results are already quite close. These results highlight the importance of higher-order impurity solvers for future GW+EDMFT calculations, where the bosonic propagator enters the GW contribution to the electronic self-energy and thus directly feeds back onto the fermionic dynamics.

\section{Conclusion}

In summary, we have presented and analyzed TCI-based strong-coupling impurity solvers employing several TT decomposition strategies for the diagrammatic expansion, and compared their efficiency in terms of both bond dimension and computational cost. Our work builds on the Keldysh time-difference parametrization introduced in Ref.~\cite{Eckstein-2024}, which provides a natural representation of nonequilibrium diagrams in terms of time differences. A quantics formulation of this Keldysh representation was subsequently introduced in Ref.~\cite{Kim-Werner-2025,Geng-Kim-Werner-2025}, where the diagrammatic expressions are evaluated in frequency space. In the present work, we develop a direct real-time algorithm for retarded convolutions in the quantics representation, systematically compare time-difference and quantics decompositions, and extend the approach to impurity models with retarded interactions.

Overall, the most efficient approaches are (i) the time-difference decomposition, in which the integrand is represented as a function of the time differences between vertices, and (ii) its quantics counterpart, where the time differences are encoded in a binary representation using either an interleaved or a fused layout. In contrast, a decomposition in terms of the absolute time arguments is generally less favorable. Among the time-difference-based approaches, the conventional time-difference decomposition yields substantially lower bond dimensions and exhibits superior convergence with the time-grid resolution owing to the use of higher-order quadrature rules. However, its sampling cost grows linearly with the number of time points $N_t$, whereas the quantics representation scales only as $\mathcal{O}(\log N_t)$ for large $N_t$. The optimal approach therefore depends on the physical regime and the maximum simulation time $t_c$ required for converged results. For the DMFT benchmarks considered here, the quantics time-difference decomposition proved to be the most efficient overall.

A key ingredient of the quantics time-difference algorithm is the ability to perform retarded convolutions directly in the quantics representation. Beyond the present application to diagram evaluation, this direct QTT convolution algorithm is of broader use in nonequilibrium methods involving Volterra- or convolution-type integral equations,  as already demonstrated in Refs.~\cite{Shinaoka-etal-2023, Inayoshi-Shinaoka-Murakami-2026, Rohshap-etal-2025}. For example, it may enable an implementation of the entire nonequilibrium DMFT cycle fully contained within the QTT formalism, eliminating the need to explicitly evaluate the pseudo-particle self-energy on the large time grid required to solve the Dyson equation.

Using the TCI-based approach,
complex diagrammatic expressions can be evaluated efficiently within steady-state nonequilibrium calculations, making intermediate expansion orders computationally accessible. As a result, the method substantially extends the range beyond the NCA and enables systematic convergence checks of the hybridization expansion up to third order. For reference, within the fused quantics time-difference parametrization, a typical Bethe DMFT iteration at the OCA level requires only a few seconds on a standard personal computer. The more demanding TOA calculations require only a few minutes per Bethe DMFT iteration on a 96-core compute node, despite the factorial increase in the number of diagrams and the four internal time integrations. These advances open the door to previously inaccessible applications. As a representative example, we demonstrated the solution of impurity models with retarded interactions, as encountered in EDMFT and GW+EDMFT, where both electronic and bosonic spectra can be computed systematically. While previous work has highlighted the limitations of the NCA for retarded interactions, the comparison with OCA and third-order results now provides a controlled estimate of the truncation error. This, in turn, paves the way for quantitatively controlled nonequilibrium GW+EDMFT simulations.

\acknowledgments

We thank  Eva Paprotzki, Leo Wehberg, Lei Geng, Andre Erpenbeck, Philipp Werner, and Denis Golež for discussions. Funding is acknowledged through the Deutsche Forschungsgemeinschaft through OPTIMAL-FOR5750 - 531215165 (Project P1) and through the Cluster of Excellence ``CUI: Advanced Imaging of Matter'' of the Deutsche Forschungsgemeinschaft (DFG) – EXC 2056 –project ID 390715994.  The authors gratefully acknowledge the computing time granted by the Resource Allocation Board and provided on the supercomputer Emmy/Grete at NHR-Nord@Göttingen as part of the NHR infrastructure. The calculations for this research were conducted with computing resources under the project \texttt{nhr\_hh\_starter\_27355}.

\input{main.bbl}

\appendix

\section{Retarded Convolution using Quantics}
\label{app:quantics}

Within this section we discuss how to solve the continuous retarded convolution of the adjacent arguments associated with logical dimensions $d$ and $d+1$,
\begin{equation}
  I_d(\dots,t,\dots)
  =\int_0^t \diff\tau\,A(\dots,t-\tau,\tau,\dots),
\end{equation}
directly within the QTT format.
Here the two input dimensions, represented by $t-\tau$ and $\tau$, are replaced by the single retained time $t$.
For each logical dimension $d=1,\dots,D$, let $m_d\in\{0,\dots,2^R-1\}$ denote the corresponding grid index (and $t_d=m_d\Delta t$ on a uniform time grid).  The main-text convention, Eq.~\eqref{eq:quantics_representation}, orders its quantics bits from most to least significant.  For the implementation-oriented traversal used throughout this appendix, we define instead
\begin{equation}
  b_{d,r}=q_{d,R-r+1},\qquad r=1,\dots,R,
\end{equation}
so that $b_{d,1}$ is the least-significant bit and
\begin{equation}
  m_d=\sum_{r=1}^{R}2^{r-1}b_{d,r}.
\end{equation}
At every site, the bits of all logical dimensions are packed into the fused index
\begin{equation}
  s_r=\sum_{d=1}^{D}2^{d-1}b_{d,r}.
\end{equation}
Thus site $r=1$ contains the least-significant bits of all coordinates, while logical dimension $d=1$ occupies the least-significant bit of each fused index. In this notation, the fused QTT is
\begin{equation}
  A(m_1,\dots,m_D)=M^{(1)}(s_1)\cdots M^{(R)}(s_R).
\end{equation}

\paragraph{Riemann Sum}

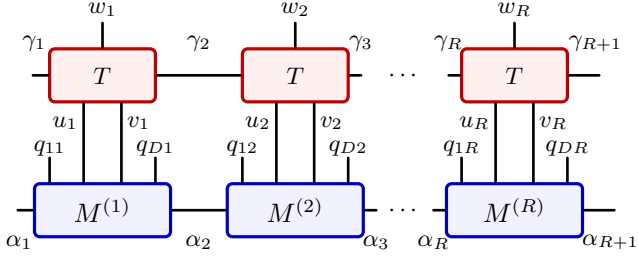
\begin{figure}[tbp]
  \centering
  \begin{tikzpicture}[
	Mcore/.style={
		ppscWideCore,
		minimum width=18mm
	},
	Tcore/.style={
		ppscTransfer,
		minimum width=14mm
	},
	bond/.style={ppscBond},
	lab/.style={ppscLegLabel},
	tinylab/.style={ppscLegLabel}
]

\def\dotStub{1.5mm}

\node[Mcore] (M1) at (0,0) {$M^{(1)}$};
\node[Mcore] (M2) at (2.55,0) {$M^{(2)}$};
\node[Mcore] (MR) at (5.45,0) {$M^{(R)}$};
\coordinate (Mdots) at ($($(M2.east)+(\dotStub,0)$)!0.5!($(MR.west)+(-\dotStub,0)$)$);

\node[Tcore] (T1) at (0,1.78) {$T$};
\node[Tcore] (T2) at (2.55,1.78) {$T$};
\node[Tcore] (TR) at (5.45,1.78) {$T$};
\coordinate (Tdots) at ($($(T2.east)+(\dotStub,0)$)!0.5!($(TR.west)+(-\dotStub,0)$)$);

\draw[bond] ($(M1.west)+(-2mm,0)$) -- (M1.west);
\node[lab] at ($(M1.west)+(-1.8mm,-4.2mm)$) {$\alpha_1$};

\draw[bond] (M1.east) -- (M2.west);
\node[lab] at ($(M1.east)!0.5!(M2.west)+(0,-4.2mm)$) {$\alpha_2$};

\draw[bond] (M2.east) -- ++(\dotStub,0);
\node[lab] at ($(M2.east)+(\dotStub,-4.2mm)$) {$\alpha_3$};

\draw[bond] (MR.west) -- ++(-\dotStub,0);
\node[ppscDots] at (Mdots) {$\cdots$};
\node[lab] at ($(MR.west)+(-\dotStub,-4.2mm)$) {$\alpha_R$};

\draw[bond] (MR.east) -- ++(4mm,0);
\node[lab] at ($(MR.east)+(3.5mm,-4.2mm)$) {$\alpha_{R+1}$};

\draw[bond] ($(T1.west)+(-2mm,0)$) -- (T1.west);
\node[lab] at ($(T1.west)+(-1.8mm,4.2mm)$) {$\gamma_1$};

\draw[bond] (T1.east) -- (T2.west);
\node[lab] at ($(T1.east)!0.5!(T2.west)+(0,4.2mm)$) {$\gamma_2$};

\draw[bond] (T2.east) -- ++(\dotStub,0);
\node[lab] at ($(T2.east)+(\dotStub,4.2mm)$) {$\gamma_3$};

\draw[bond] (TR.west) -- ++(-\dotStub,0);
\node[ppscDots] at (Tdots) {$\cdots$};
\node[lab] at ($(TR.west)+(-\dotStub,4.2mm)$) {$\gamma_R$};

\draw[bond] (TR.east) -- ++(4mm,0);
\node[lab] at ($(TR.east)+(3.5mm,4.2mm)$) {$\gamma_{R+1}$};

\draw[bond] ($(M1.north)+(-7mm,0)$) -- ++(0,3.2mm);
\node[lab] at ($(M1.north)+(-7mm,4.5mm)$) {$q_{11}$};

\draw[bond] ($(M1.north)+(-2.5mm,0)$) -- ($(T1.south)+(-2.5mm,0)$);
\draw[bond] ($(M1.north)+( 2.5mm,0)$) -- ($(T1.south)+( 2.5mm,0)$);

\node[tinylab,anchor=east,xshift=-1pt]
  at ($(M1.north)+(-2.5mm,7.8mm)$) {$u_1$};
\node[tinylab,anchor=west,xshift=1pt]
  at ($(M1.north)+( 2.5mm,7.8mm)$) {$v_1$};

\draw[bond] ($(M1.north)+(7mm,0)$) -- ++(0,3.2mm);
\node[lab] at ($(M1.north)+(7mm,4.5mm)$) {$q_{D1}$};

\draw[bond] (T1.north) -- ++(0,3.2mm);
\node[lab] at ($(T1.north)+(0,5mm)$) {$w_1$};

\draw[bond] ($(M2.north)+(-7mm,0)$) -- ++(0,3.2mm);
\node[lab] at ($(M2.north)+(-7mm,4.5mm)$) {$q_{12}$};

\draw[bond] ($(M2.north)+(-2.5mm,0)$) -- ($(T2.south)+(-2.5mm,0)$);
\draw[bond] ($(M2.north)+( 2.5mm,0)$) -- ($(T2.south)+( 2.5mm,0)$);

\node[tinylab,anchor=east,xshift=-1pt]
  at ($(M2.north)+(-2.5mm,7.8mm)$) {$u_2$};
\node[tinylab,anchor=west,xshift=1pt]
  at ($(M2.north)+( 2.5mm,7.8mm)$) {$v_2$};

\draw[bond] ($(M2.north)+(7mm,0)$) -- ++(0,3.2mm);
\node[lab] at ($(M2.north)+(7mm,4.5mm)$) {$q_{D2}$};

\draw[bond] (T2.north) -- ++(0,3.2mm);
\node[lab] at ($(T2.north)+(0,5mm)$) {$w_2$};

\draw[bond] ($(MR.north)+(-7mm,0)$) -- ++(0,3.2mm);
\node[lab] at ($(MR.north)+(-7mm,4.5mm)$) {$q_{1R}$};

\draw[bond] ($(MR.north)+(-2.5mm,0)$) -- ($(TR.south)+(-2.5mm,0)$);
\draw[bond] ($(MR.north)+( 2.5mm,0)$) -- ($(TR.south)+( 2.5mm,0)$);

\node[tinylab,anchor=east,xshift=-1pt]
  at ($(MR.north)+(-2.5mm,7.8mm)$) {$u_R$};
\node[tinylab,anchor=west,xshift=1pt]
  at ($(MR.north)+( 2.5mm,7.8mm)$) {$v_R$};

\draw[bond] ($(MR.north)+(7mm,0)$) -- ++(0,3.2mm);
\node[lab] at ($(MR.north)+(7mm,4.5mm)$) {$q_{DR}$};

\draw[bond] (TR.north) -- ++(0,3.2mm);
\node[lab] at ($(TR.north)+(0,5mm)$) {$w_R$};

\end{tikzpicture}
  \caption{Sketch of a single convolution directly implemented in the fused QTT formalism via the transfer matrix $T$ (red).}
  \label{fig:quantics_convolution_fused}
\end{figure}

On the equidistant grid $t=w\Delta t$ and $\tau=v\Delta t$, its Riemann discretization is
\begin{equation}
  I_d(\dots,w,\dots)=\Delta t\sum_{v=0}^{w}A(\dots,u=w-v,v,\dots).
\end{equation}
Defining the corresponding bits as
\begin{equation}
  u=\sum_{r=1}^{R}2^{r-1}u_r,\quad
  v=\sum_{r=1}^{R}2^{r-1}v_r,\quad
  w=\sum_{r=1}^{R}2^{r-1}w_r,
\end{equation}
binary addition is enforced locally by an incoming carry $\gamma_r\in\{0,1\}$ and an outgoing carry $\gamma_{r+1}\in\{0,1\}$.  At site $r$ they obey
\begin{equation}
  \label{eq:quantics_retarded_convolution_integer_constraint}
  u_r+v_r+\gamma_r=w_r+2\gamma_{r+1}.
\end{equation}
The associated local transfer tensor is
\begin{equation}
  T_r(u_r,v_r,w_r)_{\gamma_r,\gamma_{r+1}}
  =\delta_{u_r+v_r+\gamma_r,\,w_r+2\gamma_{r+1}}.
\end{equation}
This constitutes the rank 2 MPO depicted in Figure~\ref{fig:quantics_convolution_fused}.

Hence, the carry becomes an additional QTT bond index increasing the output bond dimension by a factor of two (compare also to Fig.~\ref{fig:quantics_convolution_fused}).  The number of QTT sites therefore remains $R$, whereas the fused local dimension decreases from $2^D$ to $2^{D-1}$.  The carry boundary conditions are
\begin{equation}
  \gamma_1=0,\qquad \gamma_{R+1}=0,
\end{equation}
such that the resulting QTT can be directly obtained via the contraction of the original QTT and the transfer MPO, as illustrated in Fig.~\ref{fig:quantics_convolution_fused}.
Subsequently, the resulting QTT has a doubled bond dimension.

\paragraph{Trapezoidal Correction}

For fixed $w\geq1$, the trapezoidal discretization of the same retarded convolution is
\begin{equation}
  \begin{aligned}
    I_d^{\mathrm{trap}}(\dots,w,\dots)
    =\Delta t\Bigg[&\frac{1}{2}A(\dots,w,0,\dots)\\
    &+\sum_{v=1}^{w-1}A(\dots,w-v,v,\dots)\\
    &+\frac{1}{2}A(\dots,0,w,\dots)\Bigg].
  \end{aligned}
\end{equation}
For the degenerate interval $w=0$, the implementation sets $I_d^{\mathrm{trap}}(\dots,0,\dots)=0$.
The endpoint conditions $v=0$ and $u=0$ are global across the QTT sites: they require $v_r=0$ or $u_r=0$, respectively, for every $r=1,\dots,R$.  They therefore cannot be inferred from a single local tuple $(u_r,v_r,w_r)$.  Instead, the implementation subtracts the two endpoint corrections from the Riemann result,
\begin{equation}
  \begin{aligned}
  I_d^{\mathrm{trap}}=I_d^{\mathrm{R}}
  &-\frac{\Delta t}{2}A(\dots,u=w,v=0,\dots)\\
  &-\frac{\Delta t}{2}A(\dots,u=0,v=w,\dots).
  \end{aligned}
\end{equation}
Both endpoint QTTs are obtained by local core selection.  
No carry channel is required for either endpoint because $u=w$ when $v=0$ and $v=w$ when $u=0$.  Denoting the corresponding QTT products by $E_{v=0}$ and $E_{u=0}$, the corrected result is the three-term TT sum
\begin{equation}
  I_d^{\mathrm{trap}}
  =\Delta t\left(N-\frac{1}{2}E_{v=0}-\frac{1}{2}E_{u=0}\right).
\end{equation}
Here $N$ denotes the unscaled carry-augmented Riemann QTT.  The implementation retains the three contributions as separately weighted QTT terms.  If they are materialized as one TT, TT addition produces a block-diagonal bond space with channel sizes $2\chi$, $\chi$, and $\chi$.  Exploiting this structure changes the leading cubic work from $(4\chi)^3=64\chi^3$ to $(2\chi)^3+\chi^3+\chi^3=10\chi^3$, compared with $8\chi^3$ for the Riemann channel alone.  The results presented in this work were obtained without intermediate recompression of these terms.  As a future optional refinement, SVD recompression may exploit the block structure by first orthogonalizing the Riemann and endpoint channels separately and then applying a truncated SVD only when their bond spaces are combined.

\paragraph{Interleaved Layout}
\label{App1C}
The convolution can also be applied to an interleaved QTT.  At site $r$, the adjacent cores of logical dimensions $d$ and $d+1$ are contracted for fixed bits $(u_r,v_r)$,
\begin{equation}
  P^{(r,d)}(u_r,v_r)
  =M^{(r,d)}(u_r)M^{(r,d+1)}(v_r).
\end{equation}
Applying the previously defined transfer MPO to this product yields the core at the given interleaved site.
Contrary to the fused protocol, this relies on the fact that we restrict the reduction to neighboring sites.
All other cores are unchanged apart from propagating the carry through an augmentation with an identity block.  The carry boundaries and trapezoidal endpoint terms are the same as in the fused construction.

Since the interleaved algorithm essentially mixes iterative fusing of adjacent dimensions with iterative convolutional reduction, we are able to substantially simplify the implementation by first fusing all logical-dimension cores at each site $r$ into a fused core
\begin{equation}
  \overline M^{(r)}(s_r)
  =\prod_{j=1}^{D}M^{(r,j)}(b_{j,r}),,
\end{equation}

  which allows to directly reuse the fused QTT algorithm without any adjustments.

\paragraph{Heaviside Function}

The same carry MPO gives an exact QTT representation of the discrete retarded mask.  Summing over the nonnegative difference $u$ yields
\begin{equation}
  \sum_{u=0}^{2^R-1}\delta_{u+v,w}=\Theta(w-v),
\end{equation}
where the no-overflow boundary condition ensures that the left-hand side is one precisely when $v\leq w$.  Define the local mask transfer matrix by summing over the difference bit,
\begin{equation}
  H_r(v_r,w_r)_{\gamma_r,\gamma_{r+1}}
  =\sum_{u_r=0}^{1}T_r(u_r,v_r,w_r)_{\gamma_r,\gamma_{r+1}}.
\end{equation}
With $e_0=(1,0)^\mathsf{T}$ selecting $\gamma_1=\gamma_{R+1}=0$, the mask is
\begin{equation}
  \Theta(w-v)
  =e_0^\mathsf{T}H_1(v_1,w_1)\cdots H_R(v_R,w_R)e_0.
\end{equation}
It is therefore an exact fused QTT with bond dimension two, using the same one-based least-to-most-significant site traversal as the convolution.

\section{Pseudo-Particle Dyson Equation in Frequency}
\label{app:dyson}
In order to use the resummed skeleton expansion for the pseudo-particle self-energy, one needs to solve the corresponding Dyson equation given in its general form in Eq.~\eqref{Dyson}.
Following \cite{Li-Eckstein-2021} and \cite{Eckstein-2024}, we can within the steady-state decompose the contributions to the retarded and lesser component
\begin{equation}
  \label{eq:app:dyson_components}
  \begin{aligned}
    (i \partial_t - H_\text{loc}) \mathcal{G}^\text{r} (t) = \int_{-\infty}^{\infty} \diff t_1 \; \Sigma^\text{r}(t - t_1) \mathcal{G}^\text{r} (t_1) \\
    \mathcal{G}^< (t) = \int_{-\infty}^{\infty} \diff t_1 \diff t_2 \; \mathcal{G}^\text{r} (t - t_1) \Sigma^<(t_1 - t_2) \mathcal{G}^\text{a} (t_2),
  \end{aligned}
\end{equation}
where $\mathcal{G}^\text{r} (t) = \theta(t) \mathcal{G}^>(t)$ and $\mathcal{G}^\text{a} (t) = - \theta(-t) \mathcal{G}^>(t)$.
Note that in the derivation of Eq. \eqref{eq:app:dyson_components} the Hermiticity of $\mathcal{G}$ and $\Sigma$ is assumed.
Proper normalization of the pseudo-particle expectation values \cite{Eckstein-Werner-2010, Li-Eckstein-2021} is guaranteed by ensuring that $Q = - i \text{tr} [ \xi \pG^<(0) ] = 1$ is valid after each update of $\pG^<$.
  This can be achieved simply by normalizing $\pG^<$ by the current value of $Q$.

Equations \eqref{eq:app:dyson_components} assume that the pseudo-particle propagators decay within the time window on which the Dyson equation is solved 
For some cases (for example the gapped single-impurity model) the intrinsic decay of $\pG(t)$ becomes very slow. It can then be helpful  to  introduce a consistent frequency-dependent broadening via an artificial symmetric contribution to the action \cite{Eckstein-2024}
\begin{equation}
  \label{eq:action_broadening}
  \mathcal{S}_\eta = - \frac{\eta}{2} \int \diff t \diff t' \; \hat{1}(t) \Delta_{\rm pp-bath}(t, t') \hat{1}(t'),
\end{equation}
representing coupling to a pseudo-particle bath. This allows a solution in the frequency domain as in \cite{Li-Eckstein-2021},
\begin{equation}
  \begin{aligned}
    \mathcal{G}^\text{r} (\omega) = (\omega + i \eta g(\omega) - H_\text{loc} - \Sigma^\text{r} (\omega))^{-1} \\
    \mathcal{G}^< (\omega) = \mathcal{G}^\text{r} (\omega) \big[ \Sigma^<(\omega) - 2 i \xi \eta g^<(\omega) \big] (\mathcal{G}^\text{r} (\omega))^\dagger,
  \end{aligned}
\end{equation}
with the regularization functions $g(\omega) = f_\beta(-\omega)$ and $g^<(\omega) = f_\beta(\omega)$ chosen to satisfy the fluctuation relation  $g^<(\omega) / g(\omega) = e^{-\beta \omega}$ expected in equilibrium.

The choice of the ratio $g(\omega)/g^<(\omega)$ can be viewed as coupling the system to a bosonic bath which enforces the temperature within the DMFT simulation, where the hybridization function is determined self-consistently and does not a priori contain information about the temperature. Alternatively, one can  directly enforce the Fermi distribution at the level of the hybridization in full analogy to the photodoped case in Sec.~\ref{ch:BetheTFLA}.

\begin{figure}[tbp]
  \centering
  \includegraphics{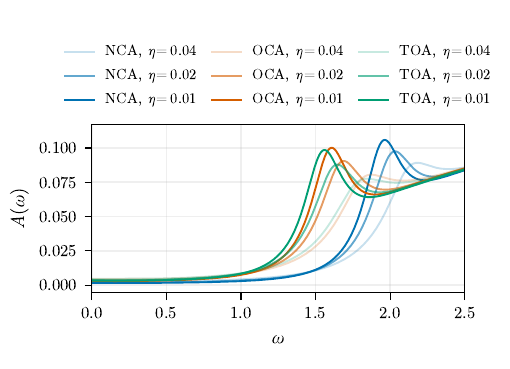}
  \caption{Dependence of the NCA, OCA, and TOA spectra of the photo-doped Bethe-lattice Hubbard model on the broadening parameter $\eta$. The parameters are the same as in Fig.~\ref{fig:dmft_tfla_broadening}.}
  \label{fig:dmft_tfla_eta_scaling}
\end{figure}

Figure~\ref{fig:dmft_tfla_eta_scaling} shows the evolution from $\eta=0.04$ to $\eta=0.01$. The quasiparticle-peak amplitude and position depend strongly on the broadening, with a clear convergence trend toward lower values of $\eta$.
The
broadening decreases the weight of sharp spectral features such as the quasiparticle peak in the metallic regime of the Hubbard model, which warrants a careful analysis of the dependence on the chosen value of $\eta$.
In practice values of around $\eta = 0.005$ to $0.02$ are often a sensible compromise between stability and convergence for the current benchmarks of the (Q)TCI solver. One can also
attempt an annealing to zero broadening during the self-consistent DMFT cycle, as in many cases the 
intrinsic decay of $\pG$ due to the self-energy $\pSigma$
can be sufficiently fast on a suitably large simulation time interval.
As the focus of the present work is the accelerated diagrammatic solver we omit this annealing. 

\begin{figure}[t]
  \centering
  \includegraphics{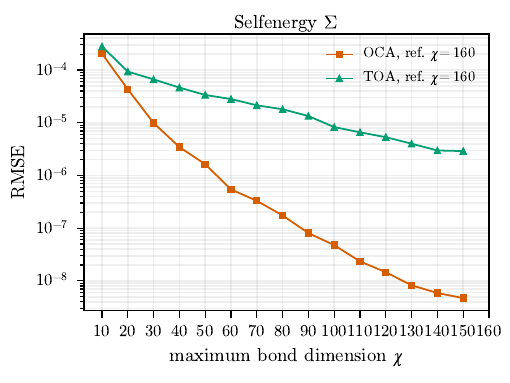}
  \caption{Bond-dimension convergence of the pseudo-particle self-energy for the photo-doped Bethe-lattice calculation at $\eta=0.01$ and $n_\mathrm{ex}=0.03$. The time-domain RMSE is computed with respect to the largest available bond dimension for each impurity expansion order and averaged over greater and lesser components.}
  \label{fig:dmft_bethe_tfla_selfconvergence}
\end{figure}

\section{Convergence of the DMFT self-consistency in the photodoped case}
\label{app:tfla_convergence}

In this section we analyze the convergence of accelerated QTCI difference scheme for the self-energy within the application to the photo-doped Hubbard model.
Since an exact solution is not possible and a direct evaluation of the diagrammatic contributions is not feasible for the necessary parameters, we resort to a self-convergence test.
This means we calculate for every selected maximum bond dimension $\chi$ the error of the self-energy with respect to the highest bond dimension reached during the scan (here $\chi = 160$), which measures the importance of the neglected ranks.
In Fig.~\ref{fig:dmft_bethe_tfla_selfconvergence} we observe convergence with $\chi$ for both OCA and TOA, albeit with a significantly slower slope for the TOA approximation. This is expected, as the TOA decomposition already required approximately four times the rank of the OCA decomposition in the synthetic test in Sec.~\ref{ch:Gaussian}.
Similarly, compared with the DMFT convergence scan in Sec.~\ref{ch:BetheEquil} the RMSE at a specific bond dimension roughly matches the error plateau of the hybridization error between different DMFT iterations.
This further validates the approach as fairly controlled and strengthens the conjectured connection between the TCI decomposition error and the DMFT convergence.

\section{EDMFT Lattice Self-Consistency}
\label{app:edmft_lattice}
In this section we follow \cite{Golez-Eckstein-Werner-2015, Golež-Eckstein-Werner-2019} in order to derive the self-consistency relations for lattice EDMFT.

\paragraph{Fermionic Lattice Self-Consistency}
Given the previous (fermionic) hybridization $\Delta (t)$ and calculated impurity Green's function $G_\text{imp} (t) = -i \langle \mathcal{T}_\mathcal{C} c (t) c^\dagger (0) \rangle$ and following \cite{Aoki-etal-2014} to avoid an explicit potentially ill-conditioned calculation of the self-energy, we first calculate the isolated impurity Green's function $g_0$
\begin{equation}
  (1 + G_\text{imp} * \Delta) * g_0 = G_\text{imp},
\end{equation}
which allows us to calculate the lattice Green's function
\begin{equation}
  (1 - g_0 * \epsilon_k) * G_k = g_0
\end{equation}
using the lattice dispersion $\epsilon_k$.
The self-consistency is then closed via the two moments
\begin{align}
  G_1 &= \frac{1}{N_k} \sum_k \epsilon_k G_k, 
  \\ G_2 &= \frac{1}{N_k} \sum_k (\epsilon_k + \epsilon_k * G_k * \epsilon_k)
\end{align}
such that
\begin{equation}
  (1 + G_1) * \Delta = G_2,
\end{equation}
where $N_k$ is the number of $k$-points.
In the steady state, these Volterra equations are efficiently solved in the frequency domain via FFT together with the associated Langreth rules for convolutions.

\paragraph{Bosonic Lattice Self-Consistency}
Here we use the more straightforward implementation directly via the bosonic self-energy $\Pi$.
Using the calculated $W_\text{imp}$ we employ $W_\text{imp} = \mathcal{U} + \mathcal{U} * \Pi_\text{imp} * W_\text{imp}$ to obtain the bosonic self-energy via
\begin{equation}
  (1 + \chi_\text{imp} * \mathcal{U}) * \Pi_\text{imp} = \chi_\text{imp}.
\end{equation}
Similar to the fermionic self-consistency, the bosonic lattice Green's function is defined via yet another Volterra equation
\begin{equation}
  (1 - \nu_k * \Pi_\text{imp}) * W_k = \nu_k,
\end{equation}
where $\nu_k = V_k + U$ is the bosonic dispersion relation $V_k$, i.e., the Fourier transform to $k$-space of $V \delta_{\langle ij \rangle}$, given, for example, by $2V (\cos(k_x) + \cos(k_y))$ on the square lattice, plus the static onsite Hubbard interaction $U$.
Using the local bosonic Green's function $W_\text{loc} = (N_k)^{-1} \sum_k W_k$, we are able to close the bosonic self-consistency
\begin{equation}
  (1 + W_\text{loc} * \Pi_\text{imp}) * \mathcal{U} = W_\text{loc},
\end{equation}
which is again efficiently solved in frequency space.
Note that for $V=0$ the bosonic Weiss field should reduce to the instantaneous part only $\mathcal{U} (t, t') = U \delta_\mathcal{C} (t, t')$, such that $\mathcal{D} (t, t') = 0$, which presents a very convenient check to verify the implementation.


\begin{thebibliography}{66}%
\makeatletter
\providecommand \@ifxundefined [1]{%
 \@ifx{#1\undefined}
}%
\providecommand \@ifnum [1]{%
 \ifnum #1\expandafter \@firstoftwo
 \else \expandafter \@secondoftwo
 \fi
}%
\providecommand \@ifx [1]{%
 \ifx #1\expandafter \@firstoftwo
 \else \expandafter \@secondoftwo
 \fi
}%
\providecommand \natexlab [1]{#1}%
\providecommand \enquote  [1]{``#1''}%
\providecommand \bibnamefont  [1]{#1}%
\providecommand \bibfnamefont [1]{#1}%
\providecommand \citenamefont [1]{#1}%
\providecommand \href@noop [0]{\@secondoftwo}%
\providecommand \href [0]{\begingroup \@sanitize@url \@href}%
\providecommand \@href[1]{\@@startlink{#1}\@@href}%
\providecommand \@@href[1]{\endgroup#1\@@endlink}%
\providecommand \@sanitize@url [0]{\catcode `\\12\catcode `\$12\catcode
  `\&12\catcode `\#12\catcode `\^12\catcode `\_12\catcode `\%12\relax}%
\providecommand \@@startlink[1]{}%
\providecommand \@@endlink[0]{}%
\providecommand \url  [0]{\begingroup\@sanitize@url \@url }%
\providecommand \@url [1]{\endgroup\@href {#1}{\urlprefix }}%
\providecommand \urlprefix  [0]{URL }%
\providecommand \Eprint [0]{\href }%
\providecommand \doibase [0]{https://doi.org/}%
\providecommand \selectlanguage [0]{\@gobble}%
\providecommand \bibinfo  [0]{\@secondoftwo}%
\providecommand \bibfield  [0]{\@secondoftwo}%
\providecommand \translation [1]{[#1]}%
\providecommand \BibitemOpen [0]{}%
\providecommand \bibitemStop [0]{}%
\providecommand \bibitemNoStop [0]{.\EOS\space}%
\providecommand \EOS [0]{\spacefactor3000\relax}%
\providecommand \BibitemShut  [1]{\csname bibitem#1\endcsname}%
\let\auto@bib@innerbib\@empty
\bibitem [{\citenamefont {Anderson}(1961)}]{Anderson-1961}%
  \BibitemOpen
  \bibfield  {author} {\bibinfo {author} {\bibfnamefont {P.~W.}\ \bibnamefont
  {Anderson}},\ }\href {https://doi.org/10.1103/PhysRev.124.41} {\bibfield
  {journal} {\bibinfo  {journal} {Physical Review}\ }\textbf {\bibinfo {volume}
  {124}},\ \bibinfo {pages} {41} (\bibinfo {year} {1961})}\BibitemShut
  {NoStop}%
\bibitem [{\citenamefont {Georges}\ \emph {et~al.}(1996)\citenamefont
  {Georges}, \citenamefont {Kotliar}, \citenamefont {Krauth},\ and\
  \citenamefont {Rozenberg}}]{Georges-etal-1996b}%
  \BibitemOpen
  \bibfield  {author} {\bibinfo {author} {\bibfnamefont {A.}~\bibnamefont
  {Georges}}, \bibinfo {author} {\bibfnamefont {G.}~\bibnamefont {Kotliar}},
  \bibinfo {author} {\bibfnamefont {W.}~\bibnamefont {Krauth}},\ and\ \bibinfo
  {author} {\bibfnamefont {M.~J.}\ \bibnamefont {Rozenberg}},\ }\href
  {https://doi.org/10.1103/RevModPhys.68.13} {\bibfield  {journal} {\bibinfo
  {journal} {Reviews of Modern Physics}\ }\textbf {\bibinfo {volume} {68}},\
  \bibinfo {pages} {13} (\bibinfo {year} {1996})}\BibitemShut {NoStop}%
\bibitem [{\citenamefont {Lu}\ and\ \citenamefont
  {Haverkort}(2017)}]{Lu-Haverkort-2017}%
  \BibitemOpen
  \bibfield  {author} {\bibinfo {author} {\bibfnamefont {Y.}~\bibnamefont
  {Lu}}\ and\ \bibinfo {author} {\bibfnamefont {M.~W.}\ \bibnamefont
  {Haverkort}},\ }\href {https://doi.org/10.1140/epjst/e2017-70042-4}
  {\bibfield  {journal} {\bibinfo  {journal} {The European Physical Journal
  Special Topics}\ }\textbf {\bibinfo {volume} {226}},\ \bibinfo {pages} {2549}
  (\bibinfo {year} {2017})}\BibitemShut {NoStop}%
\bibitem [{\citenamefont {Wilson}(1975)}]{Wilson-1975}%
  \BibitemOpen
  \bibfield  {author} {\bibinfo {author} {\bibfnamefont {K.~G.}\ \bibnamefont
  {Wilson}},\ }\href {https://doi.org/10.1103/RevModPhys.47.773} {\bibfield
  {journal} {\bibinfo  {journal} {Reviews of Modern Physics}\ }\textbf
  {\bibinfo {volume} {47}},\ \bibinfo {pages} {773} (\bibinfo {year}
  {1975})}\BibitemShut {NoStop}%
\bibitem [{\citenamefont {Bulla}\ \emph {et~al.}(2008)\citenamefont {Bulla},
  \citenamefont {Costi},\ and\ \citenamefont
  {Pruschke}}]{Bulla-Costi-Pruschke-2008}%
  \BibitemOpen
  \bibfield  {author} {\bibinfo {author} {\bibfnamefont {R.}~\bibnamefont
  {Bulla}}, \bibinfo {author} {\bibfnamefont {T.~A.}\ \bibnamefont {Costi}},\
  and\ \bibinfo {author} {\bibfnamefont {T.}~\bibnamefont {Pruschke}},\ }\href
  {https://doi.org/10.1103/RevModPhys.80.395} {\bibfield  {journal} {\bibinfo
  {journal} {Reviews of Modern Physics}\ }\textbf {\bibinfo {volume} {80}},\
  \bibinfo {pages} {395} (\bibinfo {year} {2008})}\BibitemShut {NoStop}%
\bibitem [{\citenamefont {Wolf}\ \emph {et~al.}(2015)\citenamefont {Wolf},
  \citenamefont {Go}, \citenamefont {McCulloch}, \citenamefont {Millis},\ and\
  \citenamefont {Schollw\"ock}}]{Wolf2015}%
  \BibitemOpen
  \bibfield  {author} {\bibinfo {author} {\bibfnamefont {F.~A.}\ \bibnamefont
  {Wolf}}, \bibinfo {author} {\bibfnamefont {A.}~\bibnamefont {Go}}, \bibinfo
  {author} {\bibfnamefont {I.~P.}\ \bibnamefont {McCulloch}}, \bibinfo {author}
  {\bibfnamefont {A.~J.}\ \bibnamefont {Millis}},\ and\ \bibinfo {author}
  {\bibfnamefont {U.}~\bibnamefont {Schollw\"ock}},\ }\href
  {https://doi.org/10.1103/PhysRevX.5.041032} {\bibfield  {journal} {\bibinfo
  {journal} {Phys. Rev. X}\ }\textbf {\bibinfo {volume} {5}},\ \bibinfo {pages}
  {041032} (\bibinfo {year} {2015})}\BibitemShut {NoStop}%
\bibitem [{\citenamefont {Bauernfeind}\ \emph {et~al.}(2017)\citenamefont
  {Bauernfeind}, \citenamefont {Zingl}, \citenamefont {Triebl}, \citenamefont
  {Aichhorn},\ and\ \citenamefont {Evertz}}]{Bauernfeind2017}%
  \BibitemOpen
  \bibfield  {author} {\bibinfo {author} {\bibfnamefont {D.}~\bibnamefont
  {Bauernfeind}}, \bibinfo {author} {\bibfnamefont {M.}~\bibnamefont {Zingl}},
  \bibinfo {author} {\bibfnamefont {R.}~\bibnamefont {Triebl}}, \bibinfo
  {author} {\bibfnamefont {M.}~\bibnamefont {Aichhorn}},\ and\ \bibinfo
  {author} {\bibfnamefont {H.~G.}\ \bibnamefont {Evertz}},\ }\href
  {https://doi.org/10.1103/PhysRevX.7.031013} {\bibfield  {journal} {\bibinfo
  {journal} {Phys. Rev. X}\ }\textbf {\bibinfo {volume} {7}},\ \bibinfo {pages}
  {031013} (\bibinfo {year} {2017})}\BibitemShut {NoStop}%
\bibitem [{\citenamefont {Gull}\ \emph {et~al.}(2011)\citenamefont {Gull},
  \citenamefont {Millis}, \citenamefont {Lichtenstein}, \citenamefont
  {Rubtsov}, \citenamefont {Troyer},\ and\ \citenamefont {Werner}}]{Gull2011}%
  \BibitemOpen
  \bibfield  {author} {\bibinfo {author} {\bibfnamefont {E.}~\bibnamefont
  {Gull}}, \bibinfo {author} {\bibfnamefont {A.~J.}\ \bibnamefont {Millis}},
  \bibinfo {author} {\bibfnamefont {A.~I.}\ \bibnamefont {Lichtenstein}},
  \bibinfo {author} {\bibfnamefont {A.~N.}\ \bibnamefont {Rubtsov}}, \bibinfo
  {author} {\bibfnamefont {M.}~\bibnamefont {Troyer}},\ and\ \bibinfo {author}
  {\bibfnamefont {P.}~\bibnamefont {Werner}},\ }\href
  {https://doi.org/10.1103/RevModPhys.83.349} {\bibfield  {journal} {\bibinfo
  {journal} {Rev. Mod. Phys.}\ }\textbf {\bibinfo {volume} {83}},\ \bibinfo
  {pages} {349} (\bibinfo {year} {2011})}\BibitemShut {NoStop}%
\bibitem [{\citenamefont {Werner}\ \emph {et~al.}(2006)\citenamefont {Werner},
  \citenamefont {Comanac}, \citenamefont {de’ Medici}, \citenamefont
  {Troyer},\ and\ \citenamefont {Millis}}]{Werner-etal-2006}%
  \BibitemOpen
  \bibfield  {author} {\bibinfo {author} {\bibfnamefont {P.}~\bibnamefont
  {Werner}}, \bibinfo {author} {\bibfnamefont {A.}~\bibnamefont {Comanac}},
  \bibinfo {author} {\bibfnamefont {L.}~\bibnamefont {de’ Medici}}, \bibinfo
  {author} {\bibfnamefont {M.}~\bibnamefont {Troyer}},\ and\ \bibinfo {author}
  {\bibfnamefont {A.~J.}\ \bibnamefont {Millis}},\ }\href
  {https://doi.org/10.1103/PhysRevLett.97.076405} {\bibfield  {journal}
  {\bibinfo  {journal} {Physical Review Letters}\ }\textbf {\bibinfo {volume}
  {97}},\ \bibinfo {pages} {076405} (\bibinfo {year} {2006})}\BibitemShut
  {NoStop}%
\bibitem [{\citenamefont {Rubtsov}\ \emph {et~al.}(2005)\citenamefont
  {Rubtsov}, \citenamefont {Savkin},\ and\ \citenamefont
  {Lichtenstein}}]{Rubtsov2005}%
  \BibitemOpen
  \bibfield  {author} {\bibinfo {author} {\bibfnamefont {A.~N.}\ \bibnamefont
  {Rubtsov}}, \bibinfo {author} {\bibfnamefont {V.~V.}\ \bibnamefont
  {Savkin}},\ and\ \bibinfo {author} {\bibfnamefont {A.~I.}\ \bibnamefont
  {Lichtenstein}},\ }\href {https://doi.org/10.1103/PhysRevB.72.035122}
  {\bibfield  {journal} {\bibinfo  {journal} {Phys. Rev. B}\ }\textbf {\bibinfo
  {volume} {72}},\ \bibinfo {pages} {035122} (\bibinfo {year}
  {2005})}\BibitemShut {NoStop}%
\bibitem [{\citenamefont {M\"uhlbacher}\ and\ \citenamefont
  {Rabani}(2008)}]{Muehlbacher2008}%
  \BibitemOpen
  \bibfield  {author} {\bibinfo {author} {\bibfnamefont {L.}~\bibnamefont
  {M\"uhlbacher}}\ and\ \bibinfo {author} {\bibfnamefont {E.}~\bibnamefont
  {Rabani}},\ }\href {https://doi.org/10.1103/PhysRevLett.100.176403}
  {\bibfield  {journal} {\bibinfo  {journal} {Phys. Rev. Lett.}\ }\textbf
  {\bibinfo {volume} {100}},\ \bibinfo {pages} {176403} (\bibinfo {year}
  {2008})}\BibitemShut {NoStop}%
\bibitem [{\citenamefont {Werner}\ \emph {et~al.}(2009)\citenamefont {Werner},
  \citenamefont {Oka},\ and\ \citenamefont {Millis}}]{Werner2009}%
  \BibitemOpen
  \bibfield  {author} {\bibinfo {author} {\bibfnamefont {P.}~\bibnamefont
  {Werner}}, \bibinfo {author} {\bibfnamefont {T.}~\bibnamefont {Oka}},\ and\
  \bibinfo {author} {\bibfnamefont {A.~J.}\ \bibnamefont {Millis}},\ }\href
  {https://doi.org/10.1103/PhysRevB.79.035320} {\bibfield  {journal} {\bibinfo
  {journal} {Phys. Rev. B}\ }\textbf {\bibinfo {volume} {79}},\ \bibinfo
  {pages} {035320} (\bibinfo {year} {2009})}\BibitemShut {NoStop}%
\bibitem [{\citenamefont {Gramsch}\ \emph {et~al.}(2013)\citenamefont
  {Gramsch}, \citenamefont {Balzer}, \citenamefont {Eckstein},\ and\
  \citenamefont {Kollar}}]{Gramsch-etal-2013}%
  \BibitemOpen
  \bibfield  {author} {\bibinfo {author} {\bibfnamefont {C.}~\bibnamefont
  {Gramsch}}, \bibinfo {author} {\bibfnamefont {K.}~\bibnamefont {Balzer}},
  \bibinfo {author} {\bibfnamefont {M.}~\bibnamefont {Eckstein}},\ and\
  \bibinfo {author} {\bibfnamefont {M.}~\bibnamefont {Kollar}},\ }\href
  {https://doi.org/10.1103/PhysRevB.88.235106} {\bibfield  {journal} {\bibinfo
  {journal} {Physical Review B}\ }\textbf {\bibinfo {volume} {88}},\ \bibinfo
  {pages} {235106} (\bibinfo {year} {2013})}\BibitemShut {NoStop}%
\bibitem [{\citenamefont {Wolf}\ \emph {et~al.}(2014)\citenamefont {Wolf},
  \citenamefont {McCulloch},\ and\ \citenamefont {Schollw\"ock}}]{Wolf2014}%
  \BibitemOpen
  \bibfield  {author} {\bibinfo {author} {\bibfnamefont {F.~A.}\ \bibnamefont
  {Wolf}}, \bibinfo {author} {\bibfnamefont {I.~P.}\ \bibnamefont
  {McCulloch}},\ and\ \bibinfo {author} {\bibfnamefont {U.}~\bibnamefont
  {Schollw\"ock}},\ }\href {https://doi.org/10.1103/PhysRevB.90.235131}
  {\bibfield  {journal} {\bibinfo  {journal} {Phys. Rev. B}\ }\textbf {\bibinfo
  {volume} {90}},\ \bibinfo {pages} {235131} (\bibinfo {year}
  {2014})}\BibitemShut {NoStop}%
\bibitem [{\citenamefont {Arrigoni}\ \emph {et~al.}(2013)\citenamefont
  {Arrigoni}, \citenamefont {Knap},\ and\ \citenamefont {Von
  Der~Linden}}]{Arrigoni-Knap-VonDerLinden-2013}%
  \BibitemOpen
  \bibfield  {author} {\bibinfo {author} {\bibfnamefont {E.}~\bibnamefont
  {Arrigoni}}, \bibinfo {author} {\bibfnamefont {M.}~\bibnamefont {Knap}},\
  and\ \bibinfo {author} {\bibfnamefont {W.}~\bibnamefont {Von Der~Linden}},\
  }\href {https://doi.org/10.1103/PhysRevLett.110.086403} {\bibfield  {journal}
  {\bibinfo  {journal} {Physical Review Letters}\ }\textbf {\bibinfo {volume}
  {110}},\ \bibinfo {pages} {086403} (\bibinfo {year} {2013})}\BibitemShut
  {NoStop}%
\bibitem [{\citenamefont {Chen}\ \emph {et~al.}(2024)\citenamefont {Chen},
  \citenamefont {Xu},\ and\ \citenamefont {Guo}}]{Chen2024}%
  \BibitemOpen
  \bibfield  {author} {\bibinfo {author} {\bibfnamefont {R.}~\bibnamefont
  {Chen}}, \bibinfo {author} {\bibfnamefont {X.}~\bibnamefont {Xu}},\ and\
  \bibinfo {author} {\bibfnamefont {C.}~\bibnamefont {Guo}},\ }\href
  {https://doi.org/10.1103/PhysRevB.109.045140} {\bibfield  {journal} {\bibinfo
   {journal} {Phys. Rev. B}\ }\textbf {\bibinfo {volume} {109}},\ \bibinfo
  {pages} {045140} (\bibinfo {year} {2024})}\BibitemShut {NoStop}%
\bibitem [{\citenamefont {Thoenniss}\ \emph {et~al.}(2023)\citenamefont
  {Thoenniss}, \citenamefont {Lerose},\ and\ \citenamefont
  {Abanin}}]{Thoenniss2023}%
  \BibitemOpen
  \bibfield  {author} {\bibinfo {author} {\bibfnamefont {J.}~\bibnamefont
  {Thoenniss}}, \bibinfo {author} {\bibfnamefont {A.}~\bibnamefont {Lerose}},\
  and\ \bibinfo {author} {\bibfnamefont {D.~A.}\ \bibnamefont {Abanin}},\
  }\href {https://doi.org/10.1103/PhysRevB.107.195101} {\bibfield  {journal}
  {\bibinfo  {journal} {Phys. Rev. B}\ }\textbf {\bibinfo {volume} {107}},\
  \bibinfo {pages} {195101} (\bibinfo {year} {2023})}\BibitemShut {NoStop}%
\bibitem [{\citenamefont {Nayak}\ \emph {et~al.}(2025)\citenamefont {Nayak},
  \citenamefont {Thoenniss}, \citenamefont {Sonner}, \citenamefont {Abanin},\
  and\ \citenamefont {Werner}}]{Nayak2025}%
  \BibitemOpen
  \bibfield  {author} {\bibinfo {author} {\bibfnamefont {M.}~\bibnamefont
  {Nayak}}, \bibinfo {author} {\bibfnamefont {J.}~\bibnamefont {Thoenniss}},
  \bibinfo {author} {\bibfnamefont {M.}~\bibnamefont {Sonner}}, \bibinfo
  {author} {\bibfnamefont {D.~A.}\ \bibnamefont {Abanin}},\ and\ \bibinfo
  {author} {\bibfnamefont {P.}~\bibnamefont {Werner}},\ }\href
  {https://doi.org/10.1103/xsbn-jk16} {\bibfield  {journal} {\bibinfo
  {journal} {Phys. Rev. B}\ }\textbf {\bibinfo {volume} {112}},\ \bibinfo
  {pages} {035103} (\bibinfo {year} {2025})}\BibitemShut {NoStop}%
\bibitem [{\citenamefont {Aoki}\ \emph {et~al.}(2014)\citenamefont {Aoki},
  \citenamefont {Tsuji}, \citenamefont {Eckstein}, \citenamefont {Kollar},
  \citenamefont {Oka},\ and\ \citenamefont {Werner}}]{Aoki-etal-2014}%
  \BibitemOpen
  \bibfield  {author} {\bibinfo {author} {\bibfnamefont {H.}~\bibnamefont
  {Aoki}}, \bibinfo {author} {\bibfnamefont {N.}~\bibnamefont {Tsuji}},
  \bibinfo {author} {\bibfnamefont {M.}~\bibnamefont {Eckstein}}, \bibinfo
  {author} {\bibfnamefont {M.}~\bibnamefont {Kollar}}, \bibinfo {author}
  {\bibfnamefont {T.}~\bibnamefont {Oka}},\ and\ \bibinfo {author}
  {\bibfnamefont {P.}~\bibnamefont {Werner}},\ }\href
  {https://doi.org/10.1103/RevModPhys.86.779} {\bibfield  {journal} {\bibinfo
  {journal} {Reviews of Modern Physics}\ }\textbf {\bibinfo {volume} {86}},\
  \bibinfo {pages} {779} (\bibinfo {year} {2014})}\BibitemShut {NoStop}%
\bibitem [{\citenamefont {Murakami}\ \emph {et~al.}(2025)\citenamefont
  {Murakami}, \citenamefont {Gole\ifmmode~\check{z}\else \v{z}\fi{}},
  \citenamefont {Eckstein},\ and\ \citenamefont {Werner}}]{Murakami2025}%
  \BibitemOpen
  \bibfield  {author} {\bibinfo {author} {\bibfnamefont {Y.}~\bibnamefont
  {Murakami}}, \bibinfo {author} {\bibfnamefont {D.}~\bibnamefont
  {Gole\ifmmode~\check{z}\else \v{z}\fi{}}}, \bibinfo {author} {\bibfnamefont
  {M.}~\bibnamefont {Eckstein}},\ and\ \bibinfo {author} {\bibfnamefont
  {P.}~\bibnamefont {Werner}},\ }\href {https://doi.org/10.1103/tkjh-lr83}
  {\bibfield  {journal} {\bibinfo  {journal} {Rev. Mod. Phys.}\ }\textbf
  {\bibinfo {volume} {97}},\ \bibinfo {pages} {035001} (\bibinfo {year}
  {2025})}\BibitemShut {NoStop}%
\bibitem [{\citenamefont {Keiter}\ and\ \citenamefont
  {Kimball}(1970)}]{Keiter-Kimball-1970}%
  \BibitemOpen
  \bibfield  {author} {\bibinfo {author} {\bibfnamefont {H.}~\bibnamefont
  {Keiter}}\ and\ \bibinfo {author} {\bibfnamefont {J.~C.}\ \bibnamefont
  {Kimball}},\ }\href {https://doi.org/10.1103/PhysRevLett.25.672} {\bibfield
  {journal} {\bibinfo  {journal} {Physical Review Letters}\ }\textbf {\bibinfo
  {volume} {25}},\ \bibinfo {pages} {672} (\bibinfo {year} {1970})}\BibitemShut
  {NoStop}%
\bibitem [{\citenamefont {Coleman}(1984)}]{Coleman-1984}%
  \BibitemOpen
  \bibfield  {author} {\bibinfo {author} {\bibfnamefont {P.}~\bibnamefont
  {Coleman}},\ }\href {https://doi.org/10.1103/PhysRevB.29.3035} {\bibfield
  {journal} {\bibinfo  {journal} {Physical Review B}\ }\textbf {\bibinfo
  {volume} {29}},\ \bibinfo {pages} {3035} (\bibinfo {year}
  {1984})}\BibitemShut {NoStop}%
\bibitem [{\citenamefont {Bickers}\ \emph {et~al.}(1987)\citenamefont
  {Bickers}, \citenamefont {Cox},\ and\ \citenamefont {Wilkins}}]{Bickers1987}%
  \BibitemOpen
  \bibfield  {author} {\bibinfo {author} {\bibfnamefont {N.~E.}\ \bibnamefont
  {Bickers}}, \bibinfo {author} {\bibfnamefont {D.~L.}\ \bibnamefont {Cox}},\
  and\ \bibinfo {author} {\bibfnamefont {J.~W.}\ \bibnamefont {Wilkins}},\
  }\href {https://doi.org/10.1103/PhysRevB.36.2036} {\bibfield  {journal}
  {\bibinfo  {journal} {Phys. Rev. B}\ }\textbf {\bibinfo {volume} {36}},\
  \bibinfo {pages} {2036} (\bibinfo {year} {1987})}\BibitemShut {NoStop}%
\bibitem [{\citenamefont {Haule}\ \emph {et~al.}(2001)\citenamefont {Haule},
  \citenamefont {Kirchner}, \citenamefont {Kroha},\ and\ \citenamefont
  {W\"olfle}}]{Haule2001}%
  \BibitemOpen
  \bibfield  {author} {\bibinfo {author} {\bibfnamefont {K.}~\bibnamefont
  {Haule}}, \bibinfo {author} {\bibfnamefont {S.}~\bibnamefont {Kirchner}},
  \bibinfo {author} {\bibfnamefont {J.}~\bibnamefont {Kroha}},\ and\ \bibinfo
  {author} {\bibfnamefont {P.}~\bibnamefont {W\"olfle}},\ }\href
  {https://doi.org/10.1103/PhysRevB.64.155111} {\bibfield  {journal} {\bibinfo
  {journal} {Phys. Rev. B}\ }\textbf {\bibinfo {volume} {64}},\ \bibinfo
  {pages} {155111} (\bibinfo {year} {2001})}\BibitemShut {NoStop}%
\bibitem [{\citenamefont {Kim}\ \emph {et~al.}(2023)\citenamefont {Kim},
  \citenamefont {Lenk}, \citenamefont {Li}, \citenamefont {Werner},\ and\
  \citenamefont {Eckstein}}]{Kim2023}%
  \BibitemOpen
  \bibfield  {author} {\bibinfo {author} {\bibfnamefont {A.~J.}\ \bibnamefont
  {Kim}}, \bibinfo {author} {\bibfnamefont {K.}~\bibnamefont {Lenk}}, \bibinfo
  {author} {\bibfnamefont {J.}~\bibnamefont {Li}}, \bibinfo {author}
  {\bibfnamefont {P.}~\bibnamefont {Werner}},\ and\ \bibinfo {author}
  {\bibfnamefont {M.}~\bibnamefont {Eckstein}},\ }\href
  {https://doi.org/10.1103/PhysRevLett.130.036901} {\bibfield  {journal}
  {\bibinfo  {journal} {Phys. Rev. Lett.}\ }\textbf {\bibinfo {volume} {130}},\
  \bibinfo {pages} {036901} (\bibinfo {year} {2023})}\BibitemShut {NoStop}%
\bibitem [{\citenamefont {Kim}\ \emph {et~al.}(2022)\citenamefont {Kim},
  \citenamefont {Li}, \citenamefont {Eckstein},\ and\ \citenamefont
  {Werner}}]{Kim2022}%
  \BibitemOpen
  \bibfield  {author} {\bibinfo {author} {\bibfnamefont {A.~J.}\ \bibnamefont
  {Kim}}, \bibinfo {author} {\bibfnamefont {J.}~\bibnamefont {Li}}, \bibinfo
  {author} {\bibfnamefont {M.}~\bibnamefont {Eckstein}},\ and\ \bibinfo
  {author} {\bibfnamefont {P.}~\bibnamefont {Werner}},\ }\href
  {https://doi.org/10.1103/PhysRevB.106.085124} {\bibfield  {journal} {\bibinfo
   {journal} {Phys. Rev. B}\ }\textbf {\bibinfo {volume} {106}},\ \bibinfo
  {pages} {085124} (\bibinfo {year} {2022})}\BibitemShut {NoStop}%
\bibitem [{\citenamefont {Haule}(2023)}]{Haule-2023}%
  \BibitemOpen
  \bibfield  {author} {\bibinfo {author} {\bibfnamefont {K.}~\bibnamefont
  {Haule}},\ }\href {https://arxiv.org/abs/2311.09412v1} {{\selectlanguage
  {en}\bibinfo {title} {Strong coupling quantum impurity solver on the real and
  imaginary axis}}} (\bibinfo {year} {2023})\BibitemShut {NoStop}%
\bibitem [{\citenamefont {Kaye}\ \emph
  {et~al.}(2024{\natexlab{a}})\citenamefont {Kaye}, \citenamefont {Strand},\
  and\ \citenamefont {Wentzell}}]{Kaye-Strand-Wentzell-2024}%
  \BibitemOpen
  \bibfield  {author} {\bibinfo {author} {\bibfnamefont {J.}~\bibnamefont
  {Kaye}}, \bibinfo {author} {\bibfnamefont {H.~U.~r.}\ \bibnamefont
  {Strand}},\ and\ \bibinfo {author} {\bibfnamefont {N.}~\bibnamefont
  {Wentzell}},\ }\href {https://doi.org/10.21105/joss.06297} {\bibfield
  {journal} {\bibinfo  {journal} {Journal of Open Source Software}\ }\textbf
  {\bibinfo {volume} {9}},\ \bibinfo {pages} {6297} (\bibinfo {year}
  {2024}{\natexlab{a}})}\BibitemShut {NoStop}%
\bibitem [{\citenamefont {Kaye}\ \emph {et~al.}(2022)\citenamefont {Kaye},
  \citenamefont {Chen},\ and\ \citenamefont
  {Parcollet}}]{Kaye-Chen-Parcollet-2022}%
  \BibitemOpen
  \bibfield  {author} {\bibinfo {author} {\bibfnamefont {J.}~\bibnamefont
  {Kaye}}, \bibinfo {author} {\bibfnamefont {K.}~\bibnamefont {Chen}},\ and\
  \bibinfo {author} {\bibfnamefont {O.}~\bibnamefont {Parcollet}},\ }\href
  {https://doi.org/10.1103/PhysRevB.105.235115} {\bibfield  {journal} {\bibinfo
   {journal} {Physical Review B}\ }\textbf {\bibinfo {volume} {105}},\ \bibinfo
  {pages} {235115} (\bibinfo {year} {2022})}\BibitemShut {NoStop}%
\bibitem [{\citenamefont {Huang}\ \emph {et~al.}(2025)\citenamefont {Huang},
  \citenamefont {Golez}, \citenamefont {Strand},\ and\ \citenamefont
  {Kaye}}]{Huang-etal-2025}%
  \BibitemOpen
  \bibfield  {author} {\bibinfo {author} {\bibfnamefont {Z.}~\bibnamefont
  {Huang}}, \bibinfo {author} {\bibfnamefont {D.}~\bibnamefont {Golez}},
  \bibinfo {author} {\bibfnamefont {H.~U.~R.}\ \bibnamefont {Strand}},\ and\
  \bibinfo {author} {\bibfnamefont {J.}~\bibnamefont {Kaye}},\ }\href
  {https://doi.org/10.21468/SciPostPhys.19.5.121} {\bibfield  {journal}
  {\bibinfo  {journal} {SciPost Physics}\ }\textbf {\bibinfo {volume} {19}},\
  \bibinfo {pages} {121} (\bibinfo {year} {2025})}\BibitemShut {NoStop}%
\bibitem [{\citenamefont {Kaye}\ \emph
  {et~al.}(2024{\natexlab{b}})\citenamefont {Kaye}, \citenamefont {Huang},
  \citenamefont {Strand},\ and\ \citenamefont {Golež}}]{Kaye-etal-2024}%
  \BibitemOpen
  \bibfield  {author} {\bibinfo {author} {\bibfnamefont {J.}~\bibnamefont
  {Kaye}}, \bibinfo {author} {\bibfnamefont {Z.}~\bibnamefont {Huang}},
  \bibinfo {author} {\bibfnamefont {H.~U.~R.}\ \bibnamefont {Strand}},\ and\
  \bibinfo {author} {\bibfnamefont {D.}~\bibnamefont {Golež}},\ }\href
  {https://doi.org/10.1103/PhysRevX.14.031034} {\bibfield  {journal} {\bibinfo
  {journal} {Physical Review X}\ }\textbf {\bibinfo {volume} {14}},\ \bibinfo
  {pages} {031034} (\bibinfo {year} {2024}{\natexlab{b}})}\BibitemShut
  {NoStop}%
\bibitem [{\citenamefont {Eckstein}\ and\ \citenamefont
  {Werner}(2010)}]{Eckstein-Werner-2010}%
  \BibitemOpen
  \bibfield  {author} {\bibinfo {author} {\bibfnamefont {M.}~\bibnamefont
  {Eckstein}}\ and\ \bibinfo {author} {\bibfnamefont {P.}~\bibnamefont
  {Werner}},\ }\href {https://doi.org/10.1103/PhysRevB.82.115115} {\bibfield
  {journal} {\bibinfo  {journal} {Physical Review B}\ }\textbf {\bibinfo
  {volume} {82}},\ \bibinfo {pages} {115115} (\bibinfo {year}
  {2010})}\BibitemShut {NoStop}%
\bibitem [{\citenamefont {Eckstein}\ and\ \citenamefont
  {Werner}(2011)}]{Eckstein2011}%
  \BibitemOpen
  \bibfield  {author} {\bibinfo {author} {\bibfnamefont {M.}~\bibnamefont
  {Eckstein}}\ and\ \bibinfo {author} {\bibfnamefont {P.}~\bibnamefont
  {Werner}},\ }\href {https://doi.org/10.1103/PhysRevB.84.035122} {\bibfield
  {journal} {\bibinfo  {journal} {Phys. Rev. B}\ }\textbf {\bibinfo {volume}
  {84}},\ \bibinfo {pages} {035122} (\bibinfo {year} {2011})}\BibitemShut
  {NoStop}%
\bibitem [{\citenamefont {Cohen}\ \emph {et~al.}(2015)\citenamefont {Cohen},
  \citenamefont {Gull}, \citenamefont {Reichman},\ and\ \citenamefont
  {Millis}}]{Cohen-etal-2015}%
  \BibitemOpen
  \bibfield  {author} {\bibinfo {author} {\bibfnamefont {G.}~\bibnamefont
  {Cohen}}, \bibinfo {author} {\bibfnamefont {E.}~\bibnamefont {Gull}},
  \bibinfo {author} {\bibfnamefont {D.~R.}\ \bibnamefont {Reichman}},\ and\
  \bibinfo {author} {\bibfnamefont {A.~J.}\ \bibnamefont {Millis}},\ }\href
  {https://doi.org/10.1103/PhysRevLett.115.266802} {\bibfield  {journal}
  {\bibinfo  {journal} {Physical Review Letters}\ }\textbf {\bibinfo {volume}
  {115}},\ \bibinfo {pages} {266802} (\bibinfo {year} {2015})}\BibitemShut
  {NoStop}%
\bibitem [{\citenamefont {Erpenbeck}\ \emph
  {et~al.}(2023{\natexlab{a}})\citenamefont {Erpenbeck}, \citenamefont {Gull},\
  and\ \citenamefont {Cohen}}]{Erpenbeck-Gull-Cohen-2023}%
  \BibitemOpen
  \bibfield  {author} {\bibinfo {author} {\bibfnamefont {A.}~\bibnamefont
  {Erpenbeck}}, \bibinfo {author} {\bibfnamefont {E.}~\bibnamefont {Gull}},\
  and\ \bibinfo {author} {\bibfnamefont {G.}~\bibnamefont {Cohen}},\ }\href
  {https://doi.org/10.1103/PhysRevLett.130.186301} {\bibfield  {journal}
  {\bibinfo  {journal} {Physical Review Letters}\ }\textbf {\bibinfo {volume}
  {130}},\ \bibinfo {pages} {186301} (\bibinfo {year}
  {2023}{\natexlab{a}})}\BibitemShut {NoStop}%
\bibitem [{\citenamefont {Erpenbeck}\ \emph {et~al.}(2024)\citenamefont
  {Erpenbeck}, \citenamefont {Blommel}, \citenamefont {Zhang}, \citenamefont
  {Lin}, \citenamefont {Cohen},\ and\ \citenamefont
  {Gull}}]{Erpenbeck-etal-2024}%
  \BibitemOpen
  \bibfield  {author} {\bibinfo {author} {\bibfnamefont {A.}~\bibnamefont
  {Erpenbeck}}, \bibinfo {author} {\bibfnamefont {T.}~\bibnamefont {Blommel}},
  \bibinfo {author} {\bibfnamefont {L.}~\bibnamefont {Zhang}}, \bibinfo
  {author} {\bibfnamefont {W.-T.}\ \bibnamefont {Lin}}, \bibinfo {author}
  {\bibfnamefont {G.}~\bibnamefont {Cohen}},\ and\ \bibinfo {author}
  {\bibfnamefont {E.}~\bibnamefont {Gull}},\ }\href
  {https://doi.org/10.1063/5.0226253} {\bibfield  {journal} {\bibinfo
  {journal} {The Journal of Chemical Physics}\ }\textbf {\bibinfo {volume}
  {161}},\ \bibinfo {pages} {094104} (\bibinfo {year} {2024})}\BibitemShut
  {NoStop}%
\bibitem [{\citenamefont {Savostyanov}(2014)}]{Savostyanov-2014}%
  \BibitemOpen
  \bibfield  {author} {\bibinfo {author} {\bibfnamefont {D.~V.}\ \bibnamefont
  {Savostyanov}},\ }\href {https://doi.org/10.1016/j.laa.2014.06.006}
  {\bibfield  {journal} {\bibinfo  {journal} {Linear Algebra and its
  Applications}\ }\textbf {\bibinfo {volume} {458}},\ \bibinfo {pages} {217}
  (\bibinfo {year} {2014})}\BibitemShut {NoStop}%
\bibitem [{\citenamefont {Oseledets}(2011)}]{Oseledets-2011}%
  \BibitemOpen
  \bibfield  {author} {\bibinfo {author} {\bibfnamefont {I.~V.}\ \bibnamefont
  {Oseledets}},\ }\href {https://doi.org/10.1137/090752286} {\bibfield
  {journal} {\bibinfo  {journal} {SIAM Journal on Scientific Computing}\
  }\textbf {\bibinfo {volume} {33}},\ \bibinfo {pages} {2295} (\bibinfo {year}
  {2011})}\BibitemShut {NoStop}%
\bibitem [{\citenamefont {Oseledets}\ and\ \citenamefont
  {Tyrtyshnikov}(2010)}]{Oseledets-Tyrtyshnikov-2010}%
  \BibitemOpen
  \bibfield  {author} {\bibinfo {author} {\bibfnamefont {I.}~\bibnamefont
  {Oseledets}}\ and\ \bibinfo {author} {\bibfnamefont {E.}~\bibnamefont
  {Tyrtyshnikov}},\ }\href {https://doi.org/10.1016/j.laa.2009.07.024}
  {\bibfield  {journal} {\bibinfo  {journal} {Linear Algebra and its
  Applications}\ }\textbf {\bibinfo {volume} {432}},\ \bibinfo {pages} {70}
  (\bibinfo {year} {2010})}\BibitemShut {NoStop}%
\bibitem [{\citenamefont {Núñez~Fernández}\ \emph
  {et~al.}(2022)\citenamefont {Núñez~Fernández}, \citenamefont {Jeannin},
  \citenamefont {Dumitrescu}, \citenamefont {Kloss}, \citenamefont {Kaye},
  \citenamefont {Parcollet},\ and\ \citenamefont
  {Waintal}}]{NúñezFernández-etal-2022}%
  \BibitemOpen
  \bibfield  {author} {\bibinfo {author} {\bibfnamefont {Y.}~\bibnamefont
  {Núñez~Fernández}}, \bibinfo {author} {\bibfnamefont {M.}~\bibnamefont
  {Jeannin}}, \bibinfo {author} {\bibfnamefont {P.~T.}\ \bibnamefont
  {Dumitrescu}}, \bibinfo {author} {\bibfnamefont {T.}~\bibnamefont {Kloss}},
  \bibinfo {author} {\bibfnamefont {J.}~\bibnamefont {Kaye}}, \bibinfo {author}
  {\bibfnamefont {O.}~\bibnamefont {Parcollet}},\ and\ \bibinfo {author}
  {\bibfnamefont {X.}~\bibnamefont {Waintal}},\ }\href
  {https://doi.org/10.1103/PhysRevX.12.041018} {\bibfield  {journal} {\bibinfo
  {journal} {Physical Review X}\ }\textbf {\bibinfo {volume} {12}},\ \bibinfo
  {pages} {041018} (\bibinfo {year} {2022})}\BibitemShut {NoStop}%
\bibitem [{\citenamefont {Dolgov}\ and\ \citenamefont
  {Savostyanov}(2020)}]{Dolgov-Savostyanov-2020}%
  \BibitemOpen
  \bibfield  {author} {\bibinfo {author} {\bibfnamefont {S.}~\bibnamefont
  {Dolgov}}\ and\ \bibinfo {author} {\bibfnamefont {D.}~\bibnamefont
  {Savostyanov}},\ }\href {https://doi.org/10.1016/j.cpc.2019.106869}
  {\bibfield  {journal} {\bibinfo  {journal} {Computer Physics Communications}\
  }\textbf {\bibinfo {volume} {246}},\ \bibinfo {pages} {106869} (\bibinfo
  {year} {2020})}\BibitemShut {NoStop}%
\bibitem [{\citenamefont {Khoromskij}(2011)}]{Khoromskij-2011}%
  \BibitemOpen
  \bibfield  {author} {\bibinfo {author} {\bibfnamefont {B.~N.}\ \bibnamefont
  {Khoromskij}},\ }\href {https://doi.org/10.1007/s00365-011-9131-1} {\bibfield
   {journal} {\bibinfo  {journal} {Constructive Approximation}\ }\textbf
  {\bibinfo {volume} {34}},\ \bibinfo {pages} {257} (\bibinfo {year}
  {2011})}\BibitemShut {NoStop}%
\bibitem [{\citenamefont {Fernández}\ \emph {et~al.}(2025)\citenamefont
  {Fernández}, \citenamefont {Ritter}, \citenamefont {Jeannin}, \citenamefont
  {Li}, \citenamefont {Kloss}, \citenamefont {Louvet}, \citenamefont
  {Terasaki}, \citenamefont {Parcollet}, \citenamefont {Delft}, \citenamefont
  {Shinaoka},\ and\ \citenamefont {Waintal}}]{Fernández-etal-2025}%
  \BibitemOpen
  \bibfield  {author} {\bibinfo {author} {\bibfnamefont {Y.~N.}\ \bibnamefont
  {Fernández}}, \bibinfo {author} {\bibfnamefont {M.~K.}\ \bibnamefont
  {Ritter}}, \bibinfo {author} {\bibfnamefont {M.}~\bibnamefont {Jeannin}},
  \bibinfo {author} {\bibfnamefont {J.-W.}\ \bibnamefont {Li}}, \bibinfo
  {author} {\bibfnamefont {T.}~\bibnamefont {Kloss}}, \bibinfo {author}
  {\bibfnamefont {T.}~\bibnamefont {Louvet}}, \bibinfo {author} {\bibfnamefont
  {S.}~\bibnamefont {Terasaki}}, \bibinfo {author} {\bibfnamefont
  {O.}~\bibnamefont {Parcollet}}, \bibinfo {author} {\bibfnamefont {J.~v.}\
  \bibnamefont {Delft}}, \bibinfo {author} {\bibfnamefont {H.}~\bibnamefont
  {Shinaoka}},\ and\ \bibinfo {author} {\bibfnamefont {X.}~\bibnamefont
  {Waintal}},\ }\href {https://doi.org/10.21468/SciPostPhys.18.3.104}
  {\bibfield  {journal} {\bibinfo  {journal} {SciPost Physics}\ }\textbf
  {\bibinfo {volume} {18}},\ \bibinfo {pages} {104} (\bibinfo {year} {2025})},\
  \bibinfo {note} {arXiv:2407.02454 [physics]}\BibitemShut {NoStop}%
\bibitem [{\citenamefont {Shinaoka}\ \emph {et~al.}(2023)\citenamefont
  {Shinaoka}, \citenamefont {Wallerberger}, \citenamefont {Murakami},
  \citenamefont {Nogaki}, \citenamefont {Sakurai}, \citenamefont {Werner},\
  and\ \citenamefont {Kauch}}]{Shinaoka-etal-2023}%
  \BibitemOpen
  \bibfield  {author} {\bibinfo {author} {\bibfnamefont {H.}~\bibnamefont
  {Shinaoka}}, \bibinfo {author} {\bibfnamefont {M.}~\bibnamefont
  {Wallerberger}}, \bibinfo {author} {\bibfnamefont {Y.}~\bibnamefont
  {Murakami}}, \bibinfo {author} {\bibfnamefont {K.}~\bibnamefont {Nogaki}},
  \bibinfo {author} {\bibfnamefont {R.}~\bibnamefont {Sakurai}}, \bibinfo
  {author} {\bibfnamefont {P.}~\bibnamefont {Werner}},\ and\ \bibinfo {author}
  {\bibfnamefont {A.}~\bibnamefont {Kauch}},\ }\href
  {https://doi.org/10.1103/PhysRevX.13.021015} {\bibfield  {journal} {\bibinfo
  {journal} {Physical Review X}\ }\textbf {\bibinfo {volume} {13}},\ \bibinfo
  {pages} {021015} (\bibinfo {year} {2023})}\BibitemShut {NoStop}%
\bibitem [{\citenamefont {Frankenbach}\ \emph {et~al.}(2025)\citenamefont
  {Frankenbach}, \citenamefont {Ritter}, \citenamefont {Pelz}, \citenamefont
  {Ritz}, \citenamefont {Von~Delft},\ and\ \citenamefont
  {Ge}}]{Frankenbach-etal-2025}%
  \BibitemOpen
  \bibfield  {author} {\bibinfo {author} {\bibfnamefont {M.}~\bibnamefont
  {Frankenbach}}, \bibinfo {author} {\bibfnamefont {M.~K.}\ \bibnamefont
  {Ritter}}, \bibinfo {author} {\bibfnamefont {M.}~\bibnamefont {Pelz}},
  \bibinfo {author} {\bibfnamefont {N.}~\bibnamefont {Ritz}}, \bibinfo {author}
  {\bibfnamefont {J.}~\bibnamefont {Von~Delft}},\ and\ \bibinfo {author}
  {\bibfnamefont {A.}~\bibnamefont {Ge}},\ }\href
  {https://doi.org/10.1103/jx7h-lsqk} {\bibfield  {journal} {\bibinfo
  {journal} {Physical Review Research}\ }\textbf {\bibinfo {volume} {7}},\
  \bibinfo {pages} {043032} (\bibinfo {year} {2025})}\BibitemShut {NoStop}%
\bibitem [{\citenamefont {Rohshap}\ \emph {et~al.}(2025)\citenamefont
  {Rohshap}, \citenamefont {Ritter}, \citenamefont {Shinaoka}, \citenamefont
  {von Delft}, \citenamefont {Wallerberger},\ and\ \citenamefont
  {Kauch}}]{Rohshap-etal-2025}%
  \BibitemOpen
  \bibfield  {author} {\bibinfo {author} {\bibfnamefont {S.}~\bibnamefont
  {Rohshap}}, \bibinfo {author} {\bibfnamefont {M.~K.}\ \bibnamefont {Ritter}},
  \bibinfo {author} {\bibfnamefont {H.}~\bibnamefont {Shinaoka}}, \bibinfo
  {author} {\bibfnamefont {J.}~\bibnamefont {von Delft}}, \bibinfo {author}
  {\bibfnamefont {M.}~\bibnamefont {Wallerberger}},\ and\ \bibinfo {author}
  {\bibfnamefont {A.}~\bibnamefont {Kauch}},\ }\href
  {https://doi.org/10.1103/PhysRevResearch.7.023087} {\bibfield  {journal}
  {\bibinfo  {journal} {Physical Review Research}\ }\textbf {\bibinfo {volume}
  {7}},\ \bibinfo {pages} {023087} (\bibinfo {year} {2025})}\BibitemShut
  {NoStop}%
\bibitem [{\citenamefont {Murray}\ \emph {et~al.}(2024)\citenamefont {Murray},
  \citenamefont {Shinaoka},\ and\ \citenamefont
  {Werner}}]{Murray-Shinaoka-Werner-2024}%
  \BibitemOpen
  \bibfield  {author} {\bibinfo {author} {\bibfnamefont {M.}~\bibnamefont
  {Murray}}, \bibinfo {author} {\bibfnamefont {H.}~\bibnamefont {Shinaoka}},\
  and\ \bibinfo {author} {\bibfnamefont {P.}~\bibnamefont {Werner}},\ }\href
  {https://doi.org/10.1103/PhysRevB.109.165135} {\bibfield  {journal} {\bibinfo
   {journal} {Physical Review B}\ }\textbf {\bibinfo {volume} {109}},\ \bibinfo
  {pages} {165135} (\bibinfo {year} {2024})}\BibitemShut {NoStop}%
\bibitem [{\citenamefont {Środa}\ \emph {et~al.}(2025)\citenamefont {Środa},
  \citenamefont {Inayoshi}, \citenamefont {Shinaoka},\ and\ \citenamefont
  {Werner}}]{Środa-etal-2025}%
  \BibitemOpen
  \bibfield  {author} {\bibinfo {author} {\bibfnamefont {M.}~\bibnamefont
  {Środa}}, \bibinfo {author} {\bibfnamefont {K.}~\bibnamefont {Inayoshi}},
  \bibinfo {author} {\bibfnamefont {H.}~\bibnamefont {Shinaoka}},\ and\
  \bibinfo {author} {\bibfnamefont {P.}~\bibnamefont {Werner}},\ }\href
  {https://doi.org/10.1103/dxfb-b3l5} {\bibfield  {journal} {\bibinfo
  {journal} {Physical Review Letters}\ }\textbf {\bibinfo {volume} {135}},\
  \bibinfo {pages} {226501} (\bibinfo {year} {2025})}\BibitemShut {NoStop}%
\bibitem [{\citenamefont {Erpenbeck}\ \emph
  {et~al.}(2023{\natexlab{b}})\citenamefont {Erpenbeck}, \citenamefont {Lin},
  \citenamefont {Blommel}, \citenamefont {Zhang}, \citenamefont {Iskakov},
  \citenamefont {Bernheimer}, \citenamefont {Núñez-Fernández}, \citenamefont
  {Cohen}, \citenamefont {Parcollet}, \citenamefont {Waintal},\ and\
  \citenamefont {Gull}}]{Erpenbeck-etal-2023}%
  \BibitemOpen
  \bibfield  {author} {\bibinfo {author} {\bibfnamefont {A.}~\bibnamefont
  {Erpenbeck}}, \bibinfo {author} {\bibfnamefont {W.-T.}\ \bibnamefont {Lin}},
  \bibinfo {author} {\bibfnamefont {T.}~\bibnamefont {Blommel}}, \bibinfo
  {author} {\bibfnamefont {L.}~\bibnamefont {Zhang}}, \bibinfo {author}
  {\bibfnamefont {S.}~\bibnamefont {Iskakov}}, \bibinfo {author} {\bibfnamefont
  {L.}~\bibnamefont {Bernheimer}}, \bibinfo {author} {\bibfnamefont
  {Y.}~\bibnamefont {Núñez-Fernández}}, \bibinfo {author} {\bibfnamefont
  {G.}~\bibnamefont {Cohen}}, \bibinfo {author} {\bibfnamefont
  {O.}~\bibnamefont {Parcollet}}, \bibinfo {author} {\bibfnamefont
  {X.}~\bibnamefont {Waintal}},\ and\ \bibinfo {author} {\bibfnamefont
  {E.}~\bibnamefont {Gull}},\ }\href
  {https://doi.org/10.1103/PhysRevB.107.245135} {\bibfield  {journal} {\bibinfo
   {journal} {Physical Review B}\ }\textbf {\bibinfo {volume} {107}},\ \bibinfo
  {pages} {245135} (\bibinfo {year} {2023}{\natexlab{b}})}\BibitemShut
  {NoStop}%
\bibitem [{\citenamefont {Matsuura}\ \emph {et~al.}(2025)\citenamefont
  {Matsuura}, \citenamefont {Shinaoka}, \citenamefont {Werner},\ and\
  \citenamefont {Tsuji}}]{Matsuura-etal-2025}%
  \BibitemOpen
  \bibfield  {author} {\bibinfo {author} {\bibfnamefont {S.}~\bibnamefont
  {Matsuura}}, \bibinfo {author} {\bibfnamefont {H.}~\bibnamefont {Shinaoka}},
  \bibinfo {author} {\bibfnamefont {P.}~\bibnamefont {Werner}},\ and\ \bibinfo
  {author} {\bibfnamefont {N.}~\bibnamefont {Tsuji}},\ }\href
  {https://doi.org/10.1103/PhysRevB.111.155150} {\bibfield  {journal} {\bibinfo
   {journal} {Physical Review B}\ }\textbf {\bibinfo {volume} {111}},\ \bibinfo
  {pages} {155150} (\bibinfo {year} {2025})}\BibitemShut {NoStop}%
\bibitem [{\citenamefont {Matsuura}\ \emph {et~al.}(2026)\citenamefont
  {Matsuura}, \citenamefont {Shinaoka}, \citenamefont {Werner},\ and\
  \citenamefont {Tsuji}}]{Matsuura2026}%
  \BibitemOpen
  \bibfield  {author} {\bibinfo {author} {\bibfnamefont {S.}~\bibnamefont
  {Matsuura}}, \bibinfo {author} {\bibfnamefont {H.}~\bibnamefont {Shinaoka}},
  \bibinfo {author} {\bibfnamefont {P.}~\bibnamefont {Werner}},\ and\ \bibinfo
  {author} {\bibfnamefont {N.}~\bibnamefont {Tsuji}},\ }\href
  {https://arxiv.org/abs/2607.00702} {\bibinfo {title} {Weak-coupling tensor
  cross interpolation impurity solver for nonequilibrium dynamical mean-field
  theory}} (\bibinfo {year} {2026}),\ \Eprint
  {https://arxiv.org/abs/2607.00702} {arXiv:2607.00702 [cond-mat.str-el]}
  \BibitemShut {NoStop}%
\bibitem [{\citenamefont {Li}\ and\ \citenamefont
  {Eckstein}(2021)}]{Li-Eckstein-2021}%
  \BibitemOpen
  \bibfield  {author} {\bibinfo {author} {\bibfnamefont {J.}~\bibnamefont
  {Li}}\ and\ \bibinfo {author} {\bibfnamefont {M.}~\bibnamefont {Eckstein}},\
  }\href {https://doi.org/10.1103/PhysRevB.103.045133} {\bibfield  {journal}
  {\bibinfo  {journal} {Physical Review B}\ }\textbf {\bibinfo {volume}
  {103}},\ \bibinfo {pages} {045133} (\bibinfo {year} {2021})},\ \bibinfo
  {note} {arXiv:2007.12511 [cond-mat]}\BibitemShut {NoStop}%
\bibitem [{\citenamefont {Li}\ \emph {et~al.}(2020)\citenamefont {Li},
  \citenamefont {Golez}, \citenamefont {Werner},\ and\ \citenamefont
  {Eckstein}}]{Li2020}%
  \BibitemOpen
  \bibfield  {author} {\bibinfo {author} {\bibfnamefont {J.}~\bibnamefont
  {Li}}, \bibinfo {author} {\bibfnamefont {D.}~\bibnamefont {Golez}}, \bibinfo
  {author} {\bibfnamefont {P.}~\bibnamefont {Werner}},\ and\ \bibinfo {author}
  {\bibfnamefont {M.}~\bibnamefont {Eckstein}},\ }\href
  {https://doi.org/10.1103/PhysRevB.102.165136} {\bibfield  {journal} {\bibinfo
   {journal} {Phys. Rev. B}\ }\textbf {\bibinfo {volume} {102}},\ \bibinfo
  {pages} {165136} (\bibinfo {year} {2020})}\BibitemShut {NoStop}%
\bibitem [{\citenamefont {Ray}\ \emph {et~al.}(2023)\citenamefont {Ray},
  \citenamefont {Murakami},\ and\ \citenamefont {Werner}}]{Ray2023}%
  \BibitemOpen
  \bibfield  {author} {\bibinfo {author} {\bibfnamefont {S.}~\bibnamefont
  {Ray}}, \bibinfo {author} {\bibfnamefont {Y.}~\bibnamefont {Murakami}},\ and\
  \bibinfo {author} {\bibfnamefont {P.}~\bibnamefont {Werner}},\ }\href
  {https://doi.org/10.1103/PhysRevB.108.174515} {\bibfield  {journal} {\bibinfo
   {journal} {Phys. Rev. B}\ }\textbf {\bibinfo {volume} {108}},\ \bibinfo
  {pages} {174515} (\bibinfo {year} {2023})}\BibitemShut {NoStop}%
\bibitem [{\citenamefont {Eckstein}(2024)}]{Eckstein-2024}%
  \BibitemOpen
  \bibfield  {author} {\bibinfo {author} {\bibfnamefont {M.}~\bibnamefont
  {Eckstein}},\ }\href {https://doi.org/10.48550/arXiv.2410.19707} {\bibinfo
  {title} {Solving quantum impurity models in the non-equilibrium steady state
  with tensor trains}} (\bibinfo {year} {2024}),\ \bibinfo {note}
  {arXiv:2410.19707 [cond-mat]}\BibitemShut {NoStop}%
\bibitem [{\citenamefont {Kim}\ and\ \citenamefont
  {Werner}(2025)}]{Kim-Werner-2025}%
  \BibitemOpen
  \bibfield  {author} {\bibinfo {author} {\bibfnamefont {A.~J.}\ \bibnamefont
  {Kim}}\ and\ \bibinfo {author} {\bibfnamefont {P.}~\bibnamefont {Werner}},\
  }\href {https://doi.org/10.1103/PhysRevB.111.125120} {\bibfield  {journal}
  {\bibinfo  {journal} {Physical Review B}\ }\textbf {\bibinfo {volume}
  {111}},\ \bibinfo {pages} {125120} (\bibinfo {year} {2025})}\BibitemShut
  {NoStop}%
\bibitem [{\citenamefont {Geng}\ \emph {et~al.}(2025)\citenamefont {Geng},
  \citenamefont {Kim},\ and\ \citenamefont {Werner}}]{Geng-Kim-Werner-2025}%
  \BibitemOpen
  \bibfield  {author} {\bibinfo {author} {\bibfnamefont {L.}~\bibnamefont
  {Geng}}, \bibinfo {author} {\bibfnamefont {A.~J.}\ \bibnamefont {Kim}},\ and\
  \bibinfo {author} {\bibfnamefont {P.}~\bibnamefont {Werner}},\ }\href
  {https://doi.org/10.1103/jjjl-v1pj} {\bibfield  {journal} {\bibinfo
  {journal} {Physical Review B}\ }\textbf {\bibinfo {volume} {112}},\ \bibinfo
  {pages} {245119} (\bibinfo {year} {2025})}\BibitemShut {NoStop}%
\bibitem [{\citenamefont {Golez}\ \emph {et~al.}(2015)\citenamefont {Golez},
  \citenamefont {Eckstein},\ and\ \citenamefont
  {Werner}}]{Golez-Eckstein-Werner-2015}%
  \BibitemOpen
  \bibfield  {author} {\bibinfo {author} {\bibfnamefont {D.}~\bibnamefont
  {Golez}}, \bibinfo {author} {\bibfnamefont {M.}~\bibnamefont {Eckstein}},\
  and\ \bibinfo {author} {\bibfnamefont {P.}~\bibnamefont {Werner}},\ }\href
  {https://doi.org/10.1103/PhysRevB.92.195123} {\bibfield  {journal} {\bibinfo
  {journal} {Physical Review B}\ }\textbf {\bibinfo {volume} {92}},\ \bibinfo
  {pages} {195123} (\bibinfo {year} {2015})},\ \bibinfo {note}
  {arXiv:1507.07953 [cond-mat]}\BibitemShut {NoStop}%
\bibitem [{\citenamefont {Gole\ifmmode~\check{z}\else \v{z}\fi{}}\ \emph
  {et~al.}(2017)\citenamefont {Gole\ifmmode~\check{z}\else \v{z}\fi{}},
  \citenamefont {Boehnke}, \citenamefont {Strand}, \citenamefont {Eckstein},\
  and\ \citenamefont {Werner}}]{Golez2017}%
  \BibitemOpen
  \bibfield  {author} {\bibinfo {author} {\bibfnamefont {D.}~\bibnamefont
  {Gole\ifmmode~\check{z}\else \v{z}\fi{}}}, \bibinfo {author} {\bibfnamefont
  {L.}~\bibnamefont {Boehnke}}, \bibinfo {author} {\bibfnamefont {H.~U.~R.}\
  \bibnamefont {Strand}}, \bibinfo {author} {\bibfnamefont {M.}~\bibnamefont
  {Eckstein}},\ and\ \bibinfo {author} {\bibfnamefont {P.}~\bibnamefont
  {Werner}},\ }\href {https://doi.org/10.1103/PhysRevLett.118.246402}
  {\bibfield  {journal} {\bibinfo  {journal} {Phys. Rev. Lett.}\ }\textbf
  {\bibinfo {volume} {118}},\ \bibinfo {pages} {246402} (\bibinfo {year}
  {2017})}\BibitemShut {NoStop}%
\bibitem [{\citenamefont {Werner}\ and\ \citenamefont
  {Eckstein}(2013)}]{Werner-Eckstein-2013}%
  \BibitemOpen
  \bibfield  {author} {\bibinfo {author} {\bibfnamefont {P.}~\bibnamefont
  {Werner}}\ and\ \bibinfo {author} {\bibfnamefont {M.}~\bibnamefont
  {Eckstein}},\ }\href {https://doi.org/10.1103/PhysRevB.88.165108} {\bibfield
  {journal} {\bibinfo  {journal} {Physical Review B}\ }\textbf {\bibinfo
  {volume} {88}},\ \bibinfo {pages} {165108} (\bibinfo {year}
  {2013})}\BibitemShut {NoStop}%
\bibitem [{\citenamefont {Paprotzki}\ and\ \citenamefont
  {Eckstein}(2025)}]{Paprotzki-Eckstein-2025}%
  \BibitemOpen
  \bibfield  {author} {\bibinfo {author} {\bibfnamefont {E.}~\bibnamefont
  {Paprotzki}}\ and\ \bibinfo {author} {\bibfnamefont {M.}~\bibnamefont
  {Eckstein}},\ }\href {https://doi.org/10.1103/bl6z-14m9} {\bibfield
  {journal} {\bibinfo  {journal} {Physical Review B}\ }\textbf {\bibinfo
  {volume} {112}},\ \bibinfo {pages} {205126} (\bibinfo {year}
  {2025})}\BibitemShut {NoStop}%
\bibitem [{\citenamefont {Chen}\ \emph {et~al.}(2016)\citenamefont {Chen},
  \citenamefont {Cohen}, \citenamefont {Millis},\ and\ \citenamefont
  {Reichman}}]{Chen-etal-2016}%
  \BibitemOpen
  \bibfield  {author} {\bibinfo {author} {\bibfnamefont {H.-T.}\ \bibnamefont
  {Chen}}, \bibinfo {author} {\bibfnamefont {G.}~\bibnamefont {Cohen}},
  \bibinfo {author} {\bibfnamefont {A.~J.}\ \bibnamefont {Millis}},\ and\
  \bibinfo {author} {\bibfnamefont {D.~R.}\ \bibnamefont {Reichman}},\ }\href
  {https://doi.org/10.1103/PhysRevB.93.174309} {\bibfield  {journal} {\bibinfo
  {journal} {Physical Review B}\ }\textbf {\bibinfo {volume} {93}},\ \bibinfo
  {pages} {174309} (\bibinfo {year} {2016})}\BibitemShut {NoStop}%
\bibitem [{\citenamefont {Kamenev}(2023)}]{Kamenev-2023}%
  \BibitemOpen
  \bibfield  {author} {\bibinfo {author} {\bibfnamefont {A.}~\bibnamefont
  {Kamenev}},\ }\href {https://doi.org/10.1017/9781108769266} {{\selectlanguage
  {en}\bibinfo {title} {Field theory of non-equilibrium systems}}} (\bibinfo
  {year} {2023}),\ \bibinfo {note} {iSBN: 9781108769266
  9781108488259}\BibitemShut {NoStop}%
\bibitem [{\citenamefont {Künzel}\ \emph {et~al.}(2024)\citenamefont
  {Künzel}, \citenamefont {Erpenbeck}, \citenamefont {Werner}, \citenamefont
  {Arrigoni}, \citenamefont {Gull}, \citenamefont {Cohen},\ and\ \citenamefont
  {Eckstein}}]{Künzel-etal-2024}%
  \BibitemOpen
  \bibfield  {author} {\bibinfo {author} {\bibfnamefont {F.}~\bibnamefont
  {Künzel}}, \bibinfo {author} {\bibfnamefont {A.}~\bibnamefont {Erpenbeck}},
  \bibinfo {author} {\bibfnamefont {D.}~\bibnamefont {Werner}}, \bibinfo
  {author} {\bibfnamefont {E.}~\bibnamefont {Arrigoni}}, \bibinfo {author}
  {\bibfnamefont {E.}~\bibnamefont {Gull}}, \bibinfo {author} {\bibfnamefont
  {G.}~\bibnamefont {Cohen}},\ and\ \bibinfo {author} {\bibfnamefont
  {M.}~\bibnamefont {Eckstein}},\ }\href
  {https://doi.org/10.1103/PhysRevLett.132.176501} {\bibfield  {journal}
  {\bibinfo  {journal} {Physical Review Letters}\ }\textbf {\bibinfo {volume}
  {132}},\ \bibinfo {pages} {176501} (\bibinfo {year} {2024})}\BibitemShut
  {NoStop}%
\bibitem [{\citenamefont {Golež}\ \emph {et~al.}(2019)\citenamefont {Golež},
  \citenamefont {Eckstein},\ and\ \citenamefont
  {Werner}}]{Golež-Eckstein-Werner-2019}%
  \BibitemOpen
  \bibfield  {author} {\bibinfo {author} {\bibfnamefont {D.}~\bibnamefont
  {Golež}}, \bibinfo {author} {\bibfnamefont {M.}~\bibnamefont {Eckstein}},\
  and\ \bibinfo {author} {\bibfnamefont {P.}~\bibnamefont {Werner}},\ }\href
  {https://doi.org/10.1103/PhysRevB.100.235117} {\bibfield  {journal} {\bibinfo
   {journal} {Physical Review B}\ }\textbf {\bibinfo {volume} {100}},\ \bibinfo
  {pages} {235117} (\bibinfo {year} {2019})}\BibitemShut {NoStop}%
\bibitem [{\citenamefont {Inayoshi}\ \emph {et~al.}(2026)\citenamefont
  {Inayoshi}, \citenamefont {Shinaoka},\ and\ \citenamefont
  {Murakami}}]{Inayoshi-Shinaoka-Murakami-2026}%
  \BibitemOpen
  \bibfield  {author} {\bibinfo {author} {\bibfnamefont {K.}~\bibnamefont
  {Inayoshi}}, \bibinfo {author} {\bibfnamefont {H.}~\bibnamefont {Shinaoka}},\
  and\ \bibinfo {author} {\bibfnamefont {Y.}~\bibnamefont {Murakami}},\ }\href
  {https://doi.org/10.48550/arXiv.2607.11055} {\bibinfo {title} {Generalized
  keldysh formalism for nonequilibrium correlation functions and its
  application to fluctuation dynamics}} (\bibinfo {year} {2026}),\ \bibinfo
  {note} {arXiv:2607.11055 [cond-mat.str-el]}\BibitemShut {NoStop}%
\end{thebibliography}%
\end{document}